\documentclass[showpacs,preprintnumbers,amsmath,amssymb,APSl,prd,nofootinbib,superscriptaddress,12pt]{revtex4-2}
\usepackage{hyperref}
\usepackage{amsfonts}
\usepackage{bm}
\usepackage[a4paper, margin=2cm]{geometry}
\usepackage{mathrsfs}
\usepackage{soul}
\usepackage{makecell,multirow}
\usepackage{rotating}
\usepackage[utf8]{inputenc}
\usepackage{amsmath}
\usepackage{makecell}

\usepackage{amssymb}
\usepackage{tensor}
\usepackage{graphicx}
\usepackage{bigints}
\usepackage{bbm}
\usepackage{graphicx}
\usepackage{subfigure}
\usepackage{epsf}
\usepackage{bm}
\usepackage{amsmath}
\usepackage{amsfonts}
\usepackage{amssymb}
\usepackage{graphicx}
\usepackage{tabularx}
\usepackage{multirow}
\usepackage{color}%
\providecommand{\U}[1]{\protect\rule{.1in}{.1in}}

\newcommand{\ie}{\begin{equation}}
\newcommand{\fe}{\end{equation}}

\newcommand{\mincir}{\raise
-3.truept\hbox{\rlap{\hbox{$\sim$}}\raise4.truept\hbox{$<$}\ }}
\newcommand{\magcir}{\raise
-3.truept\hbox{\rlap{\hbox{$\sim$}}\raise4.truept\hbox{$>$}\ }}

\providecommand{\U}[1]{\protect\rule{.1in}{.1in}}

\usepackage{tikz,xcolor,hyperref}

\usepackage{hyperref}             
\hypersetup{
    colorlinks=true,              
    breaklinks=true,              
    citecolor=blue,               
    linkcolor=[rgb]{0,0.5,0.9},   
    urlcolor=red,                 
    filecolor=green               
}

\definecolor{lime}{HTML}{A6CE39}
\DeclareRobustCommand{\orcidicon}{%
	\begin{tikzpicture}
	\draw[lime, fill=lime] (0,0) 
	circle [radius=0.16] 
	node[white] {{\fontfamily{qag}\selectfont \tiny ID}};
	\draw[white, fill=white] (-0.0625,0.095) 
	circle [radius=0.007];
	\end{tikzpicture}
	\hspace{-2mm}
}

\foreach \x in {A, ..., Z}{%
	\expandafter\xdef\csname orcid\x\endcsname{\noexpand\href{https://orcid.org/\csname orcidauthor\x\endcsname}{\noexpand\orcidicon}}
}

\begin{document}

\title{\Large{Black Hole Gravitational Phenomena in Higher--Order Curvature--Scalar Gravity}}


\author{A. A. Ara\'{u}jo Filho\orcidB{}}
\email{dilto@fisica.ufc.br}

\affiliation{Departamento de Física, Universidade Federal da Paraíba, Caixa Postal 5008, 58051--970, João Pessoa, Paraíba,  Brazil.}
\affiliation{Departamento de Física, Universidade Federal de Campina Grande Caixa Postal 10071, 58429-900 Campina Grande, Paraíba, Brazil.}

\author{N. Heidari\orcidA{}}
\email{heidari.n@gmail.com}

\affiliation{Departamento de Física, Universidade Federal de Campina Grande Caixa Postal 10071, 58429-900 Campina Grande, Paraíba, Brazil.}
\affiliation{Center for Theoretical Physics, Khazar University, 41 Mehseti Street, Baku, AZ-1096, Azerbaijan.}
\affiliation{School of Physics, Damghan University, Damghan, 3671641167, Iran.}


\author{Iarley P. Lobo\orcidD{}}
\email{lobofisica@gmail.com (The corresponding author)}

\affiliation{Departamento de Física, Universidade Federal da Paraíba, Caixa Postal 5008, 58051--970, João Pessoa, Paraíba,  Brazil.}
\affiliation{Departamento de Física, Universidade Federal de Campina Grande Caixa Postal 10071, 58429-900 Campina Grande, Paraíba, Brazil.}


\begin{abstract}

This work aims to explore the gravitational consequences of a recently proposed black hole solution presented in the literature [Phys. Dark Univ. 50 (2025) 102061], which incorporates quantum gravitational corrections of General Relativity. We initiate our analyzes by taking into account the horizon structure, focusing on both the event and Cauchy horizons. Subsequently, we examine the quasinormal modes by considering all types of perturbations—scalar, vector, tensor, and spinorial. To strengthen these results, we also compute the time--domain for each perturbation. Next, we turn to the study of optical properties of the black hole. In particular, we investigate null geodesics, the photon sphere and its stability, as well as the corresponding black hole shadows. Following this, we analyze gravitational lensing phenomena in two regimes: the weak--field limit, utilizing the Gauss--Bonnet theorem, and the strong deflection limit, employing Tsukamoto’s approach. In addition, we confront the lensing observables with Event Horizon Telescope (EHT) data for $Sgr A^{*}$ and $M87^{*}$. Finally, constraints on the parameter $\xi$—which is introduced by higher--order curvature--scalar gravity, thereby differing from the Schwarzschild solution—are estimated using Solar System measurements such as the precession of Mercury’s orbit, gravitational light bending, and time delay (or Shapiro effect).

\end{abstract}
\maketitle

\pagebreak

\tableofcontents

\section{Introduction}

General relativity (GR) transformed our understanding of gravity by describing it as a manifestation of spacetime curvature rather than a force acting at a distance. Its predictions have been confirmed across a broad range of astrophysical and cosmological settings, from black hole mergers to the evolution of the early Universe. Nevertheless, cosmic acceleration, the gravitational effects attributed to dark matter, and the difficulty of reconciling GR with quantum theory suggest that Einstein's description may require modification at sufficiently large or small scales
\cite{Penrose1965,Hawking1973,Joshi2012,Abbott2017,Akiyama2019,Akiyama2022,Addazi:2021xuf,AlvesBatista:2023wqm}.
These open questions have motivated numerous extensions of GR, including higher-curvature theories, scalar--tensor models, and alternative geometric formulations
\cite{DeFelice2010,Sotiriou2010,Nojiri2006,Cai2016}.

A common strategy is to modify the gravitational action itself. Representative examples include $f(R)$ gravity
\cite{Nojiri2017,Nojiri2019a,DeFelice2010,Nashed2014,Faraoni2011,Capozziello2011},
$f(T)$ gravity formulated in terms of the torsion scalar
\cite{ElHanafy2016,Nashed2002,Awad2018a,Awad2018b,nashed2015kerr,Zubair2016,Cai2016},
and extensions such as $f(R,\mathcal{T})$ and $f(T,\mathcal{T})$, where $\mathcal{T}$ is the trace of the energy--momentum tensor
\cite{Zubair2016b,Harko2011,Saleem2020}.
Models based on the Gauss--Bonnet invariant, including $f(G)$ gravity, provide another important direction
\cite{Cognola2006},
while related constructions combining modified gravity with generalized electrodynamics have also been explored
\cite{rois2025charged,araujo2025gasdsadravitational}.

The motivation for these theories is closely tied to the limitations of GR in describing cosmic acceleration, galactic rotation curves, and possible exotic structures such as wormholes and modified black hole interiors
\cite{de2015constraining,Nojiri2006,DeFelice2010,Capozziello2011}.
An early influential example was Starobinsky's quadratic-curvature model of inflation
\cite{Starobinsky1979}.
Since then, $f(R)$ gravity has been widely studied in cosmology and astrophysics, including applications to massive neutron stars
\cite{astashenok2016neutron,Astashenok2014,Astashenok2013,Astashenok2017,Astashenok2015,Nashed2019}.
Its richer field equations admit broad classes of exact and numerical solutions and have been used to study dark-energy dynamics and cosmic evolution
\cite{Nashed2021,Nojiri2006,Stabile2013,Capozziello2018,Nojiri2019a,Odintsov2019a,Odintsov2019b,Shah2019,Miranda2019,Nascimento2019,Elizalde2019a,Elizalde2019b,Chen2019,Sbisa2019,Samanta2019,Bombacigno2019,Astashenok2019}.

Among the higher-curvature extensions of GR, $f(R,G)$ gravity is particularly useful because it combines the Ricci scalar with the Gauss--Bonnet invariant $G$
\cite{Nashed2023,Nojiri2021,deHaro2023,Capozziello2023,Millano2023,Nojiri2024a,Ilyas2023}.
Although the Gauss--Bonnet term is topological in four dimensions, it can contribute dynamically when coupled to additional fields. The inclusion of scalar fields introduces further degrees of freedom that naturally arise in scalar--tensor theories, string-inspired models, and inflationary scenarios
\cite{Capozziello2011,DeFelice2010,Nojiri2006}.
The resulting curvature--scalar framework therefore provides a flexible setting for investigating strong-gravity phenomena beyond GR.

A wide variety of compact object solutions has been obtained in modified gravity. Exact vacuum solutions with spherical symmetry have been reported in $f(R)$ theories
\cite{Paul:2024rto,Nashed2018a,Nashed2018c,Multamaki2006,Nashed2018b},
while the Noether symmetry approach has been used to construct both spherical and axial geometries
\cite{Capozziello2012,Capozziello2008,Capozziello2011}.
Other nontrivial black hole configurations have been obtained for specific forms of $f(R)$
\cite{Nashed2019,Nashed2019b},
together with extensive studies of static black holes
\cite{delaCruzDombriz2009,Sultana2018,Canate2018a,Yu2018,Kehagias2015,Nelson2010,Canate2016}
and neutron stars
\cite{AparicioResco2016,Staykov2018,Doneva2016,Arapoglu2011,Feng2017,Ganguly2014,Yazadjiev2016,Yazadjiev2015,Yazadjiev2014,astashenok2013further,Orellana2013,Capozziello2016,Cooney2010}.
Moreover, $f(R)$ gravity can be recast in a Brans--Dicke-like form with a scalar potential of geometric origin
\cite{OHanlon1972,Chiba2003,Chakraborty2017a,Brans1961,Chakraborty2016},
which clarifies the role of the additional scalar degree of freedom.

More recently, Ref.~\cite{Nashed:2025ebr} constructed black hole configurations in $f(R,G)$ gravity coupled to a scalar field. In this approach, the Ricci scalar, the Gauss--Bonnet invariant, and the scalar sector are incorporated into a unified framework. The resulting static and spherically symmetric geometry deforms the Schwarzschild spacetime and provides a setting in which higher-curvature and scalar-field effects can be studied explicitly
\cite{Kanti1996,Nojiri2005}.

Recent observational advances have also turned black holes into powerful laboratories for testing gravity. The first direct detections of gravitational waves by the LIGO and Virgo collaborations opened a new window onto compact object mergers
\cite{LIGOScientific:2016aoc},
while the Event Horizon Telescope provided horizon-scale images of the supermassive black holes in $M87$ and $Sgr A^{*}$
\cite{Akiyama2019,Akiyama2022}.
These developments extend tests of gravity beyond the traditional weak-field domain
\cite{019,020,021}
and make the strong-field region particularly relevant for probing modifications of GR and additional matter sectors, including nonlinear electrodynamics
\cite{araujo2025gasdsadravitational,022,mohan2025strong}.

black hole shadows provide one of the most direct probes of the near-horizon geometry. They arise from the capture and strong deflection of photons near the unstable photon region, producing a dark silhouette against the surrounding emission. Early theoretical studies date back to Bardeen
\cite{Cunningham}
and were later developed by Falcke, Melia, and Agol
\cite{Falcke:1999pj}.
The observations of $M87^{*}$ and $Sgr A^{*}$ have since motivated numerous studies testing modified black hole geometries against observational data
\cite{Khodadi:2024ubi,Allahyari:2019jqz,Nojiri:2024txy,Afrin:2021imp,Khodadi:2021gbc,Afrin:2021wlj,Khodadi:2022pqh,Kumar:2020hgm,Afrin:2022ztr,Afrin:2023uzo,Fu:2021fxn,Ghosh:2022kit,Vagnozzi:2019apd,Liu:2024lve,Nojiri:2024qgx,Nojiri:2024nlx,Bambi:2019tjh,Afrin:2024khy}.

Gravitational lensing offers a complementary probe. The formulation developed by Virbhadra and Ellis enabled the study of highly deflected images around asymptotically flat black holes
\cite{virbhadra2000schwarzschild,031}
and provided the basis for later investigations of relativistic images and strong lensing
\cite{033,034,032}.
The subject has since been extended to a broad range of spacetimes
\cite{metcalf2019strong,grespan2023strong,Kuang:2022xjp,Pantig:2022ely,Ovgun:2018fnk},
including black holes in modified gravity
\cite{chakraborty2017strong,araujo2025antisymmetric,40,nascimento2024gravitational,heidari2023gravitational}
and geometries with nontrivial topology
\cite{ovgun2019exact,38.1,38.5,38.2,38.4,Lobo:2020jfl,38.3}.
Recent work has also emphasized image distortions and the extraction of physical information from strong-lensing observables
\cite{virbhadra2024conservation}.

Perturbations provide another sensitive probe of black hole geometry. Following a disturbance, a black hole undergoes damped oscillations described by quasinormal modes
\cite{Konoplya:2007zx,Konoplya:2013rxa,karmakar2024quasinormal,Konoplya:2019hlu,karmakar2022quasinormal,Konoplya:2011qq,Kokkotas:2010zd}.
The real part of the complex frequency determines the oscillation rate, whereas the imaginary part governs the damping time. Quasinormal modes therefore encode characteristic information about the spacetime and are closely related to both the shadow and the greybody spectrum
\cite{Jusufi:2020dhz,Konoplya:2024vuj,Konoplya:2024lir}.
Although tentative evidence for their direct detection has been reported, its statistical significance remains sensitive to uncertainty treatment
\cite{Franchini:2023eda}.

In this work, we investigate the physical properties of the black hole geometry introduced in Ref.~\cite{Nashed:2025ebr} within higher--order curvature--scalar gravity. We first determine the event and Cauchy horizons and examine their dependence on the deformation parameter $\xi$. We then study null geodesics, the photon sphere, and the black hole shadow, comparing the predicted shadow size with Event Horizon Telescope observations of $Sgr A^{*}$ and $M87^{*}$.

The dynamical response of the spacetime is examined through scalar, vector, tensor, and spinor perturbations. We derive the corresponding effective potentials and calculate the quasinormal frequencies using the WKB method, supplemented by time-domain profiles. Gravitational lensing is studied in both weak- and strong-field regimes using the Gauss--Bonnet and Tsukamoto approaches, respectively. Finally, the parameter $\xi$ is constrained using Solar-System tests, including Mercury's perihelion advance, solar light deflection, and the Shapiro time delay.

\section{Higher--order curvature gravity with a scalar field}

We briefly review the theoretical framework underlying the black hole geometry studied in this work. Following Ref.~\cite{Nashed:2025ebr}, we consider an extension of general relativity in which the gravitational action depends on the Ricci scalar $R$ and the Gauss--Bonnet invariant $G$. Rather than working with a completely general function $f(R,G)$, the theory is written in the separable form
\[
f(R,G)=f(R)+f_{1}(G).
\]
The Gauss--Bonnet sector can then be represented through a scalar field $\phi_{1}$ coupled to $G$ by a function $H(\phi_{1})$. This formulation provides a convenient way to describe the additional degrees of freedom associated with the higher--curvature sector and has been discussed in Refs.~\cite{Nojiri:2024hau,Nojiri:2023qgd,Nashed:2022mij}.

The action in a $D$-dimensional spacetime is
\begin{align}
\label{g2}
S
=
\int \mathrm{d}^{D}x\,\sqrt{-g}
\left[
\frac{1}{2\kappa^{2}}f(R)
-\frac{1}{2}\partial_{\mu}\phi_{1}\partial^{\mu}\phi_{1}
+V(\phi_{1})
+H(\phi_{1})G
\right],
\end{align}
where $V(\phi_{1})$ is the scalar-field potential and
\begin{align}
\label{eq:GB}
G
=
R^{2}
-4R_{\mu\nu}R^{\mu\nu}
+R_{\mu\nu\rho\sigma}R^{\mu\nu\rho\sigma}
\end{align}
is the Gauss--Bonnet invariant. Although the integral of $G$ is topological in four dimensions, the coupling $H(\phi_{1})G$ contributes nontrivially to the dynamics whenever $H(\phi_{1})$ is not constant. In other words, it is the scalar-dependent coupling that activates the Gauss--Bonnet sector in four-dimensional spacetime. Matter fields are assumed to couple minimally to the metric and not directly to $\phi_{1}$, thereby avoiding an additional long-range interaction.

Varying the action with respect to the scalar field gives
\begin{align}
\label{g3}
\nabla^{2}\phi_{1}
-
V'(\phi_{1})
+
H'(\phi_{1})G
=
0.
\end{align}
Variation with respect to $g_{\mu\nu}$ produces the modified gravitational field equations. These equations contain the usual $f(R)$ contributions together with terms arising from the scalar kinetic sector, its potential, and derivatives of the coupling function $H(\phi_{1})$. Since our purpose is to investigate the physical properties of the resulting black hole rather than to reproduce its full derivation, one can refere to the complete form, together with the required Bianchi identities and the corresponding trace equation given in Ref.~\cite{Nashed:2025ebr}.

An important limiting case follows when the scalar--Gauss--Bonnet coupling is constant. In that situation, the Gauss--Bonnet contribution reduces to a boundary term and drops out of the four--dimensional field equations. The nontrivial corrections considered in the following, arise from the radial variation of the scalar sector and its coupling to the curvature.

\subsection{Static and spherically symmetric configuration}
\label{S3}

We now restrict the theory to a static and spherically symmetric spacetime, described by
\begin{align}
\label{met12}
\mathrm{d}s^{2}
=
-D_{1}(r)\mathrm{d}t^{2}
+
\frac{\mathrm{d}r^{2}}{D_{2}(r)}
+
r^{2}
\left(
\mathrm{d}\theta^{2}
+
\sin^{2}\theta\,\mathrm{d}\phi^{2}
\right).
\end{align}
Here, $D_{1}(r)$ and $D_{2}(r)$ are independent metric functions. The Ricci scalar corresponding to Eq.~\eqref{met12} is
\begin{align}
\label{Ricci}
R
=
\frac{
r^{2}D_{2}D_{1}'^{\,2}
-r^{2}D_{1}D_{1}'D_{2}'
-2r^{2}D_{1}D_{2}D_{1}''
-4rD_{1}\left(D_{2}D_{1}'-D_{1}D_{2}'\right)
+4D_{1}^{2}\left(1-D_{2}\right)
}{
2r^{2}D_{1}^{2}
},
\end{align}
where a prime denotes differentiation with respect to $r$.

After imposing spherical symmetry, the modified field equations reduce to a system of radial differential equations involving
\[
D_{1}(r),~
D_{2}(r),~
F(r)=f_{R},~
H(r),~\text{and}~
\phi_{1}(r).
\]
Excluding the trace relation, only three independent equations remain for these five unknown functions. The system is therefore underdetermined, and two additional choices are required to close it.

Ref.~\cite{Nashed:2025ebr} follows a reconstruction approach. Instead of fixing $f(R)$, $H(\phi_{1})$, and $V(\phi_{1})$ in advance and then solving for the geometry, one first specifies the two metric functions. The remaining functions of the theory are subsequently reconstructed from the modified field equations. In this sense, the resulting geometry is not introduced independently of the gravitational model: it is supported by a definite scalar profile, curvature coupling, gravitational function, and potential determined by the field equations.

The general relativistic limit is recovered when the additional curvature and scalar contributions are switched off. In this case, the gravitational Lagrangian reduces to $f(R)=R$, and the two metric functions coincide with the Schwarzschild form as $D_{1}(r)=D_{2}(r)=1-\frac{2M}{r}$. 
Nonzero contributions from the higher--curvature and scalar sectors therefore generate departures from the Schwarzschild geometry which will be discussed as follows.

\subsection{Black hole solution}

After imposing static spherical symmetry, the reduced field equations involve five unknown radial functions,
$D_{1}(r),~ D_{2}(r),~ F(r)=f_{R},~ H(r),$ and $ \phi_{1}(r),$ whereas only three independent differential equations remain after excluding the trace relation. The system is therefore underdetermined and requires two additional conditions.

Following the reconstruction approach of Ref.~\cite{Nashed:2025ebr}, the metric functions $D_{1}(r)$ and $D_{2}(r)$ are specified first, while the scalar profile and the functions defining the underlying gravitational theory are subsequently reconstructed from the reduced field equations. The adopted ansatz describes a deformation of the Schwarzschild geometry controlled by the parameter $\xi$, with the Schwarzschild solution recovered smoothly in the limit $\xi\to0$. In particular, Ref.~\cite{Nashed:2025ebr} considered
\begin{align}
\label{sol}
D_{1}(r)
&=
1-\frac{c}{r}+\frac{\xi}{r^{2}},
&
D_{2}(r)
&=
1-\frac{c}{r}+\frac{c,\xi^{3/2}}{r^{4}},
\end{align}
where $c=2M$ is associated with the gravitational mass, while $\xi$ has dimensions of length squared and measures the departure from the Schwarzschild spacetime.

Once Eq.~\eqref{sol} is substituted into the reduced field equations, the corresponding functions $F(r)=f_{R}$, $H(r)$, $\phi_{1}(r)$, $f(r)$, and $V(r)$ can be reconstructed consistently. Their explicit forms contain lengthy nested expressions and are reported in Ref.~\cite{Nashed:2025ebr}. The reconstructed scalar field must also satisfy the associated reality condition; otherwise, the resulting branch would contain ghost-like degrees of freedom.

The solution is static, spherically symmetric, and asymptotically flat. Moreover, the Schwarzschild geometry is recovered smoothly in the limit $\xi\to0$. The correction terms decay more rapidly than the Schwarzschild mass term, so their effects are expected to be most relevant in the strong-field region. The different radial orders appearing in $D_{1}(r)$ and $D_{2}(r)$ should not be regarded as unrelated phenomenological corrections. They characterize the two independent metric functions permitted by spherical symmetry, while the remaining quantities of the theory are reconstructed so that the complete configuration satisfies the modified field equations.

In the present work, we take this reconstructed solution as the background geometry and investigate its causal, perturbative, and optical properties.

\section{The black hole spacetime}

Using $c=2M$, the solution in Eq.~\eqref{sol} is written as
\ie
\label{mainmetric}
\mathrm{d}s^{2}
=
-\left(
1-\frac{2M}{r}+\frac{\xi}{r^{2}}
\right)\mathrm{d}t^{2}
+
\frac{\mathrm{d}r^{2}}
{
1-\frac{2M}{r}+\frac{2M\xi^{3/2}}{r^{4}}
}
+
r^{2}
\left(
\mathrm{d}\theta^{2}
+
\sin^{2}\theta\,\mathrm{d}\phi^{2}
\right).
\fe
For convenience, we define
\ie
A(r,\xi)
\equiv
1-\frac{2M}{r}
+\frac{\xi}{r^{2}},
\fe
and
\ie
B(r,\xi)
\equiv
1-\frac{2M}{r}
+\frac{2M\xi^{3/2}}{r^{4}}.
\fe

The temporal metric function $A(r,\xi)$ resembles that of the Reissner--Nordstr\"om spacetime under the formal identification
\[
\xi\leftrightarrow Q^{2}.
\]
This resemblance concerns only the radial form of the metric component, since $\xi$ represents a higher--curvature deformation rather than an electric charge. In contrast, the radial function $B(r,\xi)$ shows a close analogy with Loop Quantum Gravity--inspired black holes \cite{Lewandowski:2022zce,Nozari:2025veb} under the formal correspondence
\ie
\alpha G^{2}M^{2}
\leftrightarrow
2M\xi^{3/2},
\fe
where
\ie
\alpha
=
16\sqrt{3}\,\pi\,\gamma^{3}l_{P},
\fe
with $l_{P}$ denoting the Planck length and $\gamma$ the Barbero--Immirzi parameter.

The geometry therefore combines an RN-like temporal component with an LQG-inspired radial component. Nevertheless, both corrections originate from the same higher--order curvature--scalar construction and are governed by the single parameter $\xi$. Since $A(r,\xi)$ and $B(r,\xi)$ have different radial dependences, they can affect the causal structure, perturbative dynamics, and optical behavior of the spacetime in distinct ways. These effects are examined in the following sections.

To begin, we examine the event horizon $r_{h}$. It is determined by imposing $1/g_{rr} = 0$ and solving the resulting equation. This, therefore, yields
\ie
\begin{split}
\label{evenhhooo}
r_{h} = & \, \frac{M}{2} + \frac{1}{2} \sqrt{\left(\frac{2}{3}\right)^{2/3} \sqrt[3]{\eta }+M^2+\frac{4 \sqrt[3]{\frac{2}{3}} M \xi ^{3/2}}{\sqrt[3]{\eta }}}\\
& +\frac{1}{2} \sqrt{-\left(\frac{2}{3}\right)^{2/3} \sqrt[3]{\eta }+2 M^2+\frac{2 M^3}{\sqrt{\left(\frac{2}{3}\right)^{2/3} \sqrt[3]{\eta }+M^2+\frac{4 \sqrt[3]{\frac{2}{3}} M \xi ^{3/2}}{\sqrt[3]{\eta }}}}-\frac{4 \sqrt[3]{\frac{2}{3}} M \xi ^{3/2}}{\sqrt[3]{\eta }}}\\
\approx & \, \, 2 M-\frac{\xi ^{3/2}}{4 M^2}.
\end{split}
\fe
For the above short notation, we assumed $\xi$ to be small; here, $\eta \equiv 9 M^{3}\xi^{3/2} + \sqrt{81 M^{6}\xi^{3} - 96 M^{3}\xi^{9/2}}$. As will be shown in the following sections, the assumption that $\xi$ is small is fully supported by the bounds derived in the final section of the manuscript. Notably, the horizon radius $r_{h}$ in this case is smaller than in the Schwarzschild solution. Since $\xi$ carries dimensions of $[\mathrm{L}^{2}]$, it must be strictly positive; otherwise, $r_{h}$ would become imaginary, as is it straightforwardly verified from Eq. (\ref{evenhhooo}).

\begin{figure}
    \centering
    \includegraphics[scale=0.55]{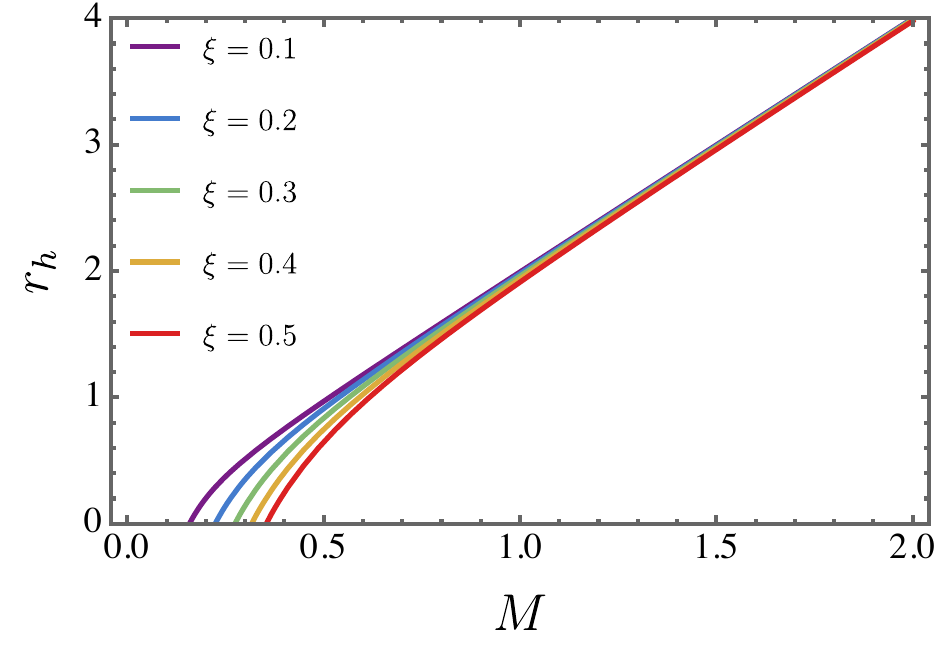}
    \caption{The dependence of the event horizon $r_{h}$ on the mass $M$ is illustrated for various configurations of the deformed parameter $\xi$.}
    \label{plotevent}
\end{figure}

To corroborate our results, we examine their behavior for different configurations of $\xi$ and the mass $M$. For this purpose, we provide both a plot and a table to offer qualitative and quantitative perspectives. Fig.~\ref{plotevent} illustrates the dependence of the event horizon $r_{h}$ on the mass $M$ for various values of the deformed parameter $\xi$, while Table~\ref{tabevent} reports the corresponding quantitative values for different choices of $M$ and $\xi$. In general lines, this latter parameter tends to decrease the magnitude of the event horizon, as anticipated by the shape of Eq. (\ref{evenhhooo}).

\begin{figure}
    \centering
    \includegraphics[scale=0.55]{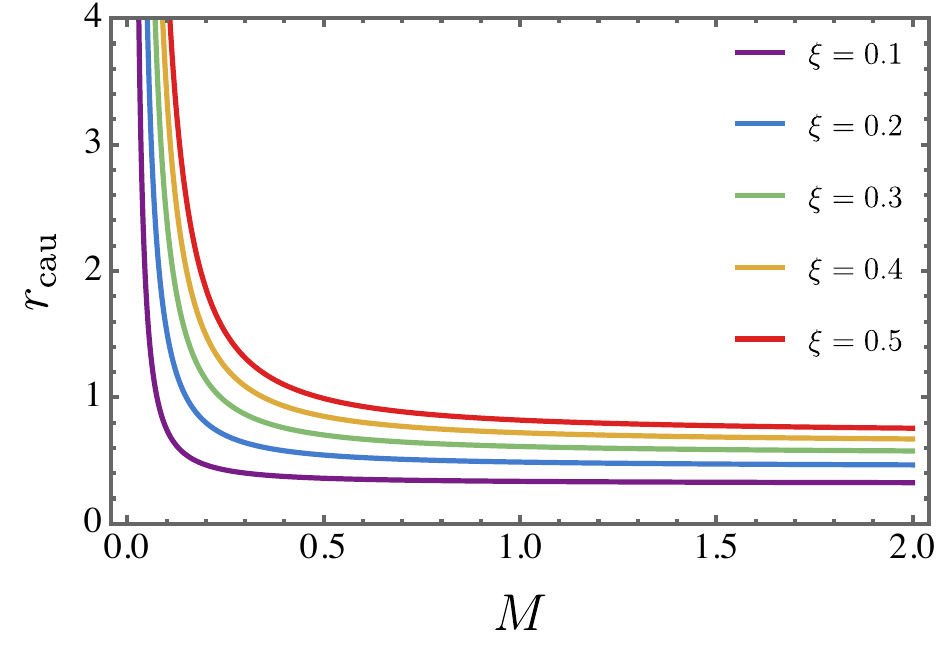}
    \caption{The behavior of the Cauchy horizon $r_{\text{cau}}$ as a function of the mass $M$ is shown for different configurations of the deformed parameter $\xi$.}
    \label{plotcau}
\end{figure}

Another comment is worthy to be pointed out: besides the solution ascribed to the event horizon, there also exists another one (real and positive defined), which is related to the Cauchy horizon, $r_{\text{cau}}$:
\ie
r_{\text{cau}} \approx \, \, \sqrt{\xi } + \frac{\xi }{6 M}+ \frac{\xi ^{3/2}}{12 M^2} .
\fe
Unlike the behavior of $r_{h}$, the Cauchy horizon $r_{\text{cau}}$ increases as either $\xi$ or $M$ increase, as shown in Fig. \ref{plotcau} and Tab. \ref{tabecau}.

\begin{table}[!h]
\begin{center}
\begin{tabular}{c c c || c c c} 
 \hline\hline\hline
 $\xi$ & $M$ &  $r_{h}$ & $\xi$ & $M$ &  $r_{h}$ \\ [0.2ex] 
 \hline
 0.1  & 1.0 & 1.99209 & 0.1  & 1.1 & 2.19347  \\ 

 0.2  & 1.0 & 1.97764 & 0.1  & 1.2 & 2.39451  \\
 
 0.3  & 1.0 & 1.95892 & 0.1  & 1.3 & 2.59532  \\
 
 0.4  & 1.0 & 1.93675 & 0.1  & 1.4 & 2.79597  \\
 
 0.5  & 1.0 & 1.91161 & 0.1  & 1.5 & 2.99649  \\ 
 [0.2ex] 
 \hline \hline \hline
\end{tabular}
\caption{\label{tabevent} The quantitative values of the event horizon are presented for several configurations of $M$ and $\xi$.}
\end{center}
\end{table}

\begin{table}[!h]
\begin{center}
\begin{tabular}{c c c || c c c} 
 \hline\hline\hline
 $\xi$ & $M$ &  $r_{\text{cau}}$ & $\xi$ & $M$ &  $r_{\text{cau}}$ \\ [0.2ex] 
 \hline
 0.1  & 1.0 & 0.33553 & 0.1  & 1.1 & 0.333557  \\ 

 0.2  & 1.0 & 0.48800 & 0.1  & 1.2 & 0.331947  \\
 
 0.3  & 1.0 & 0.611416 & 0.1  & 1.3 & 0.330608  \\
 
 0.4  & 1.0 & 0.720204 & 0.1  & 1.4 & 0.329477  \\
 
 0.5  & 1.0 & 0.819903 & 0.1  & 1.5 & 0.328510  \\ 
 [0.2ex] 
 \hline \hline \hline
\end{tabular}
\caption{\label{tabecau} Table entries report the numerical values of the Cauchy horizon $r_{\text{cau}}$ corresponding to various choices of $M$ and $\xi$.}
\end{center}
\end{table}


\subsection{Domain where the spacetime is a black hole}

The existence of a black hole structure in the present geometry is determined by the behavior
of the radial metric function $B(r,\xi)$, since the horizon locations follow from the condition
$g_{rr}^{-1}=B(r,\xi)=0$. Using the explicit form of the metric, this condition leads to
\begin{equation}
B(r,\xi)=1-\frac{2M}{r}+\frac{2M\xi^{3/2}}{r^{4}}=0.
\end{equation}

Introducing the dimensionless variables $x=r/M$ and $z=\xi/M^{2}$, the horizon equation can
be rewritten as
\begin{equation}
x^{4}-2x^{3}+2z^{3/2}=0.
\end{equation}
For a given value of $z$, the spacetime represents a black hole if this equation admits at
least one real and positive root. When two such roots exist, the larger one corresponds to the
event horizon $r_{h}$, while the smaller one defines the Cauchy horizon $r_{cau}$.

By solving this equation numerically, one finds that the black hole configuration persists
within a finite interval of the deformation parameter $\xi$. In particular, the existence of
real and positive horizons requires
\begin{equation}
0\leq \frac{\xi}{M^{2}}\leq \left(\frac{351}{256}\right)^{2/3},
\end{equation}
within which the condition $r_{h}>0$ is always satisfied. In this domain, the spacetime
maintains a well-defined black hole structure with $r_{h}>r_{cau}$. Outside this interval,
the horizons cease to exist as real solutions, and the geometry no longer represents a black
hole.

In the perturbative regime $\xi\ll 1$, the exact numerical solution for the event
horizon smoothly reduces to the approximate expression
\begin{equation}
r_{h}\approx 2M-\frac{\xi^{3/2}}{4M^{2}},
\end{equation}
which is the form adopted for the sake of a more compact notation.


\section{Quasinormal mode spectra }

Here, it is devoted to the study of quasinormal oscillations of the black hole background. The analysis is performed for a wide range of perturbations, including scalar, vector, tensor, and spinor fields. The procedure begins with the field equations in the curved geometry, which are decomposed through a separation of variables to yield the radial master equations. From these, the effective potential associated with each perturbing field is identified. The resulting Schrödinger--like equations are then solved by applying the WKB approximation, allowing, therefore, the computation of the complex quasinormal frequencies that characterize the dissipative response of the spacetime.


\subsection{Scalar field fluctuations }

One of the most commonly used techniques for calculating quasinormal mode spectra is the Wentzel--Kramers--Brillouin (WKB) method. This semi--analytic approach, first developed by Will and Iyer \cite{iyer1987black,iyer1987black1} and, after that, generalized to higher orders by Konoplya \cite{konoplya2003quasinormal}, provided a reliable way to approximate the solutions by taking into account their perturbation versions. In the present investigation, we focus on scalar field fluctuations and examine the propagation of the Klein--Gordon field for the metric tensor of Eq. (\ref{mainmetric}). Solving the resulting radial equation within this framework yields the complex quasinormal frequencies that encode the characteristic damped oscillations of the system. To do so, let us, initially, start with
\ie
\frac{1}{\sqrt{-g}}\partial_{\mu}(g^{\mu\nu}\sqrt{-g} \,\partial_{\nu} \, \phi) = 0.\label{kkgg}
\fe
At this point, it is imporntat to mention that while the inclusion of backreaction effects would bring about a more complete description of the system, such contributions lie beyond the scope of this work. Here, the scalar field is regarded purely as a probe, evolving on a fixed spacetime geometry without influencing the background metric. Under this assumption, the dynamics reduce to Eq.~(\ref{kkgg}), which serves as the master equation of such a field, governing therefore its corresponding evolution. In other words, Eq. (\ref{kkgg}) can be decomposed as
\ie
\label{decnmmm}
\begin{split}
-& \frac{1}{\left(1-\frac{2M}{r}+\frac{\xi}{r^{2}}\right)} \frac{\partial^{2} \phi}{\partial t^{2}} + \frac{1}{r^{2}} \left\{  \frac{\partial}{\partial r} \left[  \left( 1-\frac{2M}{r}+\frac{2M\xi^{3/2}}{r^{4}}  \right) \, r^{2}  \frac{\partial \phi}{\partial r}  \right]  \right\} \\  + & \frac{1}{r^{2} \sin \theta}  \left[  \frac{\partial }{\partial \theta} \left( \sin \theta \frac{\partial}{\partial \theta} \phi   \right)        \right] 
 +  \frac{1}{r^{2} \sin^{2}\theta}\frac{\partial^{2} \phi}{\partial \varphi^{2}} = 0.
\end{split}
\fe

Taking advantage of the spherical symmetry of the spacetime, the scalar field can be expanded into angular and radial parts through a variable separation procedure. This step transforms the field equation into an ordinary differential equation for the radial component. To achieve this, we write

\ie
\label{fggggg}
\phi(t, r, \theta, \varphi) = \sum_{\ell=0}^{\infty} \sum_{m=-\ell}^{\ell}  \mathcal{Y}_{\ell m}(\theta, \varphi) \frac{e^{-i\omega t}\psi(r)}{r}.
\fe
By taking into account the above expression and expressing the angular part via the spherical harmonics $\mathcal{Y}_{\ell m}(\theta,\varphi)$, the field equation simplifies considerably.




Moreover, considering the tortoise coordinate defined as $r^{*} = 1/ \sqrt{A(r,\xi) B(r,\xi)}$, the system simplifies to a single radial equation that admits a Schrödinger--like representation as


\ie
\frac{\partial^{2} \psi}{\partial r^{*2}} - \left[  \omega^{2} - \mathcal{V}^{s}(r,\ell,\xi)\right]\psi = 0.\label{sserffs}
\fe

where the effective potential reads the following expression
\ie
\begin{split}
\mathcal{V}^{s}(r,\ell,\xi) & = A(r,\xi)\left[\frac{{\ell(\ell + 1)}}{{{r^2}}} + \frac{1}{{r\sqrt {{A(r,\xi)}{B(r,\xi)^{ - 1}}} }}\frac{\mathrm{d}}{{\mathrm{d}r}}\sqrt {{A(r,\xi)}B(r,\xi)}\right]\\
& = \frac{\left(-2 M r+\xi +r^2\right) \left(\frac{\ell (\ell+1)}{r^2}+\frac{M^2 \left(10 \xi ^{3/2} r-4 r^4\right)+M \left(3 \xi +2 r^2\right) \left(r^3-2 \xi ^{3/2}\right)+\xi  \left(-r^4\right)}{r^6 \sqrt{\frac{-2 M r+\xi +r^2}{2 M \left(\xi ^{3/2}-r^3\right)+r^4}} \sqrt{\left(-2 M r+\xi +r^2\right) \left(2 M \left(\xi ^{3/2}-r^3\right)+r^4\right)}}\right)}{r^2}.
\end{split}
\fe

\begin{figure}
    \centering
    \includegraphics[scale=0.51]{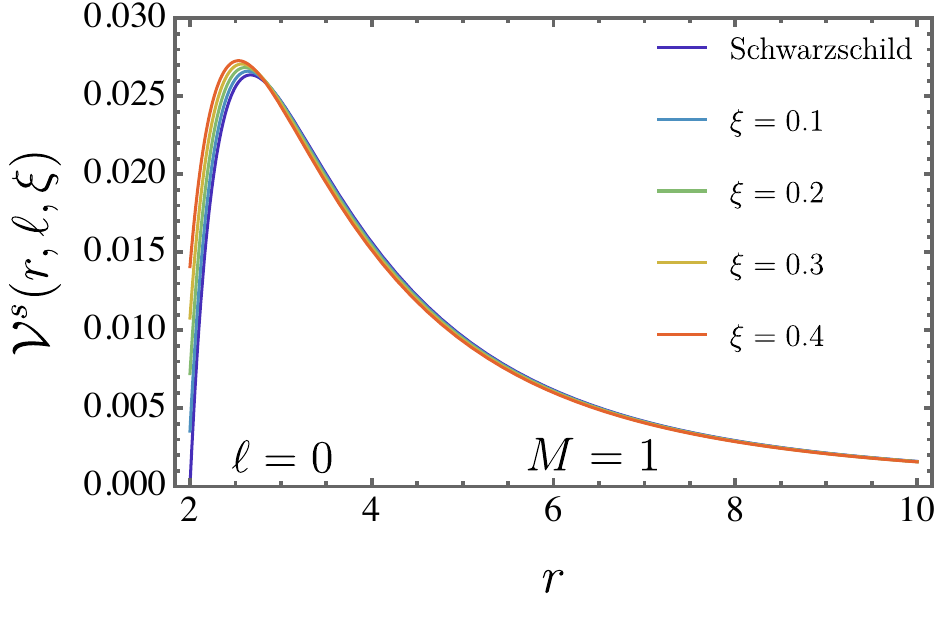}
    \includegraphics[scale=0.51]{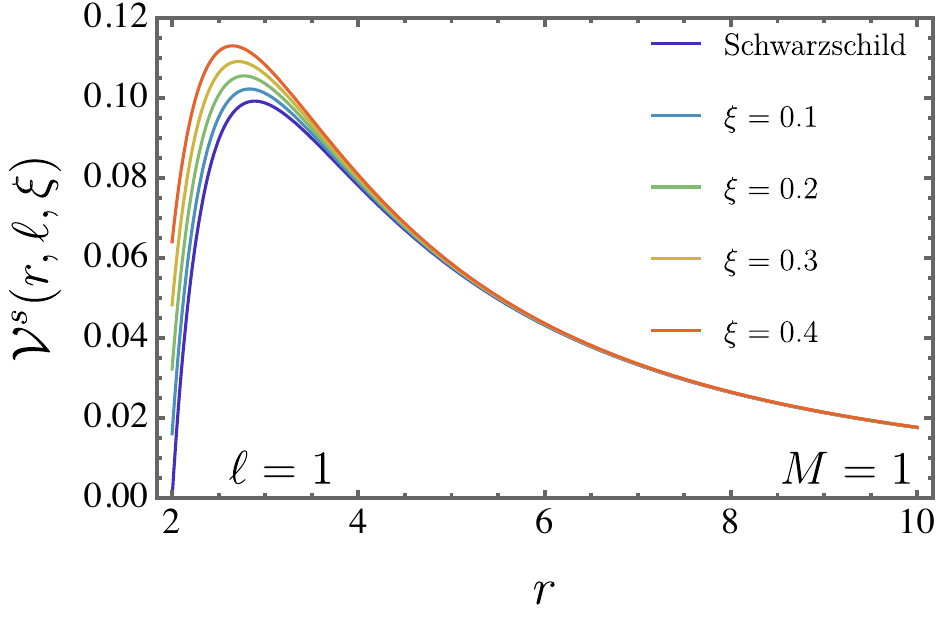}
    \includegraphics[scale=0.51]{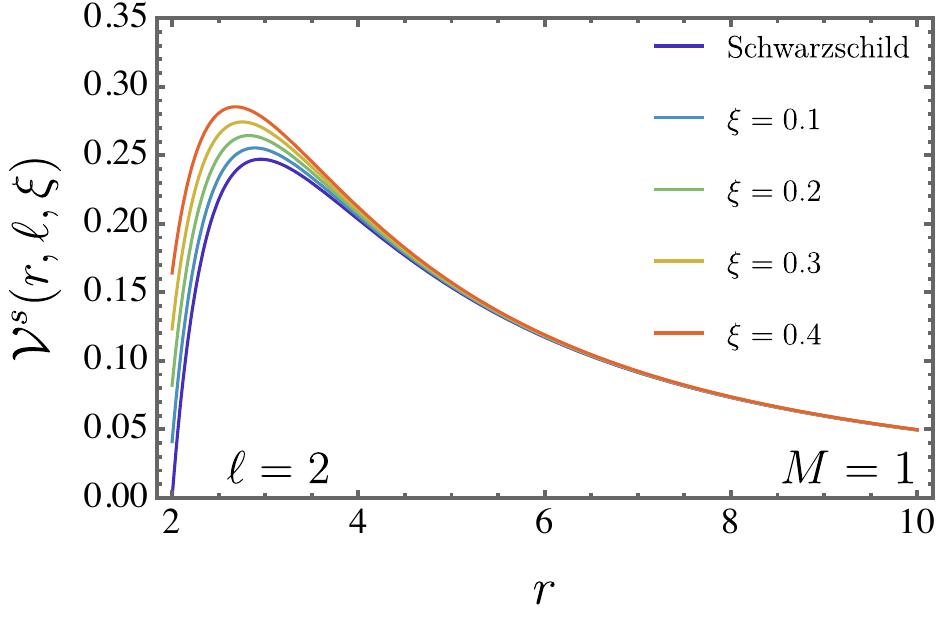}
    \caption{The behavior of the scalar potential $\mathcal{V}^{s}(r,\ell,\xi)$ is presented as a function of radial coordinate $r$, exploring various combinations of parameter $\xi$ and the multipole number $\ell$ to highlight how both quantities influence its overall shape and amplitude.}
    \label{solkkmm}
\end{figure}

In the limit $\xi \to 0$, we recover the familiar Schwarzschild result for the potential, as expected. Fig.~\ref{solkkmm} shows the behavior of $\mathcal{V}^{s}(r,\ell,\xi)$ plotted against  $r$ for a variety multipole numbers $\ell$ and $\xi$. Increasing $\xi$ (or $\ell$) leads to a taller potential barrier. This behavior has a direct impact on the resulting quasinormal mode spectrum and also affects the time--domain profiles, as will be demonstrated in the following analysis. It is also worth noting that, due to the asymptotically flat nature of the spacetime, the potential vanishes as $r \to \infty$.


The quasinormal mode spectrum is obtained by constructing the solution in the vicinity of the maximum of the effective potential, which plays naturally an essential role in identifying the classical turning points. Expanding the wave function around this peak and applying a WKB matching procedure leads to a semi--analytic expression for the complex frequencies that describe black hole ringdown. Using Konoplya’s higher--order extension of the WKB method, the frequencies satisfy
\ie
\frac{i\bigl(\omega^{2}-V_{0}\bigr)}{\sqrt{-2V_{0}^{''}}}-\sum_{j=2}^{6}\Lambda_{j}=n+\frac{1}{2},
\fe
where $V_{0}^{''}$ is the second derivative of the potential evaluated at its maximum $r_{0}$. The terms $\Lambda_{j}$ represent successive higher--order corrections, constructed from $V_{0}$ and its derivatives, and are crucial for improving the precision of the computed quasinormal frequencies.

For clarity, we quantify the shift of the quasinormal frequencies from the Schwarzschild case by defining the relative deviations of their real and imaginary parts as
\ie
\Delta\omega_{R/I}=\frac{|\omega_{R/I}^{Sch}|-|\omega_{R/I}(\xi)|}{|\omega_{R/I}^{Sch}|}\times 100
\fe
where the subscripts $R$ and $I$ refer to the real and imaginary parts, respectively, while $\omega_{R/I}^{Sch}$ denotes the corresponding Schwarzschild value obtained at $\xi=0$.

\begin{figure}
    \centering
    \includegraphics[scale=0.32]{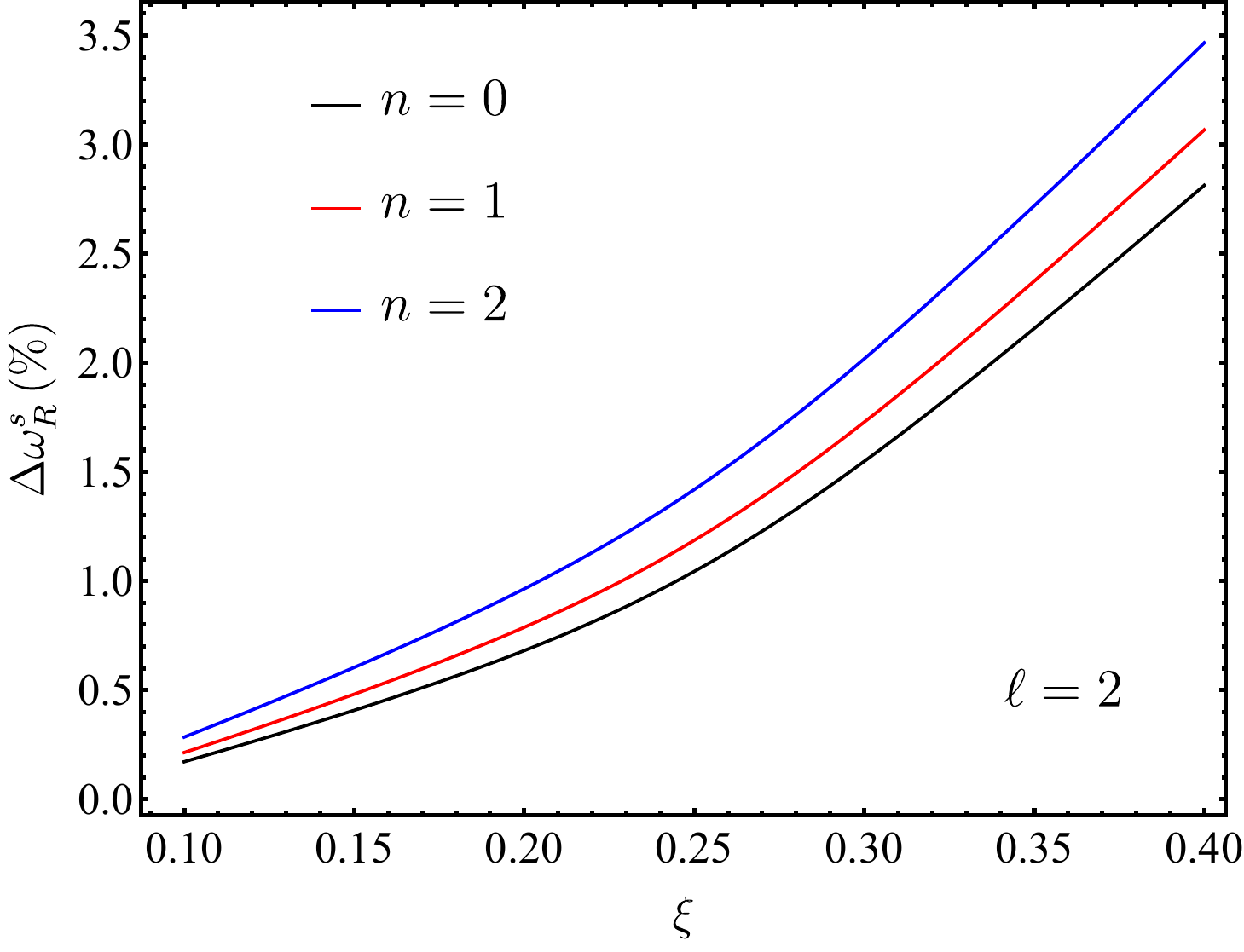}
    \includegraphics[scale=0.33]{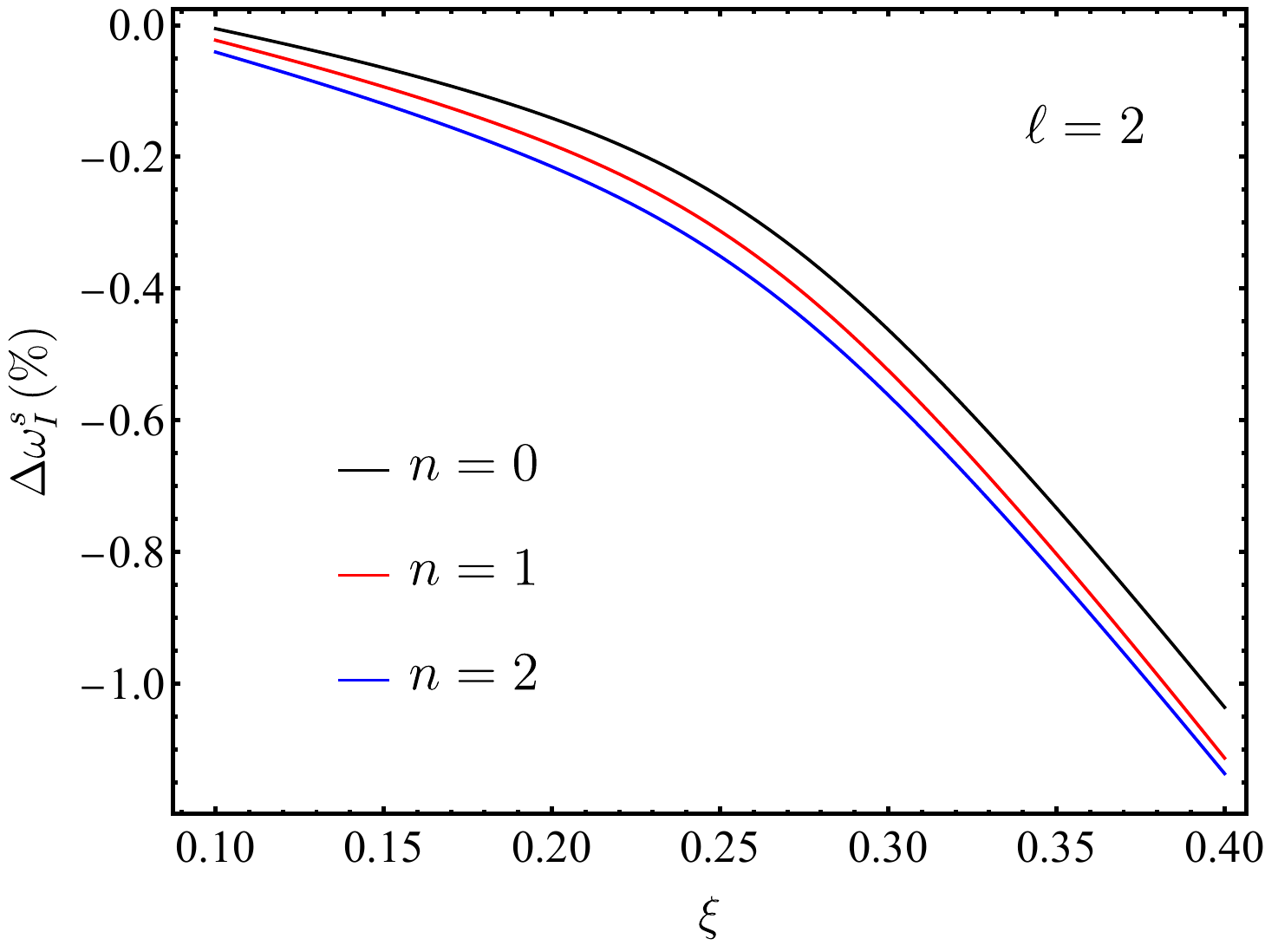}
    \caption{Relative percentage deviations of the real (left panel) and imaginary (right panel) parts of the scalar quasinormal frequencies from their Schwarzschild values are shown as functions of $\xi$ for $\ell=2$ and corresponding overtones.}
    \label{fig:delscalar}
\end{figure}

Fig.~\ref{solkkmm} illustrates the scalar potential $\mathcal{V}^{s}(r,\ell,\xi)$ plotted against the radial coordinate $r$ for several combinations of $\ell$ and $\xi$. The figure reveals that larger values of either parameter lead to a taller and broader potential barrier, effectively shifting the corresponding peak. Since the background spacetime is asymptotically flat, the potential smoothly decays to zero as $r \to \infty$, as expected.
Complementing this analysis, Table \ref{Tab:scalarQNMS} lists the quasinormal frequencies obtained for a range of $\xi$, and $\ell$. Moreover, the real and imaginary relative deviations are shown in Fig.~\ref{fig:delscalar}. Across all cases studied, the data indicate that increasing $\xi$ systematically  increases the wave frequency and reduces the damping rate, leading to longer--lived oscillations.

\begin{table}[]
\caption{The quasinormal mode spectra related to scalar perturbations for the case $\ell = 1,2,3$ and corresponding overtone number $n$, presented as a function of the deformation parameter $\xi$.}
\label{Tab:scalarQNMS}

\begin{tabular}{|cc|c|c|c|c|}
\hline
\multicolumn{2}{|c|}{} &
  $\xi = 0.0$ &
  $\xi = 0.1$ &
  $\xi = 0.2$ &
  $\xi = 0.3$ \\ \hline
\multicolumn{1}{|c|}{\multirow{2}{*}{\begin{tabular}[c]{@{}c@{}}$l=1$\end{tabular}}} &
  $n = 0$ &
  {$0.291114-0.0980014i$} &
  {$0.291622-0.0979666i$} &
  {$0.293090-0.0977933i$} &
  {$0.295561-0.0974385i$} \\ \cline{2-6} 
\multicolumn{1}{|c|}{} &
  $n = 1$ &
  {$0.262212-0.307432i$} &
  {$0.262885-0.307234i$} &
  {$0.264517-0.306600i$} &
  {$0.267050-0.305436i$} \\ \hline
\multicolumn{1}{|c|}{\multirow{3}{*}{\begin{tabular}[c]{@{}c@{}}$l=2$\end{tabular}}} &
  $n = 0$ &
  {$0.483211-0.0968049i$} &
  {$0.484038-0.0967998i$} &
  {$0.486499-0.0966682i$} &
  {$0.490690-0.0963572i$} \\ \cline{2-6} 
\multicolumn{1}{|c|}{} &
  $n = 1$ &
  {$0.463192-0.295810i$} &
  {$0.464180-0.295743i$} &
  {$0.466838-0.295273i$} &
  {$0.471194-0.294259i$} \\ \cline{2-6} 
\multicolumn{1}{|c|}{} &
  $n = 2$ &
  {$0.431660-0.503433i$} &
  {$0.432887-0.503228i$} &
  {$0.435819-0.502352i$} &
  {$0.440370-0.500605i$} \\ \hline
\multicolumn{1}{|c|}{\multirow{4}{*}{\begin{tabular}[c]{@{}c@{}}$l=3$\end{tabular}}} &
  $n = 0$ &
  {$0.675206-0.0965121i$} &
  {$0.676351-0.0965133i$} &
  {$0.679793-0.0963893i$} &
  {$0.685676-0.0960846i$} \\ \cline{2-6} 
\multicolumn{1}{|c|}{} &
  $n = 1$ &
  {$0.660414-0.292344i$} &
  {$0.661691-0.292318i$} &
  {$0.665302-0.291903i$} &
  {$0.671340-0.290940i$} \\ \cline{2-6} 
\multicolumn{1}{|c|}{} &
  $n = 2$ &
  {$0.634839-0.494118i$} &
  {$0.636336-0.494004i$} &
  {$0.640215-0.493225i$} &
  {$0.646474-0.491541i$} \\ \cline{2-6}
\multicolumn{1}{|c|}{} &
  $n = 3$ &
  {$0.602182-0.701053i$} &
  {$0.603949-0.700803i$} &
  {$0.608158-0.699626i$} &
  {$0.614683-0.697225i$} \\ 
\hline
\end{tabular}

\end{table}


\subsection{Vector field fluctuations }

The electromagnetic perturbations are analyzed within the tetrad formalism, following Refs.~\cite{chandrasekhar1998mathematical,Bouhmadi-Lopez:2020oia,Gogoi:2023kjt}. We introduce a set of tetrad vectors $e_{\mu}^{a}$ adapted to the background metric $g_{\mu\nu}$, satisfying
\ie
\begin{split}
&e^{a}_{\mu} e^{\mu}_{b} = \delta^{a}_{b},
\qquad
e^{a}_{\mu} e^{\nu}_{a} = \delta^{\nu}_{\mu},\\
&e^{a}_{\mu} = g_{\mu\nu}\eta^{ab}e^{\nu}_{b},
\qquad
g_{\mu\nu} = \eta_{ab}e^{a}_{\mu}e^{b}_{\nu}
= e_{a\mu}e^{a}_{\nu}.
\end{split}
\fe

Applying the Bianchi identity,
\[
\mathcal{F}_{[ab|c]}=0,
\]
to the electromagnetic field--strength tensor gives
\begin{align}
\left(r\sqrt{A(r,\xi)}\,\mathcal{F}_{t\phi}\right)_{,r}
+r\sqrt{B(r,\xi)}\,\mathcal{F}_{\phi r,t}
&=0,
\label{ffedefm1}\\
\left(r\sqrt{A(r,\xi)}\,\mathcal{F}_{t\phi}\sin\theta\right)_{,\theta}
+r^{2}\sin\theta\,\mathcal{F}_{\phi r,t}
&=0.
\label{ffedefm2}
\end{align}

The corresponding source--free Maxwell equation is
\ie
\eta^{bc}\left(\mathcal{F}_{ab}\right)_{|c}=0.
\fe
Here, the vertical bar denotes the covariant derivative in the tetrad frame, while the comma indicates differentiation along the corresponding tetrad direction. The antisymmetrization is defined as
\[
\mathcal{F}_{[ab|c]}
=
\mathcal{F}_{ab|c}
+\mathcal{F}_{bc|a}
+\mathcal{F}_{ca|b}.
\]

In spherical polar coordinates, the relevant Maxwell equation becomes
\ie
\label{ffedefm3}
\left(r\sqrt{A(r,\xi)}\,\mathcal{F}_{\phi r}\right)_{,r}
+\sqrt{A(r,\xi)B(r,\xi)}\,\mathcal{F}_{\phi\theta,\theta}
+r\sqrt{B(r,\xi)}\,\mathcal{F}_{t\phi,t}
=0.
\fe

Taking the time derivative of Eq.~\eqref{ffedefm3} and using Eqs.~\eqref{ffedefm1} and \eqref{ffedefm2}, we obtain
\ie
\label{ffedefm4}
\begin{split}
&\left[
\sqrt{A(r,\xi)B(r,\xi)^{-1}}
\left(
r\sqrt{A(r,\xi)}\,\mathcal{F}
\right)_{,r}
\right]_{,r}\\
&+\frac{A(r,\xi)\sqrt{B(r,\xi)}}{r}
\left(
\frac{\mathcal{F}_{,\theta}}{\sin\theta}
\right)_{,\theta}
\sin\theta
-r\sqrt{B(r,\xi)}\,\mathcal{F}_{,tt}
=0.
\end{split}
\fe

We then define
\[
F\equiv\mathcal{F}_{t\phi}\sin\theta
\]
and perform a Fourier decomposition in time, $\partial_t\rightarrow-i\omega$. The angular dependence is separated as
\[
F(r,\theta)
=
F(r)\frac{Y_{,\theta}}{\sin\theta},
\]
where $Y(\theta)$ is the Gegenbauer function \cite{baruah2025quasinormal,Gogoi:2023kjt,g1,g2,g3,g5,g6}, satisfying
\ie
\frac{1}{\sin\theta}
\frac{\mathrm{d}}{\mathrm{d}\theta}
\left(
\sin\theta\frac{\mathrm{d}Y}{\mathrm{d}\theta}
\right)
+\ell(\ell+1)Y
=0.
\fe

With these definitions, Eq.~\eqref{ffedefm4} reduces to
\ie
\label{ffedefm5}
\begin{split}
&\left[
\sqrt{A(r,\xi)B(r,\xi)^{-1}}
\left(
r\sqrt{A(r,\xi)}\,F
\right)_{,r}
\right]_{,r}\\
&+\omega^{2}r\sqrt{B(r,\xi)}\,F
-\frac{A(r,\xi)\sqrt{B(r,\xi)}}{r}
\ell(\ell+1)F
=0.
\end{split}
\fe

Finally, introducing the master variable
\[
\psi^{\text{vec}}
=
r\sqrt{A(r,\xi)}\,F,
\]
and the tortoise coordinate $r^{*}$, Eq.~\eqref{ffedefm5} can be written in the Schr\"odinger--like form
\ie
\partial^{2}_{r_*}\psi^{\text{vec}}
+\omega^{2}\psi^{\text{vec}}
=
\mathcal{V}^{v}(r,\ell,\xi)\psi^{\text{vec}},
\fe
where
\ie
\mathcal{V}^{v}(r,\ell,\xi)
=
A(r,\xi)\frac{\ell(\ell+1)}{r^{2}}.
\fe

\begin{figure}
    \centering
    \includegraphics[scale=0.51]{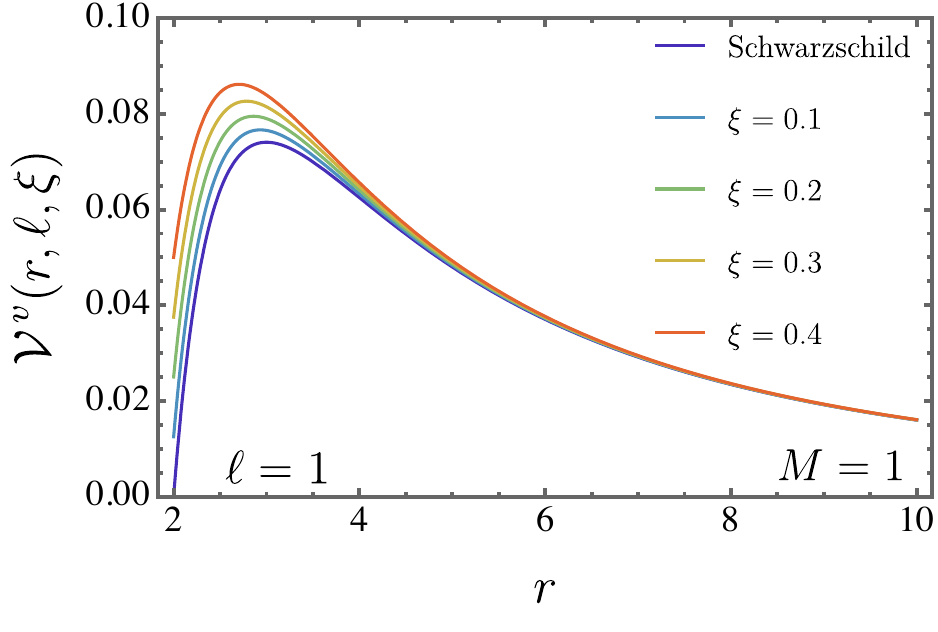}
    \includegraphics[scale=0.51]{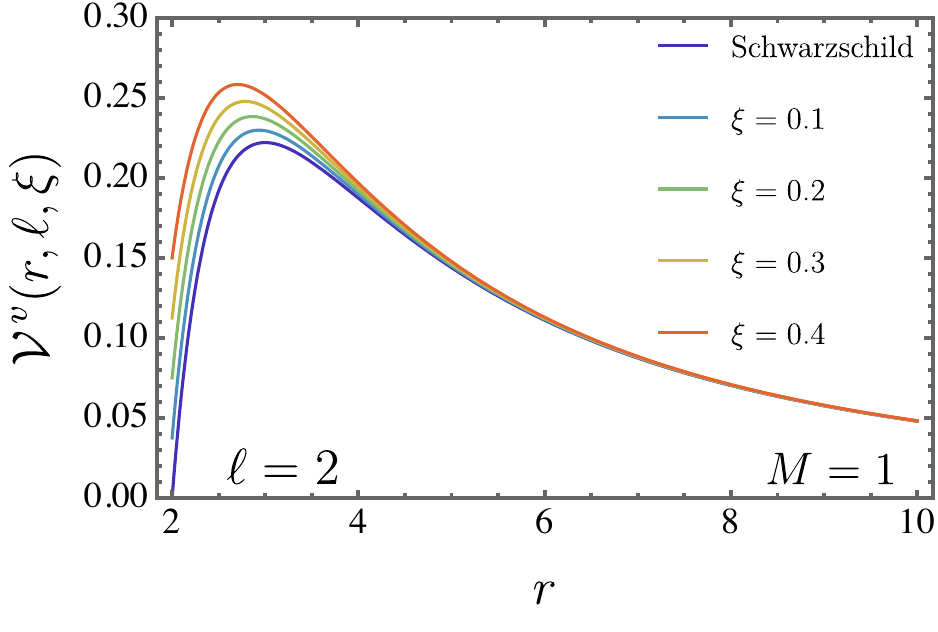}
    \includegraphics[scale=0.51]{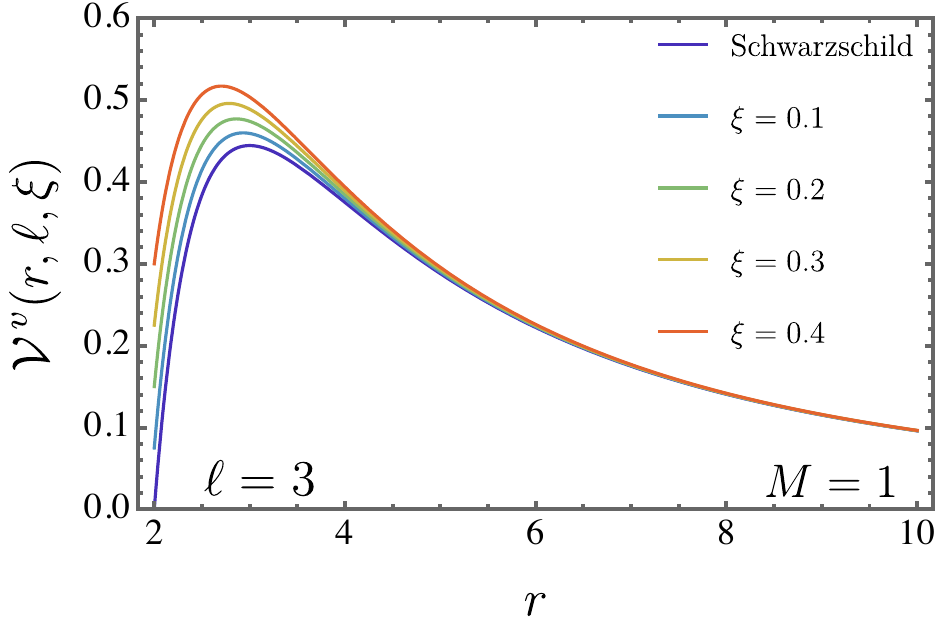}
    \caption{The profile of the vector perturbation potential $\mathcal{V}^{\,v}(r,\ell,\xi)$ is shown as a function of the radial coordinate $r$, considering different values of $\ell$ and $\xi$ to illustrate how these parameters affect both the height and structure of the potential barrier.}
    \label{vecvecpot}
\end{figure}

\begin{figure}
    \centering
    \includegraphics[scale=0.40]{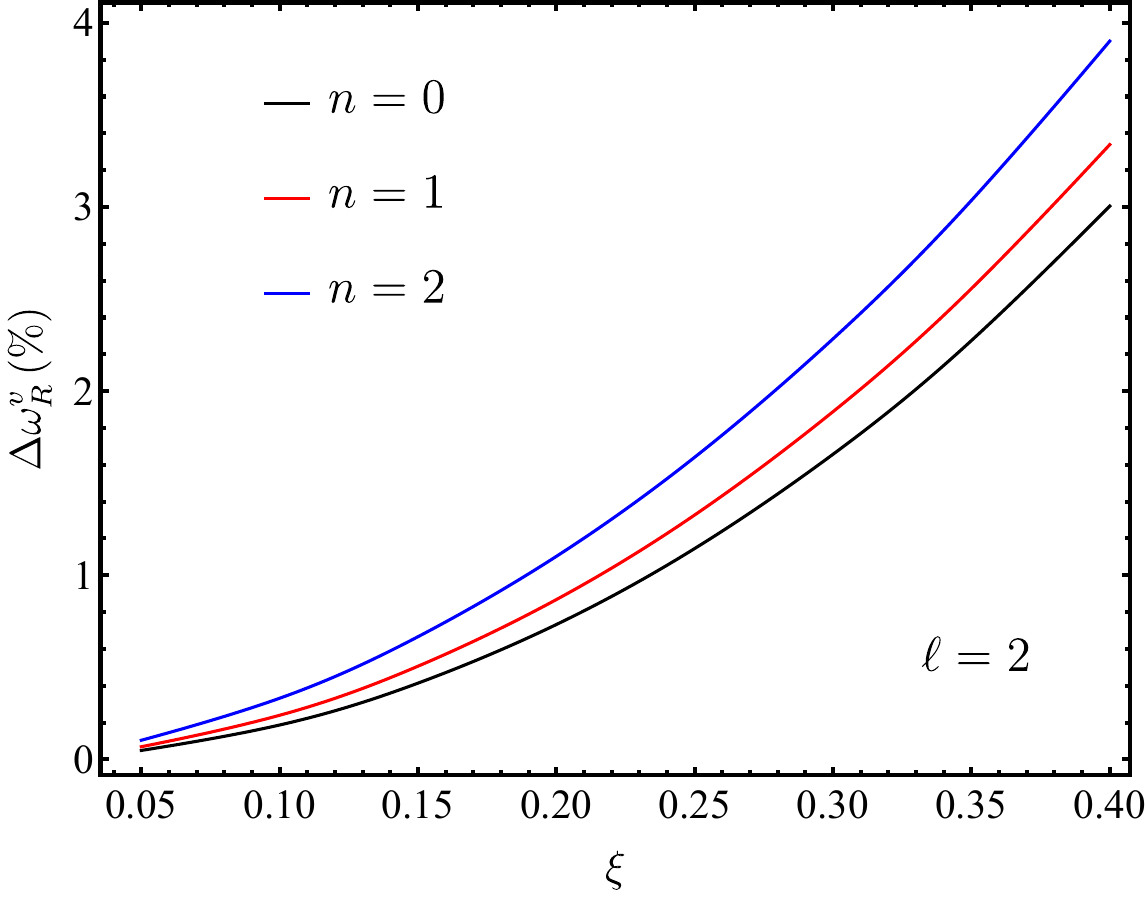}
    \includegraphics[scale=0.42]{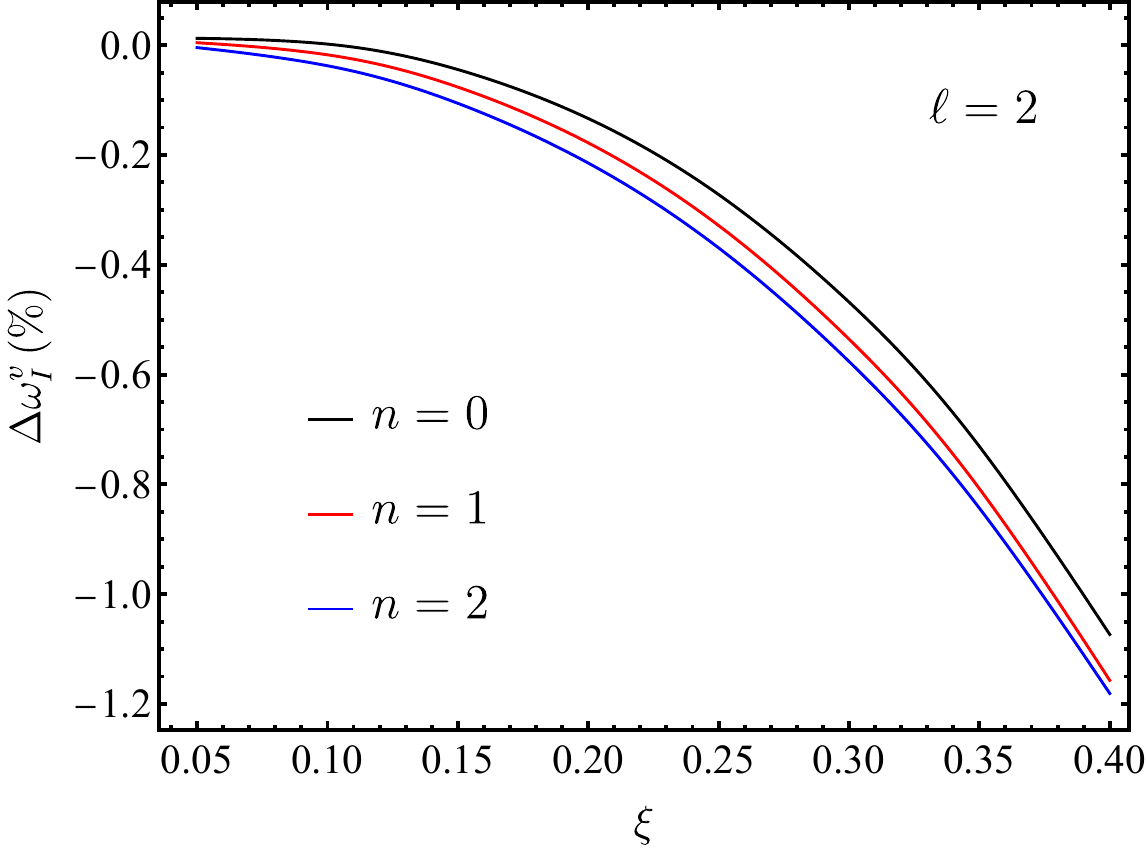}
    \caption{Dependence of the relative vector frequency shifts with respect to the Schwarzschild solution (real part on the left, imaginary part on the right) on the parameter $\xi$ for $\ell=2$. The distinct curves correspond to the radial overtone numbers $n=0,1,2$.}
    \label{fig:delvector}
\end{figure}

In Fig. \ref{vecvecpot}, it is displayed the vector perturbation potential $\mathcal{V}^{\,v}(r,\ell,\xi)$ as a function of $r$, explicitly comparing multiple values of $\ell$ and $\xi$. The plot clearly shows that increasing either $\xi$ or $\ell$ parameters increase the potential barrier. Another important observation is that, because the spacetime treated here is asymptotically flat, the potential approaches zero in the limit $r \to \infty$.
In addition, Table \ref{vectabqnm2} presents the quasinormal frequencies for different choices of $\xi$ and $\ell$. Furthermore, the real and imaginary relative deviations demonstrated in Fig.~\ref{fig:delvector} show that for all vector perturbation cases considered ($\ell=1,2,3$ with the corresponding overtones), increasing $\xi$ leads to a monotonic increase in the oscillation frequency ($\omega_R$) and a decrease in the damping rate ($|\omega_I|$), resulting in longer-lived ringdown modes.

\begin{table}[]
\caption{\label{vectabqnm2} The table presents the quasinormal mode spectra regarding vector fluctuations with $\ell = 1,2,3$, highlighting how these modes change as the parameter $\xi$ varies. The mass is fixed at $M = 1$.}

\begin{tabular}{|cc|c|c|c|c|}
\hline
\multicolumn{2}{|c|}{} &
  $\xi = 0.0$ &
  $\xi = 0.1$ &
  $\xi = 0.2$ &
  $\xi = 0.3$ \\ \hline
\multicolumn{1}{|c|}{\multirow{2}{*}{\begin{tabular}[c]{@{}c@{}}$l=1$\end{tabular}}} &
  $n = 0$ &
  {$0.291114-0.0980014i$} &
  {$0.291622-0.0979666i$} &
  {$0.293090-0.0977933i$} &
  {$0.295561-0.0974385i$} \\ \cline{2-6} 
\multicolumn{1}{|c|}{} &
  $n = 1$ &
  {$0.262212-0.307432i$} &
  {$0.262885-0.307234i$} &
  {$0.264517-0.306600i$} &
  {$0.267050-0.305436i$} \\ \hline
\multicolumn{1}{|c|}{\multirow{3}{*}{\begin{tabular}[c]{@{}c@{}}$l=2$\end{tabular}}} &
  $n = 0$ &
  {$0.314866-0.151628i$} &
  {$0.315354-0.151612i$} &
  {$0.316675-0.151405i$} &
  {$0.318846-0.150934i$} \\ \cline{2-6} 
\multicolumn{1}{|c|}{} &
  $n = 1$ &
  {$0.361537-0.386194i$} &
  {$0.362144-0.386139i$} &
  {$0.363507-0.385916i$} &
  {$0.365546-0.385435i$} \\ \cline{2-6} 
\multicolumn{1}{|c|}{} &
  $n = 2$ &
  {$0.374275-0.589655i$} &
  {$0.375054-0.589453i$} &
  {$0.376763-0.589111i$} &
  {$0.379266-0.588571i$} \\ \hline
\multicolumn{1}{|c|}{\multirow{4}{*}{\begin{tabular}[c]{@{}c@{}}$l=3$\end{tabular}}} &
  $n = 0$ &
  {$0.339306-0.196909i$} &
  {$0.339784-0.196923i$} &
  {$0.341017-0.196749i$} &
  {$0.342997-0.196292i$} \\ \cline{2-6} 
\multicolumn{1}{|c|}{} &
  $n = 1$ &
  {$0.435568-0.454491i$} &
  {$0.436184-0.454520i$} &
  {$0.437484-0.454464i$} &
  {$0.439359-0.454213i$} \\ \cline{2-6} 
\multicolumn{1}{|c|}{} &
  $n = 2$ &
  {$0.485523-0.662549i$} &
  {$0.486310-0.662453i$} &
  {$0.487920-0.662319i$} &
  {$0.490190-0.662071i$} \\ \cline{2-6}
\multicolumn{1}{|c|}{} &
  $n = 3$ &
  {$0.504339-0.858598i$} &
  {$0.505262-0.858287i$} &
  {$0.507237-0.857947i$} &
  {$0.510080-0.857556i$} \\ 
\hline
\end{tabular}

\end{table}

\subsection{Tensor field fluctuations }

Unlike scalar and electromagnetic perturbations, tensor perturbations require a consistent treatment of both the metric and the matter sector supporting the background geometry. Since no specific underlying theory is assumed here, we follow an effective description in which the spacetime is regarded as a solution of Einstein's equations sourced by an effective stress--energy tensor. This approach has been widely used in the study of modified and quantum--corrected black holes \cite{ashtekar2018quantum2,ashtekar2018quantum,asasas2,baruah2025quasinormal}.

For a static and spherically symmetric background, the effective source can be modeled as an anisotropic fluid,
\ie
\mathcal{T}_{\mu\nu}
=
\left(\rho+p_{2}\right)u_{\mu}u_{\nu}
+
\left(p_{1}-p_{2}\right)x_{\mu}x_{\nu}
+
p_{2}g_{\mu\nu},
\label{afluid}
\fe
where $\rho$, $p_{1}$, and $p_{2}$ denote the energy density, radial pressure, and tangential pressure, respectively. The timelike and radial unit vectors satisfy
\ie
u_{\mu}u^{\mu}=-1,
\qquad
x_{\mu}x^{\mu}=1,
\qquad
u_{\mu}x^{\mu}=0.
\label{fourvelocity}
\fe
{
Following Ref.~\cite{chen2019gravitational}, we work in the comoving frame of the effective fluid, in which the background vectors take the form $u^\mu=(u^t,0,0,0)$ and $x^\mu=(0,x^r,0,0)$. This choice is consistent with the static and spherically symmetric background geometry and specifies the anisotropic fluid ansatz employed in the perturbation analysis.
}

In the fluid rest frame, the nonvanishing mixed components of the background stress--energy tensor are
\begin{align}
\mathcal{T}^{t}_{t}&=-\rho,
&
\mathcal{T}^{r}_{r}&=p_{1},
&
\mathcal{T}^{\theta}_{\theta}
&=
\mathcal{T}^{\varphi}_{\varphi}
=p_{2}.
\end{align}
The background geometry fixes their radial dependence through the corresponding components of the Einstein tensor.

To study the axial, or odd--parity, gravitational perturbations, we adopt the standard axially symmetric line element \cite{chen2019gravitational},
\begin{align}
\mathrm{d}s^{2}
={}&
-e^{2\nu}\left(\mathrm{d}x^{0}\right)^{2}
+e^{2\psi}
\left(
\mathrm{d}x^{1}
-\sigma\,\mathrm{d}x^{0}
-q_{2}\,\mathrm{d}x^{2}
-q_{3}\,\mathrm{d}x^{3}
\right)^{2}
\nonumber\\
&+
e^{2\mu_{2}}\left(\mathrm{d}x^{2}\right)^{2}
+
e^{2\mu_{3}}\left(\mathrm{d}x^{3}\right)^{2},
\label{dmdedtdrdidcg}
\end{align}
where $x^{0}=t$, $x^{1}=\varphi$, $x^{2}=r$, and $x^{3}=\theta$. The functions $\sigma$, $q_{2}$, and $q_{3}$ vanish in the static background and therefore enter only at linear order.

Following Ref.~\cite{chen2019gravitational}, the perturbation equations are projected onto an orthonormal tetrad adapted to Eq.~\eqref{dmdedtdrdidcg}. In this frame, the perturbed anisotropic stress--energy tensor can be written as
\begin{align}
\delta\mathcal{T}_{ab}
={}&
(\rho+p_{2})\delta(u_{a}u_{b})
+
(\delta\rho+\delta p_{2})u_{a}u_{b}
\nonumber\\
&+
(p_{1}-p_{2})\delta(x_{a}x_{b})
+
(\delta p_{1}-\delta p_{2})x_{a}x_{b}
+
\delta p_{2}\eta_{ab}.
\end{align}
Using the normalization and orthogonality conditions in Eq.~\eqref{fourvelocity}, its axial components vanish,
\ie
\delta\mathcal{T}_{10}
=
\delta\mathcal{T}_{12}
=
\delta\mathcal{T}_{13}
=
0.
\fe
Thus, although the effective matter sector must be perturbed consistently, it does not directly source the odd--parity modes. This simplification is specific to the axial sector and does not generally apply to polar perturbations.

The axial field equations therefore reduce to
\ie
R_{ab}\big|_{\mathrm{axial}}=0.
\fe
Combining the relevant tetrad components as described in Ref.~\cite{chen2019gravitational}, one obtains a Schr\"odinger--like master equation for the tensor perturbations. The corresponding effective potential is \cite{baruah2025quasinormal,araujo2025does}
\ie
\mathcal{V}^{\,t}(r,\ell,\xi)
=
A(r,\xi)
\left[
\frac{2}{r^{2}}
\left(
B(r,\xi)-1
\right)
+
\frac{\ell(\ell+1)}{r^{2}}
-
\frac{1}{
r\sqrt{A(r,\xi)B(r,\xi)^{-1}}
}
\frac{\mathrm{d}}{\mathrm{d}r}
\sqrt{A(r,\xi)B(r,\xi)}
\right].
\fe
For the metric given in Eq.~\eqref{mainmetric}, this potential takes the explicit form
\ie
\label{tensorpotentialmetric}
\begin{split}
\mathcal{V}^{\,t}(r,\ell,\xi)
={}&
\left(
1-\frac{2M}{r}+\frac{\xi}{r^{2}}
\right)
\Bigg[
\frac{\ell(\ell+1)}{r^{2}}
+
\frac{4M\left(\xi^{3/2}-r^{3}\right)}{r^{6}}
\\
&+
\frac{
6M\xi^{7/2}
+
2Mr^{4}(2M-r)
+
\xi^{2}r^{3}(r-3M)
+
2M\xi^{3/2}r(2r-5M)
}{
r^{6}
\sqrt{
\frac{-2Mr+\xi^{2}+r^{2}}
{2M\left(\xi^{3/2}-r^{3}\right)+r^{4}}
}
\sqrt{
\left(-2Mr+\xi^{2}+r^{2}\right)
\left[
2M\left(\xi^{3/2}-r^{3}\right)+r^{4}
\right]
}
}
\Bigg].
\end{split}
\fe

It is immediate to see that, in the limit $\xi \to 0$, the effective potential reproduces the familiar result obtained for the Schwarzschild spacetime.

\begin{figure}
    \centering
    \includegraphics[scale=0.51]{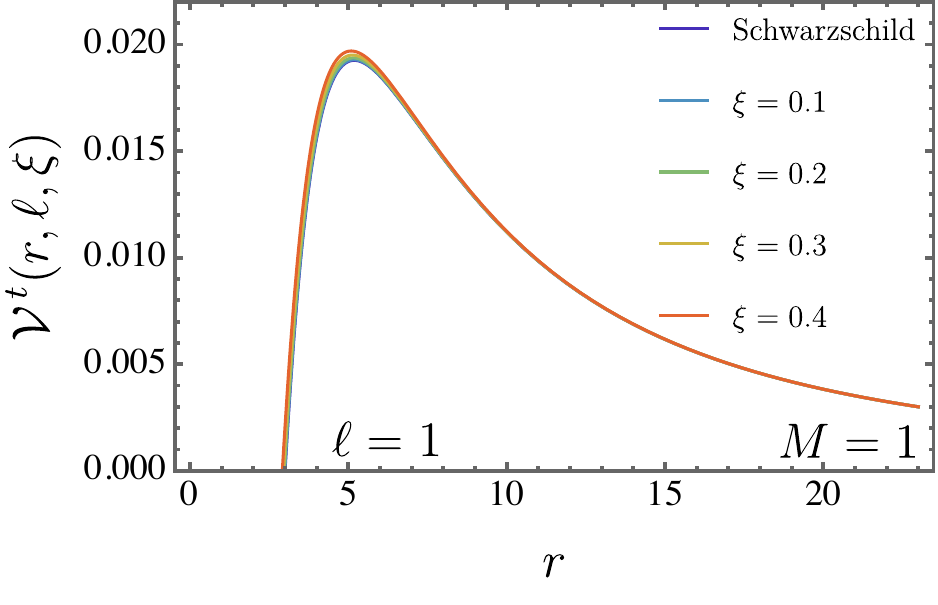}
    \includegraphics[scale=0.51]{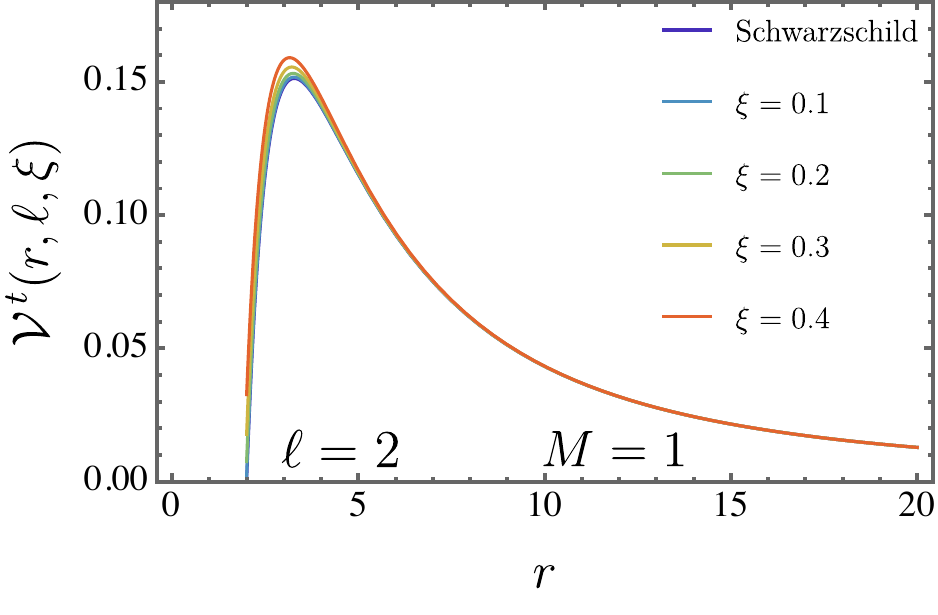}
    \includegraphics[scale=0.51]{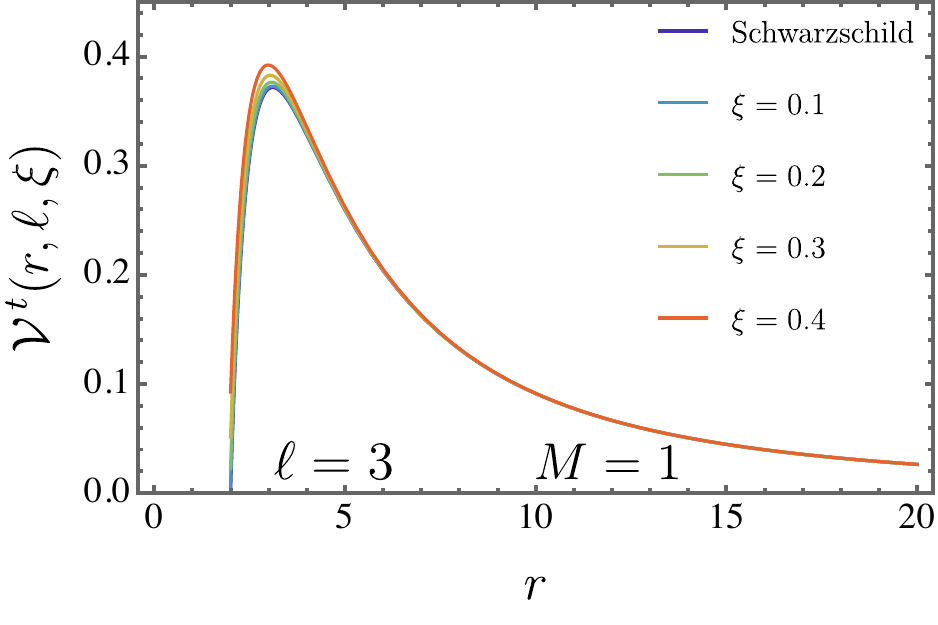}
    \caption{The tensor perturbation potential $\mathcal{V}^{\,t}(r,\ell,\xi)$ is plotted as a function of $r$, with several choices of $\ell$ and $\xi$, demonstrating pictorially how variations in these parameters modify the barrier’s shape, peak position, and overall amplitude. The radial component is displayed over the interval $r \in [2,20]$.}
    \label{tentenpot}
\end{figure}

Fig.~\ref{tentenpot} shows the behavior of the tensor potential $\mathcal{V}^{t}(r,\ell,\xi)$ as a function of the radial coordinate $r$ for several values of $\ell$ and $\xi$. The curves demonstrate that larger $\xi$ or $\ell$ shift the peak upward and broaden the barrier, indicating stronger confinement of the perturbations. Since the geometry is asymptotically flat, the potential smoothly falls to zero as $r \to \infty$, in accordance with the expected behavior at spatial infinity.

\begin{figure}
    \centering
    \includegraphics[scale=0.40]{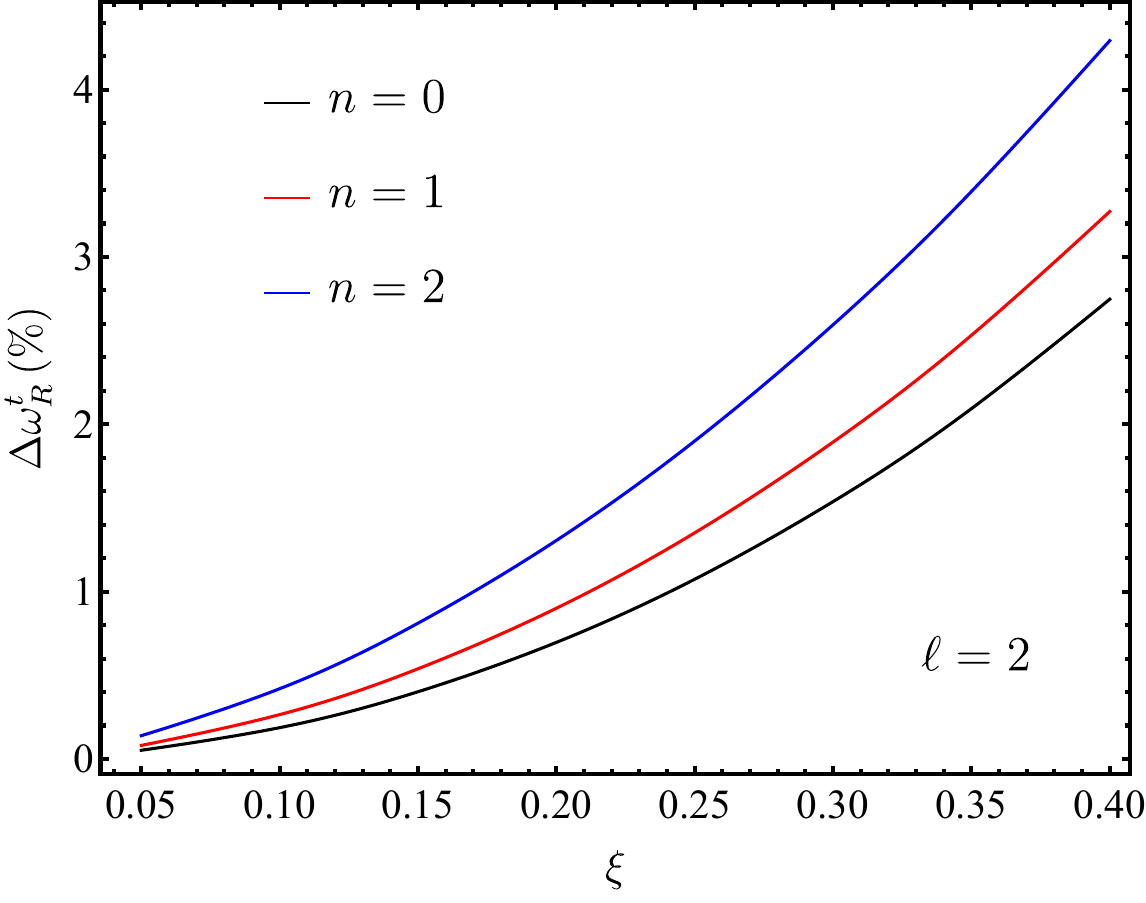}
    \includegraphics[scale=0.42]{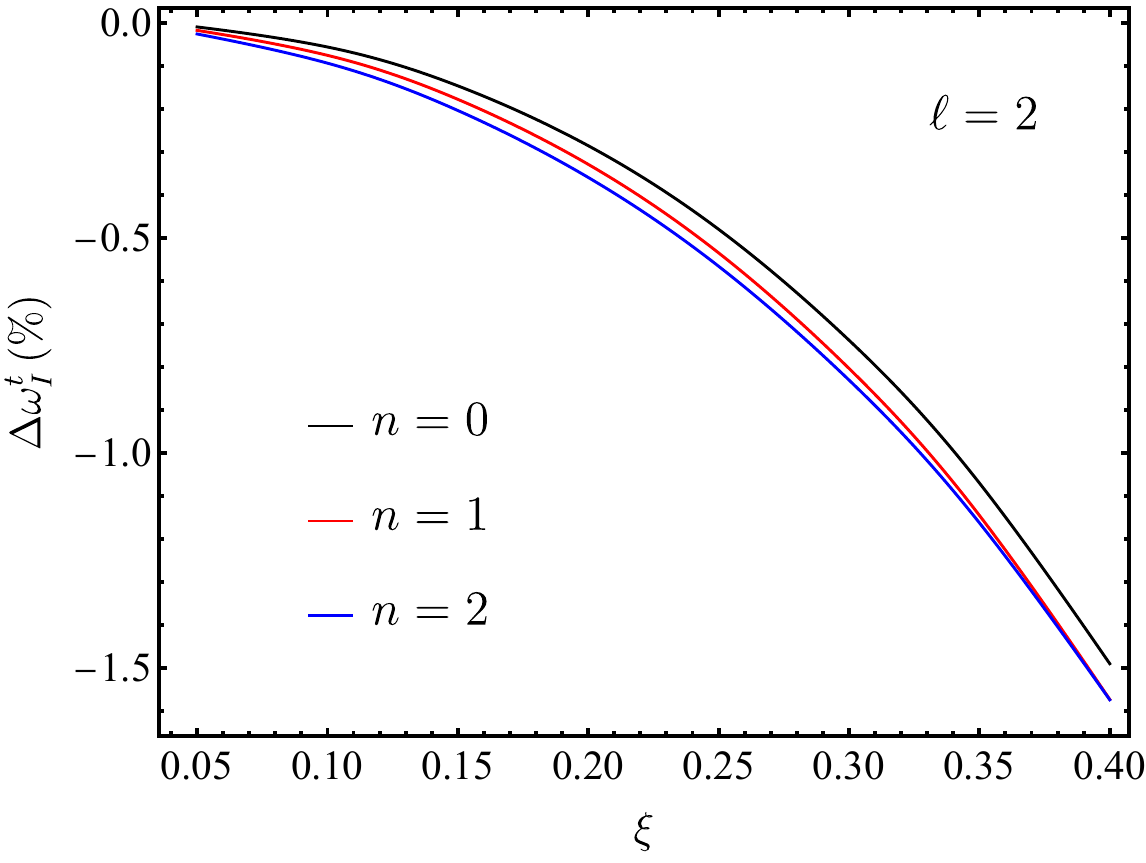}
    \caption{Relative shifts in the real (left) and imaginary (right) parts of the tensor quasinormal frequencies with respect to the Schwarzschild case ($\xi=0$) as functions of $\xi$ for $\ell=2$. The curves correspond to $n=0,1,$ and $2$.}
    \label{fig:deltensor}
\end{figure}

The associated quasinormal resonances are summarized in Table \ref{tab:tensorqnm} for different combinations of $\xi$, and $\ell$. A tensor QNM deviation are represented in Fig.~\ref{fig:deltensor}. As $\xi$ increases, the real part, $\omega_R$ of the quasinormal frequency shifts toward larger values, whereas $|\omega_I|$ decreases. The tensor modes therefore oscillate more rapidly and decay more slowly than in the Schwarzschild limit.

\begin{table}[]
\caption{\label{tab:tensorqnm} The table presents the quasinormal mode spectra regarding tensor fluctuations for $\ell = 2,3,4$, highlighting how these modes change as the parameter $\xi$ varies. The mass is fixed at $M = 1$.}

\begin{tabular}{|cc|c|c|c|c|}
\hline
\multicolumn{2}{|c|}{} &
  $\xi = 0.0$ &
  $\xi = 0.1$ &
  $\xi = 0.2$ &
  $\xi = 0.3$ \\ \hline
\multicolumn{1}{|c|}{\multirow{3}{*}{\begin{tabular}[c]{@{}c@{}}$\ell=2$\end{tabular}}} &
  $n = 0$ &
  {$0.373162-0.0892174i$} &
  {$0.373860-0.0891673i$} &
  {$0.375761-0.0889624i$} &
  {$0.378901-0.0885588i$} \\ \cline{2-6} 
\multicolumn{1}{|c|}{} &
  $n = 1$ &
  {$0.346017-0.274915i$} &
  {$0.346934-0.274707i$} &
  {$0.349134-0.274010i$} &
  {$0.352569-0.272705i$} \\ \cline{2-6} 
\multicolumn{1}{|c|}{} &
  $n = 2$ &
  {$0.302935-0.471064i$} &
  {$0.304206-0.470622i$} &
  {$0.306890-0.469370i$} &
  {$0.310794-0.467150i$} \\ \hline
\multicolumn{1}{|c|}{\multirow{4}{*}{\begin{tabular}[c]{@{}c@{}}$\ell=3$\end{tabular}}} &
  $n = 0$ &
  {$0.599265-0.0927284i$} &
  {$0.600330-0.0927209i$} &
  {$0.603415-0.0925844i$} &
  {$0.608625-0.0922654i$} \\ \cline{2-6} 
\multicolumn{1}{|c|}{} &
  $n = 1$ &
  {$0.582355-0.281406i$} &
  {$0.583581-0.281353i$} &
  {$0.586892-0.280898i$} &
  {$0.592338-0.279889i$} \\ \cline{2-6} 
\multicolumn{1}{|c|}{} &
  $n = 2$ &
  {$0.553200-0.476684i$} &
  {$0.554706-0.476524i$} &
  {$0.558399-0.475677i$} &
  {$0.564224-0.473913i$} \\ \cline{2-6}
\multicolumn{1}{|c|}{} &
  $n = 3$ &
  {$0.515747-0.677429i$} &
  {$0.517618-0.677111i$} &
  {$0.521818-0.675833i$} &
  {$0.528157-0.673312i$} \\ \hline
\multicolumn{1}{|c|}{\multirow{5}{*}{\begin{tabular}[c]{@{}c@{}}$\ell=4$\end{tabular}}} &
  $n = 0$ &
  {$0.809098-0.0941711i$} &
  {$0.810500-0.0941707i$} &
  {$0.814652-0.0940442i$} &
  {$0.821714-0.0937360i$} \\ \cline{2-6} 
\multicolumn{1}{|c|}{} &
  $n = 1$ &
  {$0.796499-0.284366i$} &
  {$0.798023-0.284347i$} &
  {$0.802341-0.283939i$} &
  {$0.809573-0.282981i$} \\ \cline{2-6} 
\multicolumn{1}{|c|}{} &
  $n = 2$ &
  {$0.773636-0.478974i$} &
  {$0.775379-0.478891i$} &
  {$0.779989-0.478143i$} &
  {$0.787505-0.476478i$} \\ \cline{2-6}
\multicolumn{1}{|c|}{} &
  $n = 3$ &
  {$0.743312-0.678300i$} &
  {$0.745343-0.678106i$} &
  {$0.750335-0.676973i$} &
  {$0.758213-0.674573i$} \\ \cline{2-6}
\multicolumn{1}{|c|}{} &
  $n = 4$ &
  {$0.707214-0.881266i$} &
  {$0.709589-0.880926i$} &
  {$0.715046-0.879383i$} &
  {$0.723375-0.876257i$} \\ 
\hline
\end{tabular}

\end{table}


\subsection{Spinor field fluctuations }

This part of the study focuses on the evolution of massless Dirac fields propagating in a static, spherically symmetric black hole geometry. The dynamics of spin-$\tfrac{1}{2}$ perturbations are examined using the Newman--Penrose formalism, which provides a convenient framework for handling spinor equations in curved spacetime. Within this approach, the governing Dirac equations take the form \cite{newman1962approach, chandrasekhar1984mathematical}:
\begin{align}
(D + \epsilon - \rho) \psi_1 +( \bar{\delta} + \pi - \alpha) \psi_2 &= 0, \\
(\Delta + \mu - \gamma) \psi_2 + (\delta + \beta - \mathrm{t}) \psi_1 &= 0.
\end{align}
Within this formalism, two spinor components, $\psi_{1}$ and $\psi_{2}$, are defined, and their evolution is described using directional derivatives taken along the null tetrad vectors: $D = l^{\mu}\partial_{\mu}$, $\Delta = n^{\mu}\partial_{\mu}$, $\delta = m^{\mu}\partial_{\mu}$, and $\bar{\delta} = \bar{m}^{\mu}\partial_{\mu}$.

The next step consists of building the null tetrad itself from the underlying metric, specifying the vectors $l^{\mu}$, $n^{\mu}$, $m^{\mu}$, and $\bar{m}^{\mu}$ explicitly, which provides the basis required to express the Dirac equations in this spacetime.
\ie
\begin{split}
l^\mu &= \left(\frac{1}{A(r,\xi)}, \sqrt{\frac{B(r,\xi)}{A(r,\xi)}}, 0, 0\right), \quad n^\mu  = \frac{1}{2} \left(1, -\sqrt{A(r,\xi) B(r,\xi)}, 0, 0\right), \\
m^\mu &= \frac{1}{\sqrt{2} r} \left(0, 0, 1, \frac{i}{\sin \theta}\right), \quad \bar{m}^\mu = \frac{1}{\sqrt{2} r} \left(0, 0, 1, \frac{-i}{\sin \theta}\right).
\end{split}
\fe

Using these definitions, one can compute the spin coefficients, obtaining the following non--vanishing terms
\ie
\begin{split}
 & \rho = -\frac{1}{r} \frac{B(r,\xi)}{A(r,\xi)},  \quad
\mu = -\frac{\sqrt{A(r,\xi) B(r,\xi)}}{2r}, \\ & \gamma = \frac{A(r,\xi)'}{4}\sqrt{\frac{B(r,\xi)}{A(r,\xi)}},  \quad
\beta = -\alpha = \frac{\cot{\theta}}{2\sqrt{2}r}. 
\end{split}
\fe

Decoupling the coupled Dirac equations leads to a single differential equation that governs the evolution of $\psi_{1}$, fully characterizing the dynamics of the massless spin--$\tfrac{1}{2}$ field in this spacetime
\ie
\left[(D - 2\rho)(\Delta + \mu - \gamma) - (\delta + \beta) (\bar{\delta}+\beta)\right] \psi_1 = 0.
\fe

After inserting the explicit expressions for the directional derivatives and the computed spin coefficients into the Dirac equation, it can be reformulated and expressed in the form
\ie
\begin{split}
&\left[ \frac{1}{2A(r,\xi)} \partial_t^2 - \left( \frac{\sqrt{A(r,\xi)B(r,\xi)}}{2r} +\frac{A(r,\xi)'}{4}\sqrt{\frac{B(r,\xi)}{A(r,\xi)}}\right)\frac{1}{A(r,\xi)}\partial_t \right. \\
& \left. - \frac{\sqrt{A(r,\xi)B(r,\xi)}}{2} \sqrt{\frac{B(r,\xi)}{A(r,\xi)}}\partial_r^2 \right. \\
& \left. -\sqrt{\frac{B(r,\xi)}{A(r,\xi)}} \partial_r \left( \frac{\sqrt{A(r,\xi)B(r,\xi)}}{2} + \frac{A(r,\xi)'}{4}{\sqrt{\frac{B(r,\xi)}{A(r,\xi)}}} \right) \right] \psi_1 \\ + 
&\left[ \frac{1}{\sin^2\theta} \partial_\phi^2 + i \frac{\cot \theta}{\sin \theta}\partial_\phi \right. \\
& \left.+ \frac{1}{\sin \theta}\partial_\theta \left( \sin \theta \partial_\theta \right) - \frac{1}{4} \cot^2 \theta + \frac{1}{2} \right] \psi_1 = 0.
\end{split}
\fe

In order to separate the Dirac equation into its radial and angular components, the spinor field is expressed as $\psi_1 = \Psi(r) Y_{lm}(\theta, \phi) e^{-i \omega t}$, so that
\begin{align}
&\left[  \frac{-\omega^2}{2A(r,\xi)} - \left(\frac{\sqrt{A(r,\xi)B(r,\xi)}}{2r}+\frac{A(r,\xi)'}{4} + \sqrt{\frac{B(r,\xi)}{A(r,\xi)}}\right)\frac{- i\omega}{A(r,\xi)} \right. \\
& \left. - \frac{\sqrt{A(r,\xi)B(r,\xi)}}{2} \sqrt{\frac{B(r,\xi)}{A(r,\xi)}}\partial_r^2 -\lambda_{lm} \right. \\
& \left. - \sqrt{\frac{B(r,\xi)}{A(r,\xi)}}\partial_r \left(\frac{\sqrt{A(r,\xi)B(r,\xi)}}{2r} + \frac{A(r,\xi)'}{4}\sqrt{\frac{B(r,\xi)}{A(r,\xi)}}\right) \right] \Psi(r) = 0.
\end{align}

Here, $\lambda_{lm}$ acts as the separation constant connecting the angular and radial parts of the solution. By introducing $r^{*}$, we verify that the radial equation can be recast into a Schrödinger--type wave equation, taking the form:
\begin{align}
\left[\frac{\mathrm{d}^2 }{\mathrm{d}r_*^2} +( \omega^2 - \mathcal{V}^{\,spin \pm}(r,\ell,\xi)) \right]\Psi_{\pm}(r) = 0.
\end{align}
In addition, the effective potentials $\mathcal{V}^{\,\text{spin}\pm}(r,\ell,\xi)$ governing the propagation of the massless spin-$\tfrac{1}{2}$ field can be introduced, taking the form given in \cite{albuquerque2023massless, al2024massless,arbey2021hawking}
\ie\label{Vpm}
\mathcal{V}^{\,spin \pm}(r,\ell,\xi) = \frac{(\ell + \frac{1}{2})^2}{r^2} A(r,\xi)
\pm \left(\ell + \frac{1}{2}\right) \sqrt{A(r,\xi) B(r,\xi)} \partial_r \left(\frac{\sqrt{A(r,\xi)}}{r}\right).
\fe

\begin{figure}
    \centering
    \includegraphics[scale=0.51]{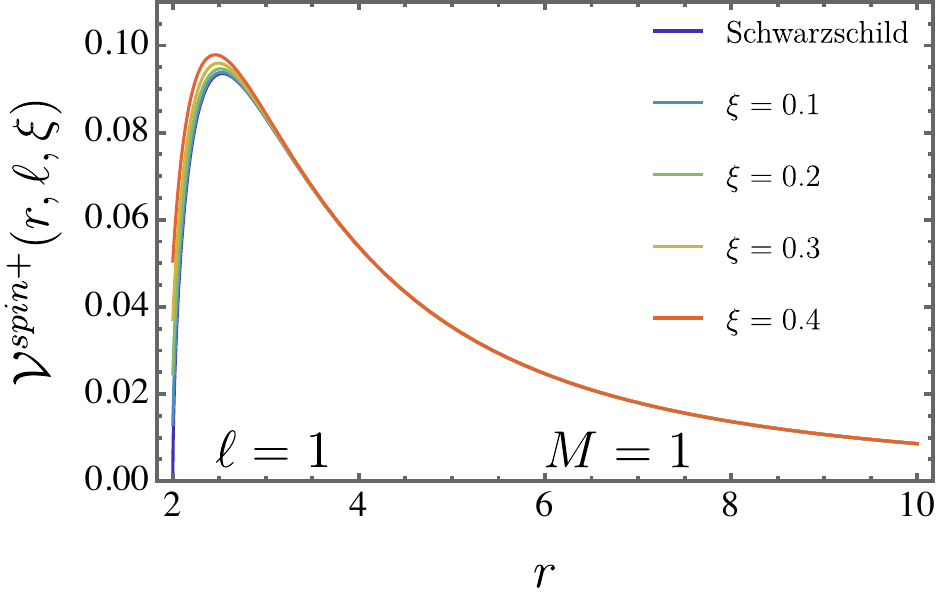}
    \includegraphics[scale=0.51]{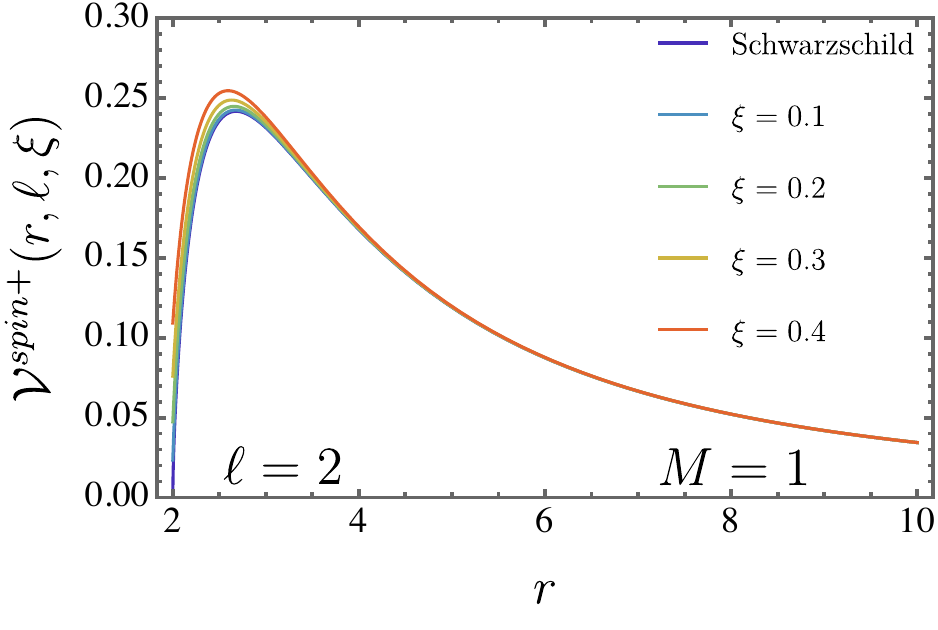}
    \includegraphics[scale=0.51]{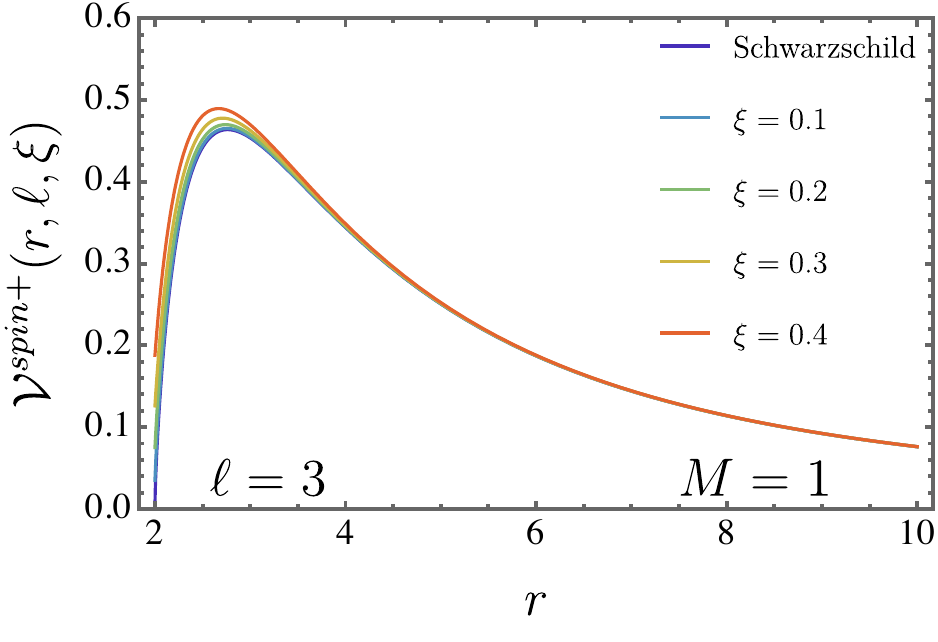}
    \caption{The spinor perturbation potential $\mathcal{V}^{\,\text{spin+}}(r,\ell,\xi)$ is shown as a function $r$ for distinct values $\xi$ and $\ell$, clearly illustrating how changes in these parameters affect the potential barrier.}
    \label{spinplot}
\end{figure}

In what follows, the potential $\mathcal{V}^{\,\text{spin+}}(r,\ell,\xi)$ is selected as the representative case, since the corresponding $\mathcal{V}^{\,\text{spin-}}(r,\ell,\xi)$ exhibits a qualitatively similar profile \cite{albuquerque2023massless,devi2020quasinormal,araujo2025does}. Hence, the analysis concentrates on $\mathcal{V}^{\,\text{spin+}}(r,\ell,\xi)$. Its behavior is depicted in Fig.~\ref{spinplot}, where the potential is shown as a function of $r$ for different parameter choices. Consistent with the asymptotically flat nature of the spacetime, $\mathcal{V}^{\,\text{spin+}}(r,\ell,\xi)$ tends to zero as $r \to \infty$.

Fig.~\ref{spinplot} highlights the spinor potential $\mathcal{V}^{\,\text{spin+}}(r,\ell,\xi)$ plotted versus $r$ for a range of $\ell$ and $\xi$ values. The plots reveal that increasing either parameter raises the maximum of the potential and widens the barrier region, which enhances the trapping of perturbations near the black hole. Because the spacetime approaches flatness at large distances, the potential decays to zero as $r \to \infty$, which is naturally consistent with the expected asymptotic behavior.

\begin{figure}
    \centering
    \includegraphics[scale=0.40]{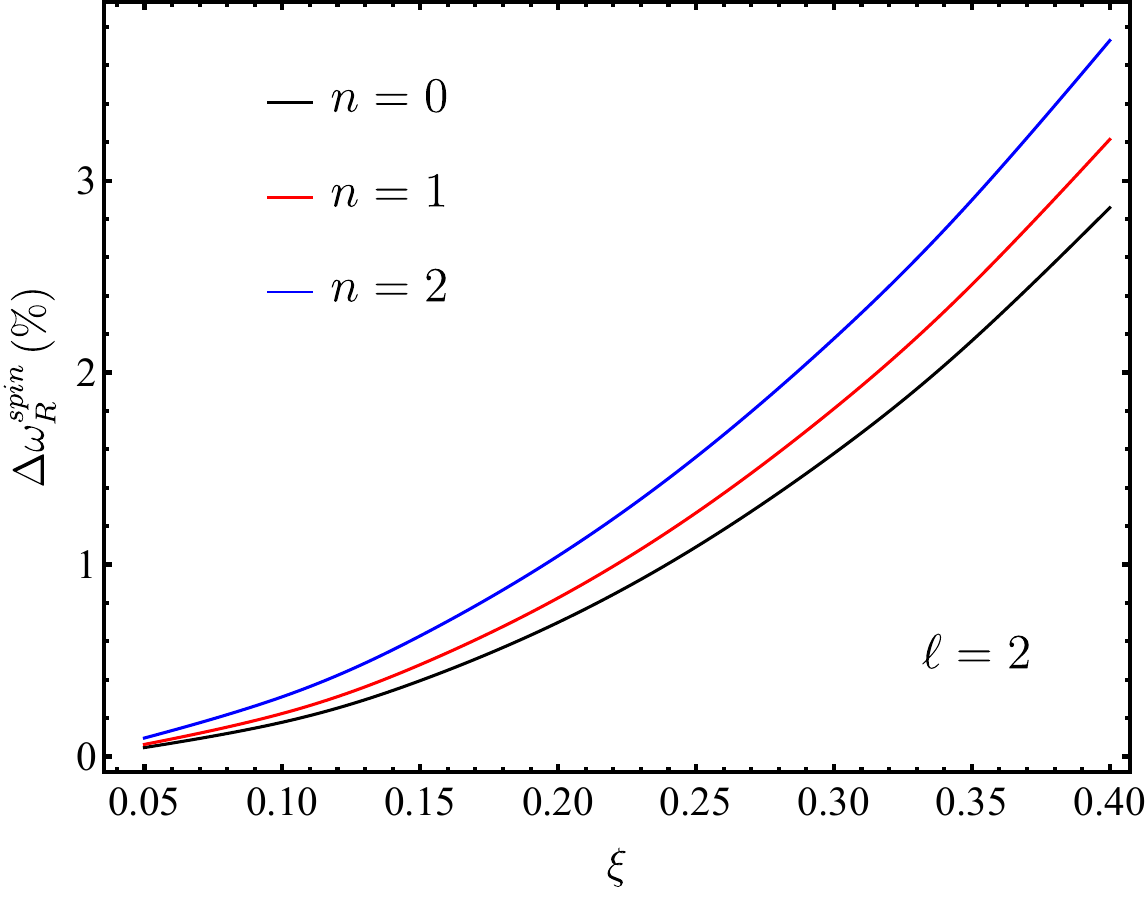}
    \includegraphics[scale=0.42]{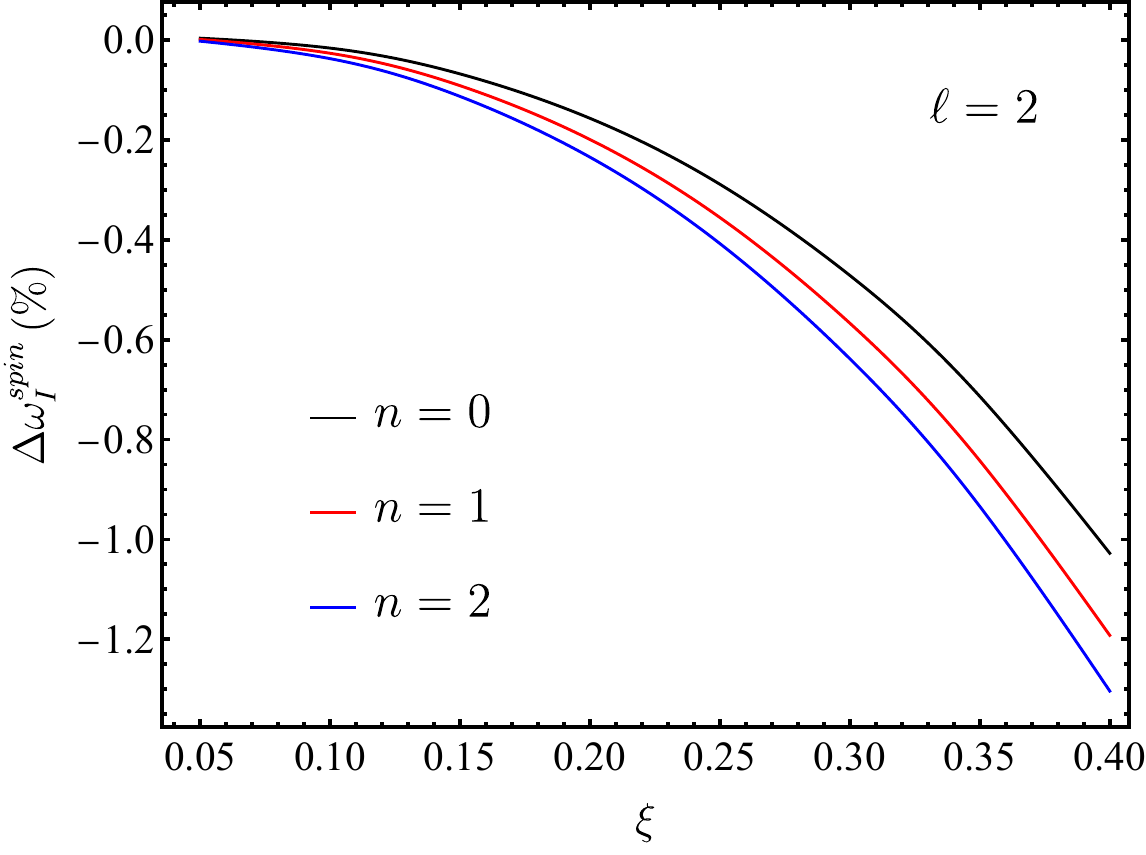}
    \caption{Relative shifts of the spinor quasinormal modes as functions of parameter $\xi$ for $\ell=2$. The left and right panels show the real and imaginary parts, respectively. Results are shown for the fundamental mode $n = 0$ and the first two overtones $n=1,2$.}
    \label{fig:delspin}
\end{figure}

The quasinormal frequencies obtained from this potential are collected in Table \ref{tab:spinorqnm}, covering several choices of $\xi$ and $\ell$. The relative deviations shown in Fig.~\ref{fig:delspin} indicate that the spinor spectrum is also sensitive to $\xi$. Larger values of the parameter $\xi$ increase the oscillation frequency while weakening the damping, thereby extending the characteristic lifetime of the spinor modes.

\begin{table}[]
\caption{\label{tab:spinorqnm} The table reports the quasinormal spectra for spinor fluctuations in the case $\ell = 1,2,3$, presenting how the frequencies evolve as $ \ xi$ grows. The mass is fixed at $M=1$.}

\begin{tabular}{|cc|c|c|c|c|}
\hline
\multicolumn{2}{|c|}{} &
$\xi=0.0$ &
$\xi=0.1$ &
$\xi=0.2$ &
$\xi=0.3$ \\
\hline

\multicolumn{1}{|c|}{\multirow{2}{*}{$\ell=1$}}
& $n=0$
& $0.279242-0.0971428i$
& $0.279821-0.0971098i$
& $0.281415-0.0969493i$
& $0.284049-0.0966185i$
\\ \cline{2-6}

\multicolumn{1}{|c|}{}
& $n=1$
& $0.249534-0.306401i$
& $0.250419-0.306287i$
& $0.252412-0.305497i$
& $0.255415-0.303858i$
\\
\hline

\multicolumn{1}{|c|}{\multirow{3}{*}{$\ell=2$}}
& $n=0$
& $0.476498-0.0963845i$
& $0.477348-0.0963685i$
& $0.479827-0.0962330i$
& $0.484020-0.0959299i$
\\ \cline{2-6}

\multicolumn{1}{|c|}{}
& $n=1$
& $0.455761-0.294988i$
& $0.456782-0.294909i$
& $0.459526-0.294400i$
& $0.464017-0.293315i$
\\ \cline{2-6}

\multicolumn{1}{|c|}{}
& $n=2$
& $0.423533-0.502794i$
& $0.424851-0.502604i$
& $0.427958-0.501613i$
& $0.432752-0.499584i$
\\
\hline

\multicolumn{1}{|c|}{\multirow{4}{*}{$\ell=3$}}
& $n=0$
& $0.670533-0.0962939i$
& $0.671692-0.0962879i$
& $0.675135-0.0961606i$
& $0.681001-0.0958597i$
\\ \cline{2-6}

\multicolumn{1}{|c|}{}
& $n=1$
& $0.655450-0.291800i$
& $0.656739-0.291759i$
& $0.660373-0.291328i$
& $0.666445-0.290352i$
\\ \cline{2-6}

\multicolumn{1}{|c|}{}
& $n=2$
& $0.629465-0.493464i$
& $0.630981-0.493342i$
& $0.634919-0.492523i$
& $0.641276-0.490761i$
\\ \cline{2-6}

\multicolumn{1}{|c|}{}
& $n=3$
& $0.596416-0.700441i$
& $0.598232-0.700199i$
& $0.602540-0.698948i$
& $0.609208-0.696369i$
\\
\hline
\end{tabular}

\end{table}

\section{Time-Domain solution }

To obtain the time evolution of scalar, vector, and tensor fluctuations, it is essential to perform a full dynamical analysis rather than relying solely on frequency--domain methods. This approach makes it possible to study how quasinormal modes influence scattering and decay processes. Because the effective potentials governing these perturbations are generally complicated, an accurate numerical scheme is required to evolve the system. Thereby, the characteristic integration method, first proposed by Gundlach et al.~\cite{Gundlach:1993tp}, is employed.

Following the procedure outlined in \cite{Skvortsova:2024wly,Bolokhov:2024ixe,Guo:2023nkd,Yang:2024rms,Baruah:2023rhd,Gundlach:1993tp,Shao:2023qlt}, the problem is reformulated in double-null coordinates defined by $u = t - r^{*}$ and $v = t + r^{*}$. This change of variables simplifies the wave equation and casts it into a form suitable for numerical integration. In terms of these coordinates, the equation is given by:
\ie
\left(4 \frac{\partial^{2}}{\partial u \, \partial v} + V(u,v)\right) \Tilde{\psi} (u,v) = 0.
\fe

A common manner to solve the resulting equation numerically is to discretize the domain through a finite--difference approach, allowing the wave function to be evolved step by step across the grid
\ie
\Tilde{\psi}(N) = -\Tilde{\psi}(S) + \Tilde{\psi}(W) + \Tilde{\psi}(E) - \frac{h^{2}}{8}V(S)\Big[\Tilde{\psi}(W) + \Tilde{\psi}(E)\Big] + \mathcal{O}(h^{4}).
\fe

Such a procedure begins by defining a grid on the $(u,v)$ plane, where $h$ represents the step size. Each cell is identified by four points: the starting position $S = (u,v)$, its neighbors $W = (u+h,v)$ and $E = (u,v+h)$, and the forward point $N = (u+h,v+h)$. The evolution is initialized along the characteristic lines $u = u_{0}$ and $v = v_{0}$, which act as the initial boundaries for the integration scheme. To specify the initial data, a Gaussian pulse centered at $v = v_{c}$ with width $\sigma$ is prescribed along $u = u_{0}$, providing the initial wave profile used to propagate the solution across the grid
\ie
\Tilde{\psi}(\Tilde{u} = u_{0},v) = A e^{-(v-v_{0})^{2}}/2\sigma^{2}, \,\,\,\,\,\, \Tilde{\psi}(u,v_{0}) = \Tilde{\psi}_{0}.
\fe

The numerical evolution starts by specifying the initial data along $v = v_{0}$ through the condition $\tilde{\psi}(u,v_{0}) = \tilde{\psi}_{0}$, which is set to zero to simplify the setup. The algorithm then advances the solution step by step along constant-$u$ slices, updating the values as $v$ increases according to the null--grid arrangement. For clarity and efficiency, the analysis is restricted to massless acrros all perturbations encountered in this paper with $M = 1$. The initial configuration is taken as a Gaussian wave packet centered at $v = 0$, with width $\sigma = 1$ and vanishing starting amplitude. The integration domain is discretized uniformly over $u, v \in [0,1000]$ with a grid spacing of $h = 0.1$, providing the resolution needed to track the propagation and decay of the signal.


\subsection{Scalar field }

\begin{figure}
    \centering
    \includegraphics[scale=0.53]{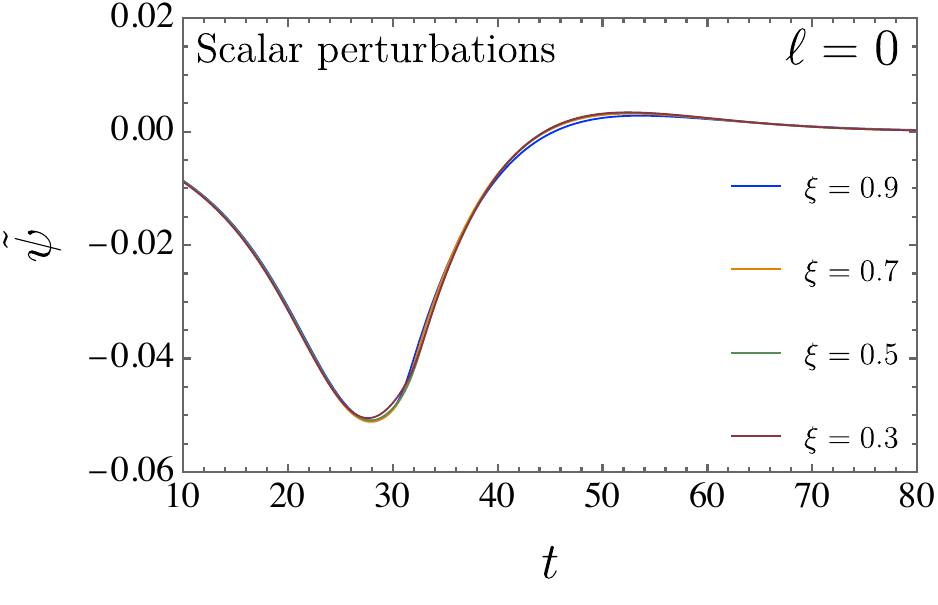}
    \includegraphics[scale=0.53]{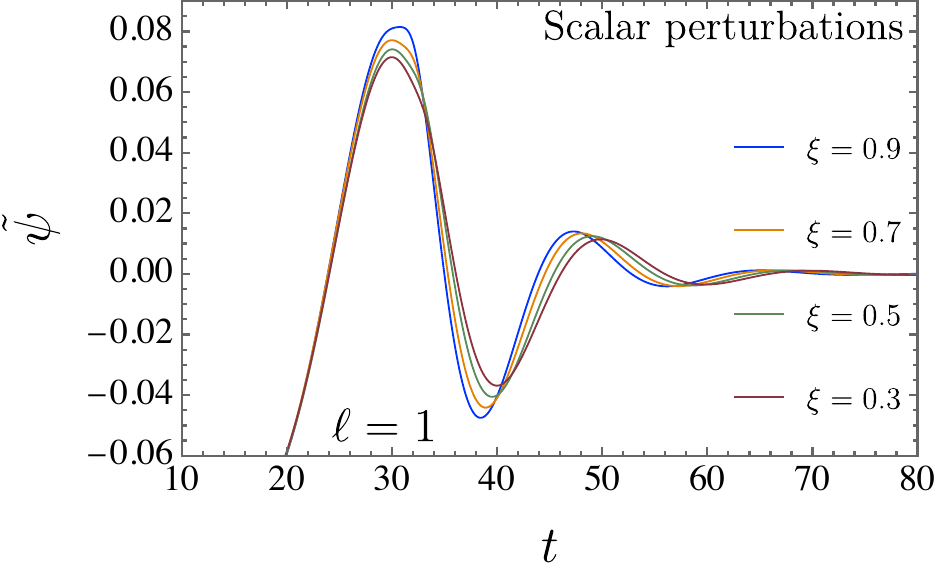}
    \includegraphics[scale=0.53]{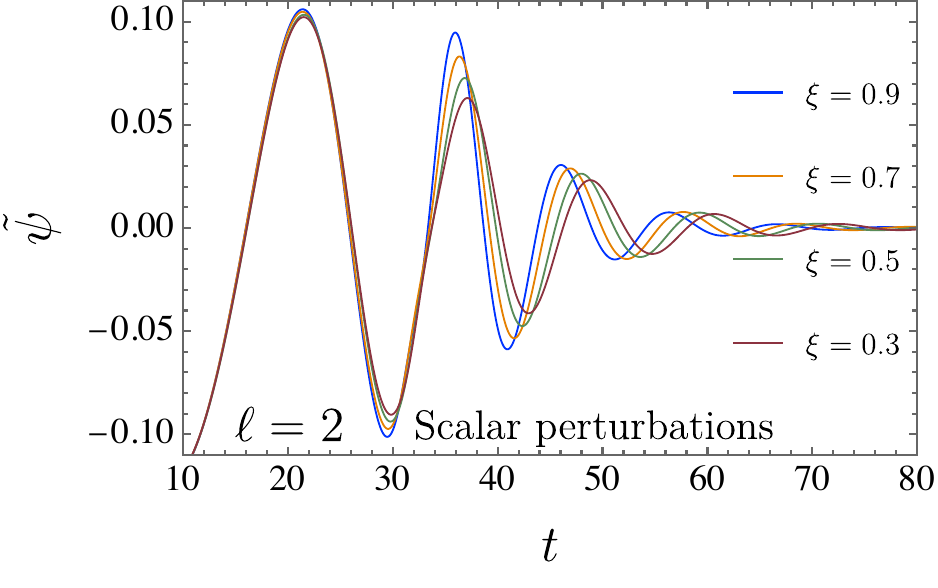}
    \caption{The dynamics of scalar perturbations are shown by evolving the waveform $\tilde{\psi}$ in time for a fixed black hole mass $M = 1$ and several values of the parameter $\xi$. The analysis considers $\xi = 0.3, 0.5, 0.7,$ and $0.9$, and the resulting waveforms are displayed according to the angular mode: the top--left panel corresponds to $\ell = 0$, the top--right to $\ell = 1$, and the bottom panel to $\ell = 2$.}
    \label{timedomainspsi}
\end{figure}

The temporal evolution of scalar fluctuations for the the black hole backgrond considered here is investigated in this part of the study. In Fig.~\ref{timedomainspsi}, the field $\tilde{\psi}$ is evolved for a fixed mass $M = 1$ while the parameter $\xi$ is varied across $0.3$, $0.5$, $0.7$, and $0.9$. The results are displayed according to the angular mode: $\ell = 0$ (top left), $\ell = 1$ (top right), and $\ell = 2$ (bottom). Each curve exhibits exponentially damped oscillations, clearly signaling the presence of the quasinormal ringing phase characteristic of perturbed black hole spacetime considered in this paper (\ref{mainmetric}) for the scalar perturbations.

\begin{figure}
    \centering
    \includegraphics[scale=0.53]{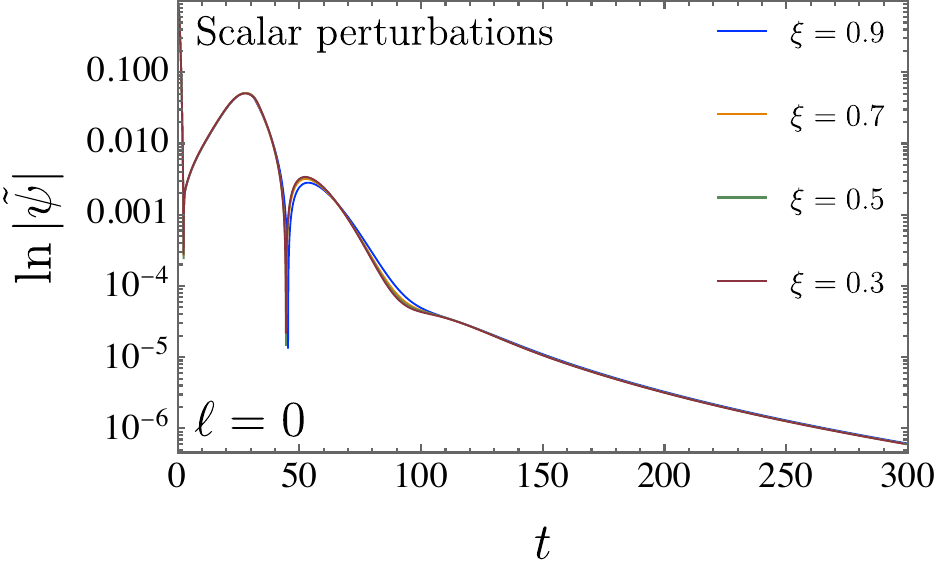}
    \includegraphics[scale=0.53]{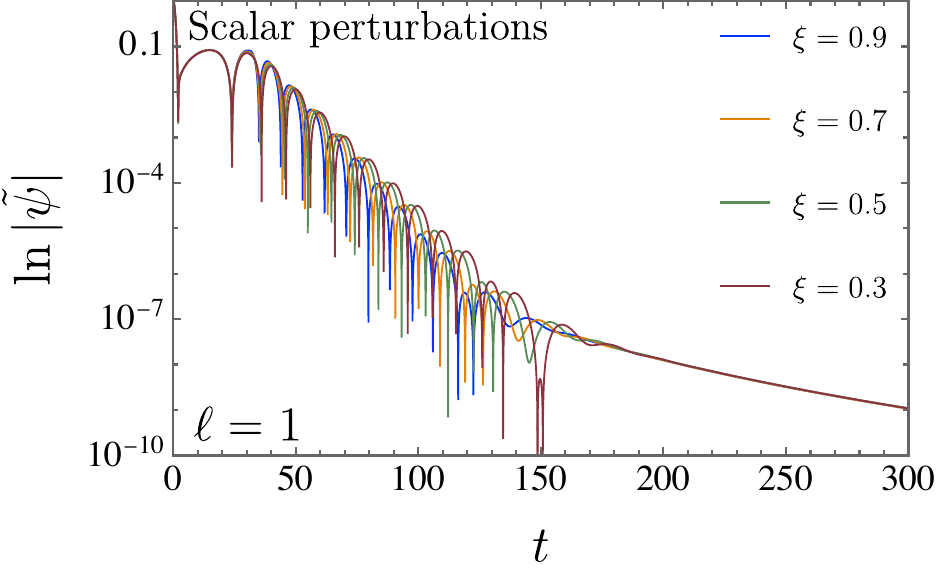}
    \includegraphics[scale=0.53]{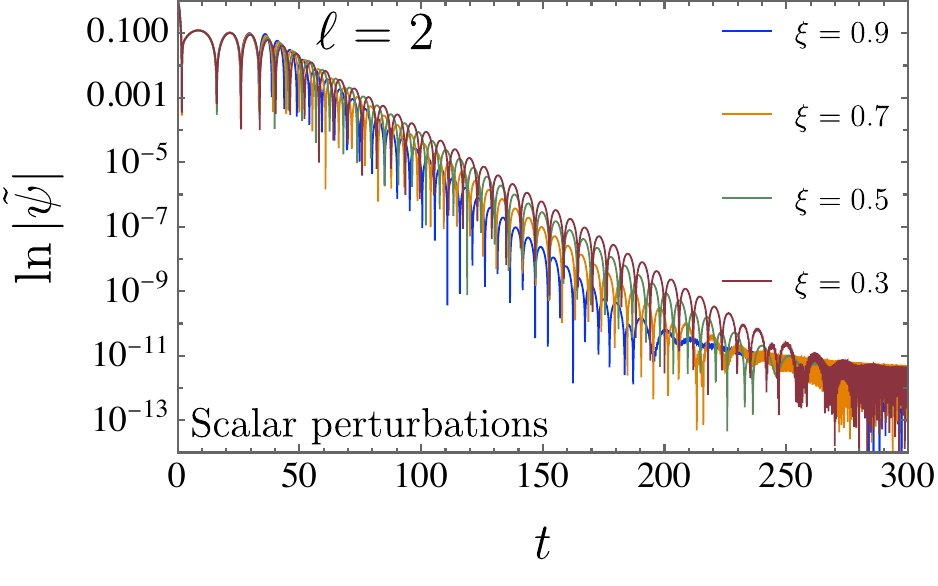}
    \caption{The logarithmic evolution of the scalar field amplitude, $\ln|\tilde{\psi}|$, is presented as a function of time $t$ for a black hole with $M = 1$ and charge parameter values $\xi = 0.3, 0.5, 0.7,$ and $0.9$. The plots are displayed by angular index, with $\ell = 0$ shown in the upper--left panel, $\ell = 1$ in the upper--right, and $\ell = 2$ in the bottom panel.}
    \label{timedomainlnspsi}
\end{figure}

In addition, a closer look at the damping behavior is provided in Fig.~\ref{timedomainlnspsi}, where the quantity $\ln|\tilde{\psi}|$ is plotted for the same set of $\xi$ and $\ell$ values. These curves make the quasinormal phase more evident by displaying the exponential decay on a straight line, and they also highlight the point at which the system exits the oscillatory stage and enters the characteristic power--law tail regime at late times.

\begin{figure}
    \centering
    \includegraphics[scale=0.53]{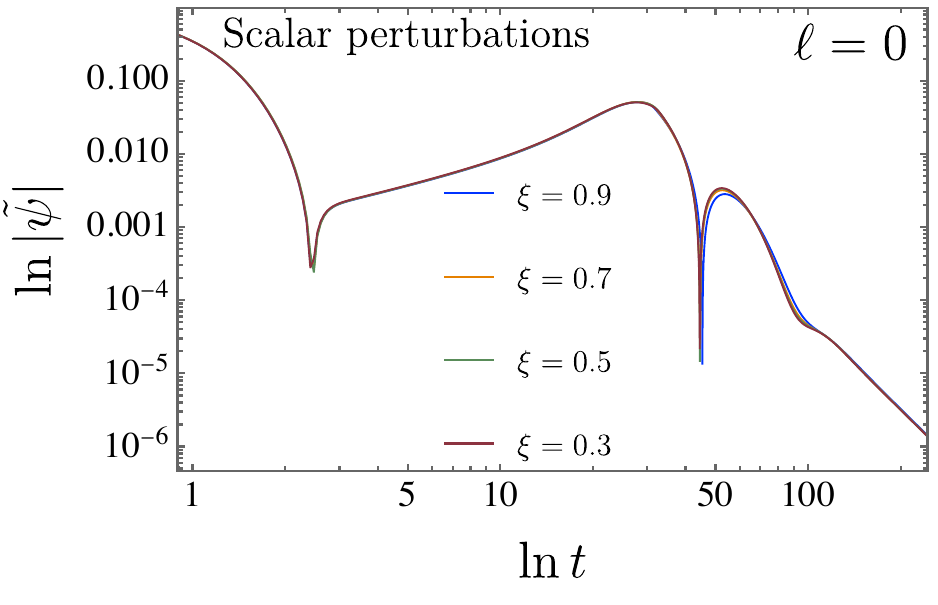}
    \includegraphics[scale=0.53]{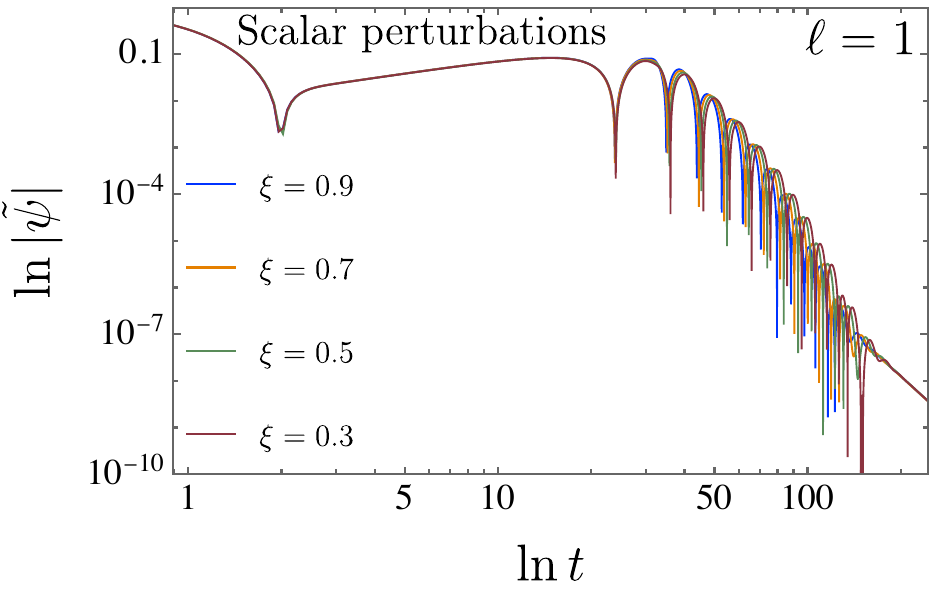}
    \includegraphics[scale=0.53]{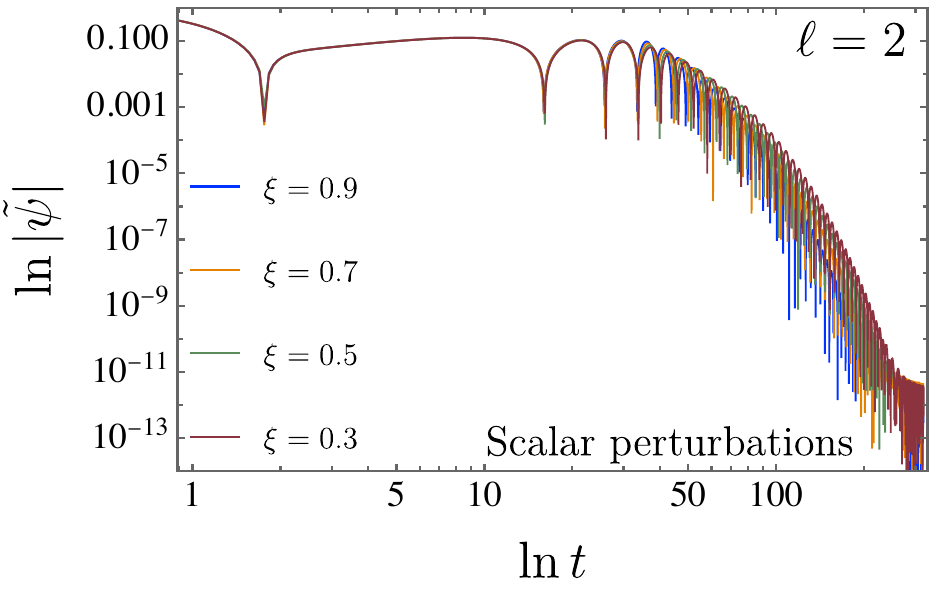}
    \caption{This figure shows the late--time behavior of the scalar field on a double--logarithmic scale, plotting $\ln|\tilde{\psi}|$ versus $\ln t$ for a fixed mass parameter $M = 1$ and charge values $\xi = 0.3, 0.5, 0.7,$ and $0.9$. The panels are displayed by multipole number, with $\ell = 0$ in the top--left plot, $\ell = 1$ in the top--right, and $\ell = 2$ displayed in the bottom panel, which highlights the power--law decay characteristic of the tail regime.}
    \label{timedomainlnlnspsi}
\end{figure}

Lastly, Fig.~\ref{timedomainlnlnspsi} presents the evolution of $\tilde{\psi}$ as a function of time on a logarithmic--logarithmic scale, using the same panel layout as before. This representation highlights the asymptotic regime, making the power--law decay at late times explicit and confirming the appearance of the characteristic tails that succeed the quasinormal ringing stage, as one should naturally expect.

\subsection{Vector field }

\begin{figure}
    \centering
    \includegraphics[scale=0.53]{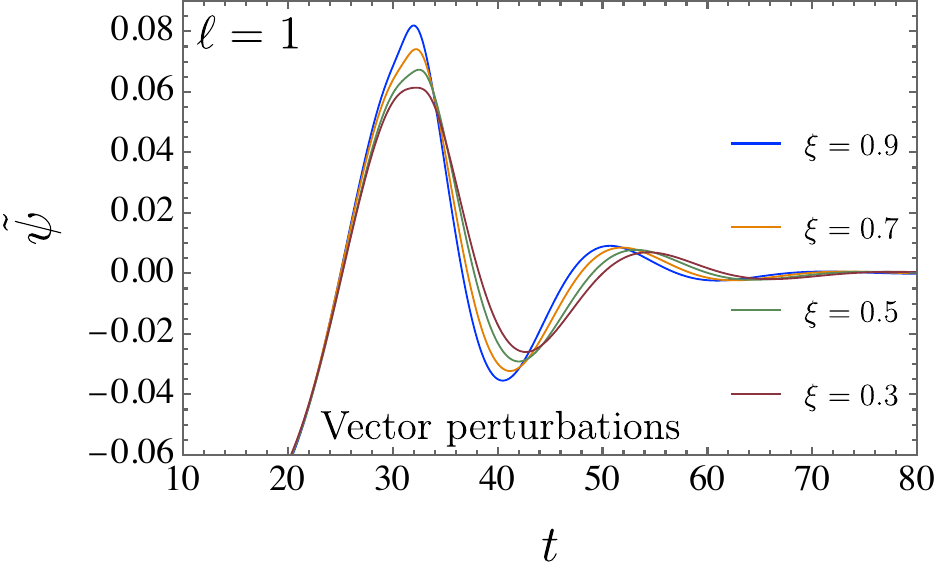}
    \includegraphics[scale=0.53]{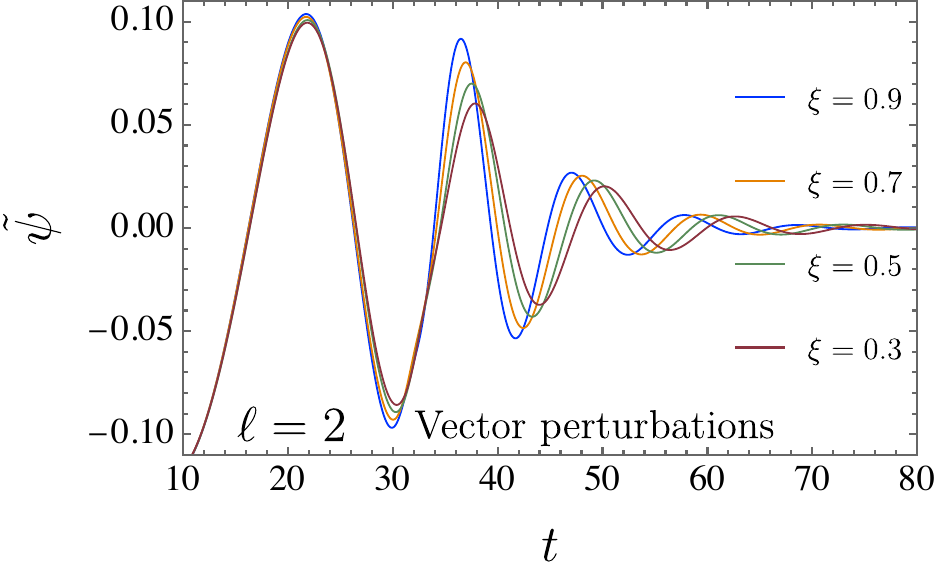}
    \includegraphics[scale=0.53]{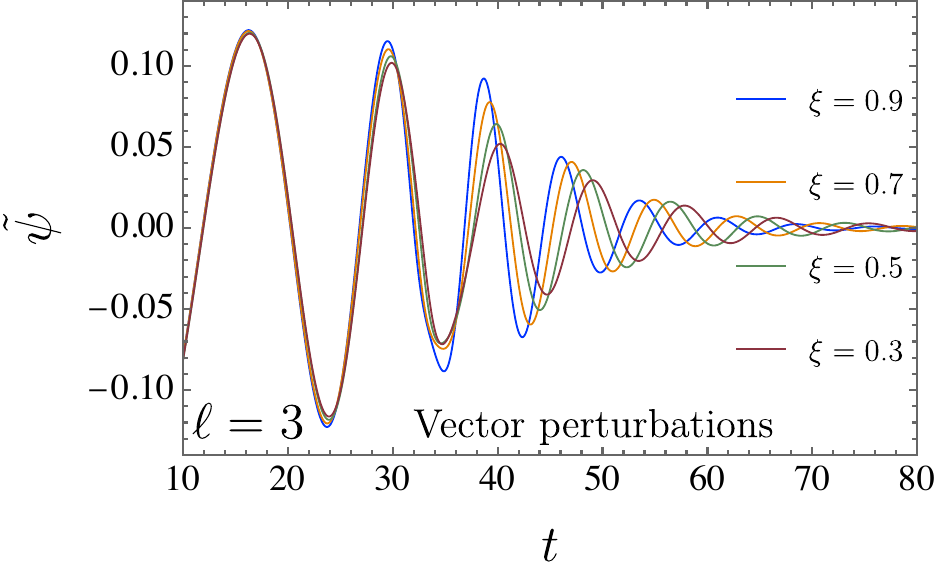}
    \caption{The temporal evolution of the vector perturbation $\tilde{\psi}$ concerning the time $t$ is shown for a black hole with $M = 1$ and charge parameters $\xi = 0.3, 0.5, 0.7,$ and $0.9$. The panels correspond to $\ell = 1$ (left), $\ell = 2$ (right), and $\ell = 3$ (bottom), illustrating the waveform’s decay pattern for each multipole configuration.}
    \label{vpsi}
\end{figure}

The evolution of vector--type perturbations is analyzed by propagating the waveform $\tilde{\psi}$ over time for a fixed black hole mass $M = 1$ and several choices of the parameter $\xi$. The outcomes, shown in Fig.~\ref{vpsi} for $\xi = 0.3, 0.5, 0.7,$ and $0.9$, are shown by angular index: the left panel displays $\ell = 1$ while the right panel corresponds to $\ell = 2$ and bottom one correspond to $\ell = 3$. Across all cases, the profiles reveal oscillatory behavior with amplitudes that steadily decay, characteristic of the quasinormal ringing phase.

\begin{figure}
    \centering
    \includegraphics[scale=0.53]{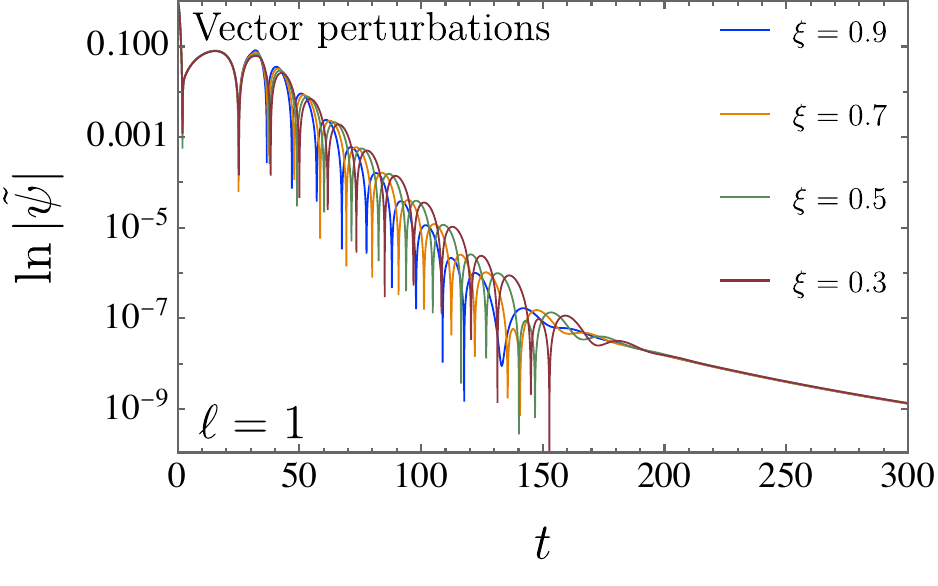}
    \includegraphics[scale=0.53]{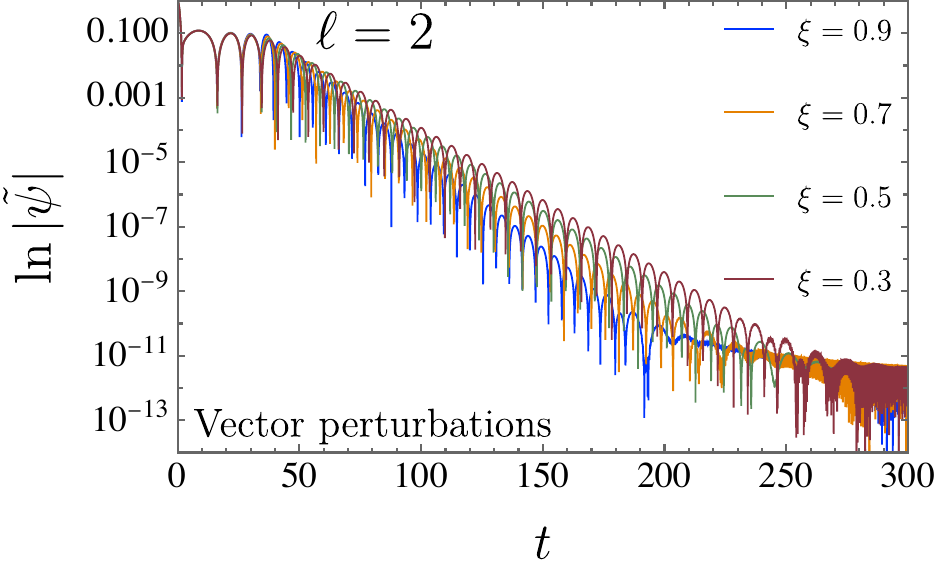}
    \includegraphics[scale=0.53]{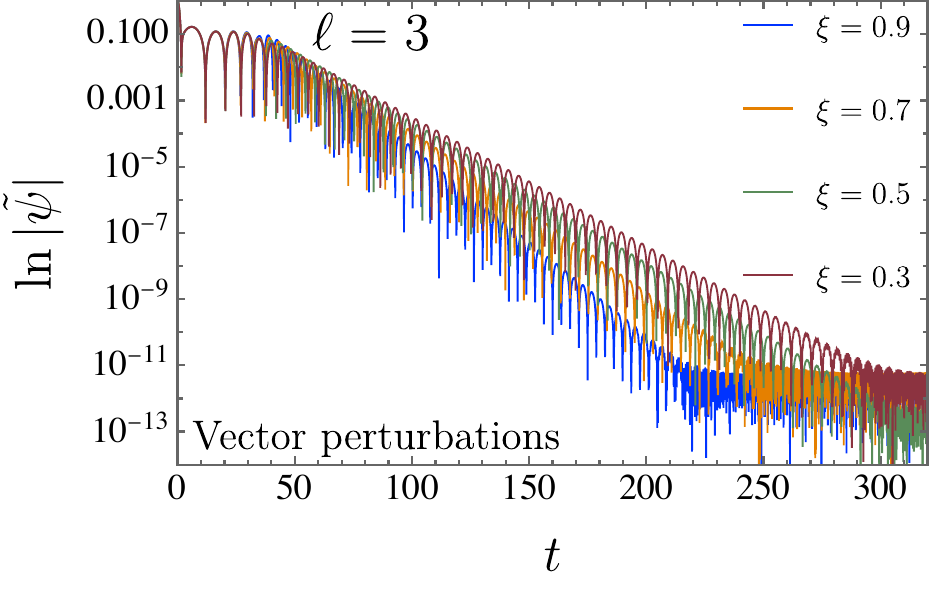}
    \caption{The time evolution of $\ln|\tilde{\psi}|$ with respect to the time $t$ for vector perturbations in a background with $M = 1$. The analysis is performed for four distinct values of the parameter $\xi$ ($0.3$, $0.5$, $0.7$, and $0.9$). Results are shown by angular momentum number, with $\ell = 1$ shown in the left panel, $\ell = 2$ in the right panel, and $\ell = 3$ in the bottom panel}
    \label{lnvpsi}
\end{figure}

Fig.~\ref{lnvpsi} displays the quantity $\ln|\tilde{\psi}|$ versus time, providing a clearer view of how the signal decays. The curves show that the evolution begins with a regime of exponential damping, after which the decay rate slows and follows a power--law profile. This crossover indicates the development of the late--time tail, a well--known effect caused by the scattering of perturbations by the spacetime considered here.

\begin{figure}
    \centering
    \includegraphics[scale=0.53]{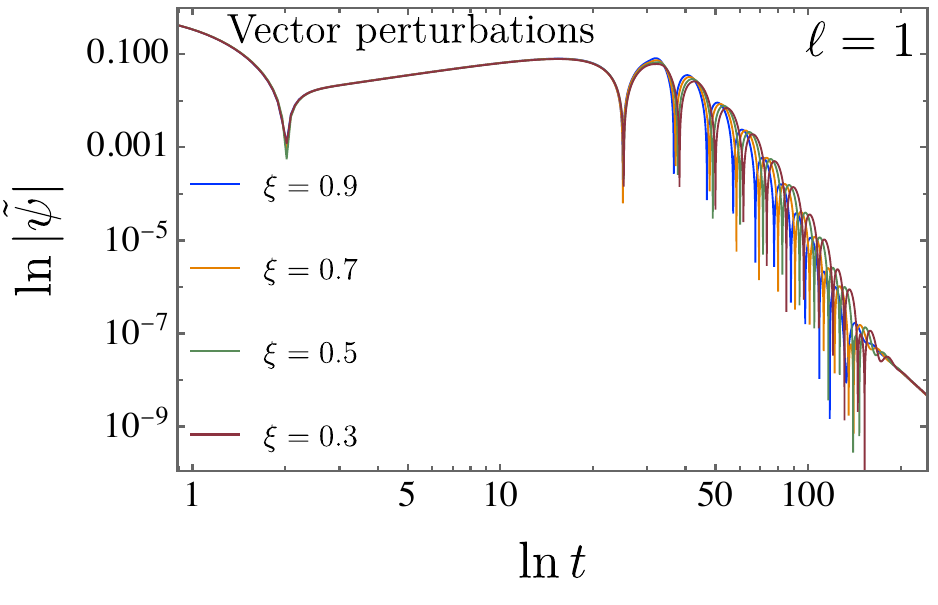}
    \includegraphics[scale=0.53]{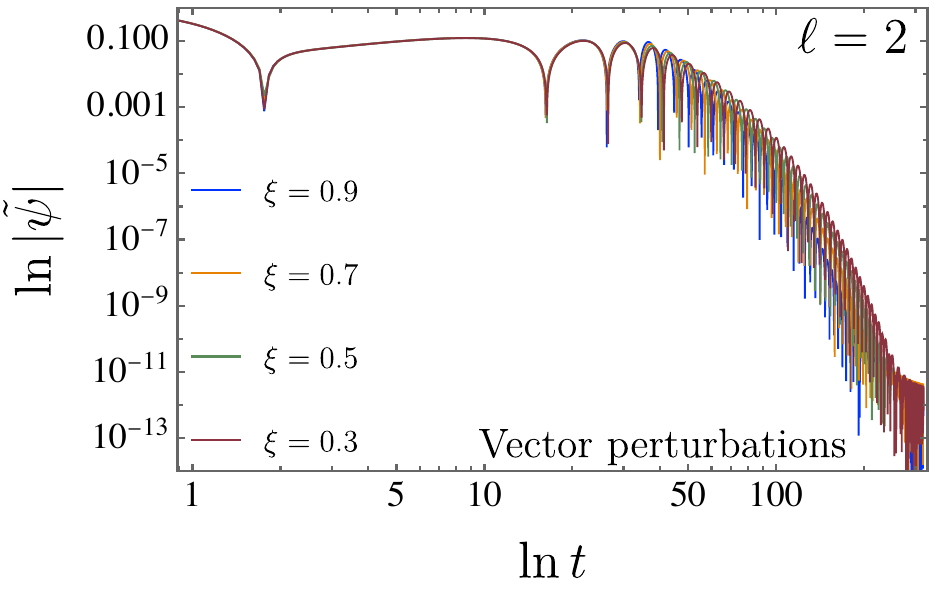}
    \includegraphics[scale=0.53]{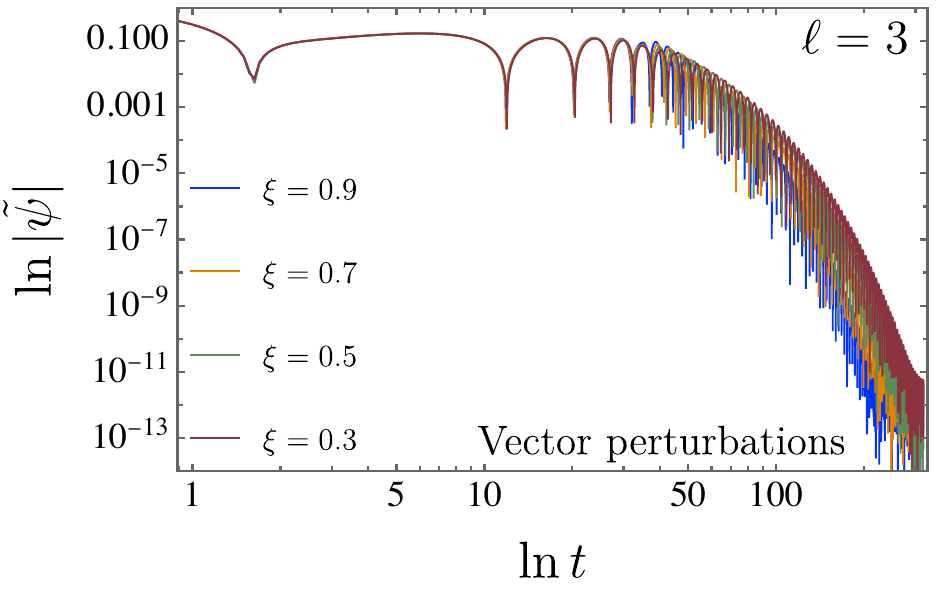}
    \caption{The double--logarithmic representation of $\tilde{\psi}$, plotting $\ln|\tilde{\psi}|$ against $\ln t$ for vector perturbations with $M = 1$. The analysis includes four values of the parameter $\xi$ ($0.3$, $0.5$, $0.7$, and $0.9$). The panels correspond to the three angular modes: $\ell = 1$ (left), $\ell = 2$ (right), and $\ell = 3$ (bottom), emphasizing the late--time power--law decay that characterizes the tail behavior.}
    \label{lnlnvpsi}
\end{figure}

In addition, Fig.~\ref{lnlnvpsi} presents the waveform evolution on a double--logarithmic scale, where $\ln|\tilde{\psi}|$ is plotted against $\ln t$. This representation highlights the late--time dynamics, clearly showing that the signal settles into a power--law decay. The slope of this tail remains nearly unchanged for different values of $\xi$ and $\ell$, indicating that the asymptotic behavior is largely universal.

\subsection{Tensor field }

\begin{figure}
    \centering
    \includegraphics[scale=0.53]{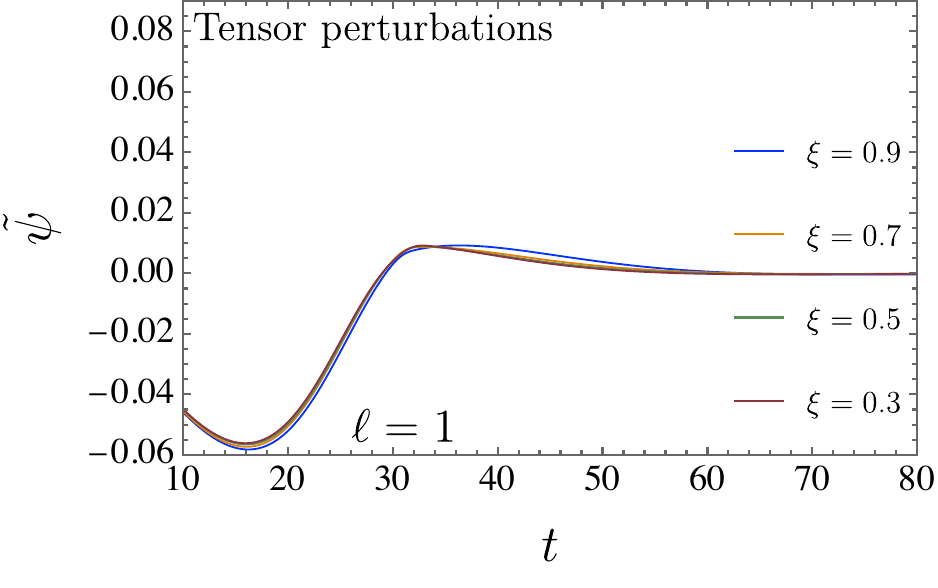}
    \includegraphics[scale=0.53]{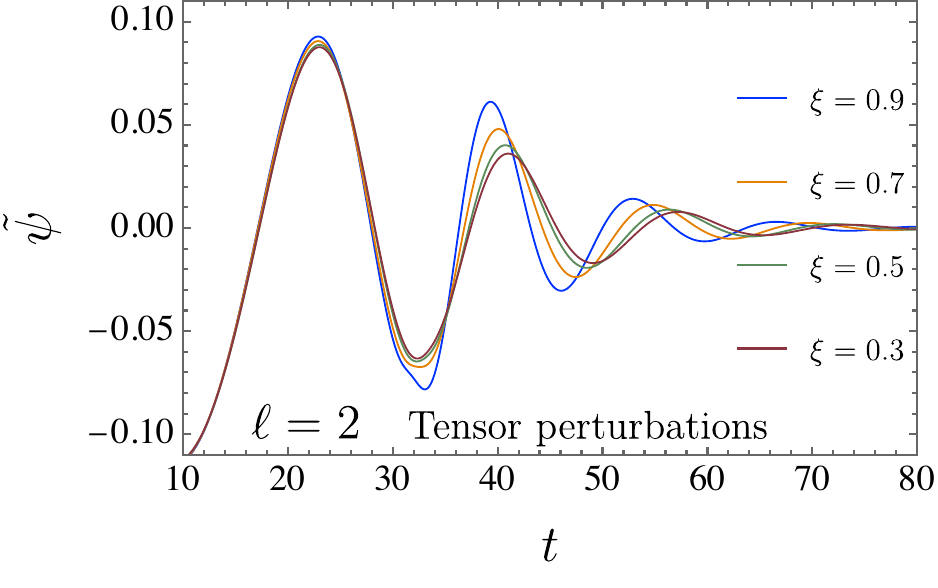}
    \includegraphics[scale=0.53]{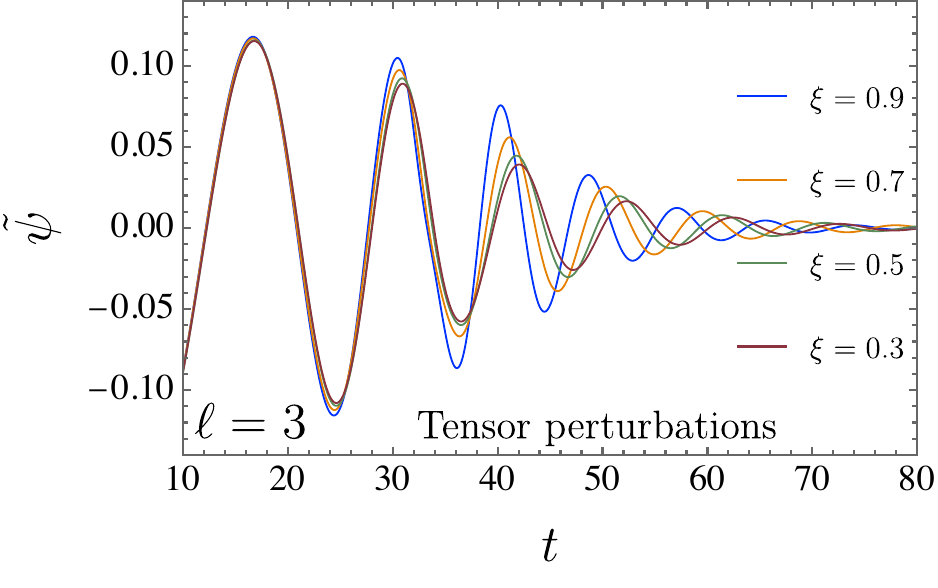}
    \caption{It is shown the temporal evolution of the tensor perturbation $\tilde{\psi}$ {with respect to the time $t$,} for a black hole with $M = 1$ and charge parameters $\xi = 0.3, 0.5, 0.7,$ and $0.9$. The results are grouped by angular index, with $\ell = 1$ displayed in the left panel, $\ell = 2$ in the right panel, and $\ell = 3$ in the bottom panel, illustrating how the waveform evolves for each mode.}
    \label{tpsi}
\end{figure}

\begin{figure}
    \centering
    \includegraphics[scale=0.53]{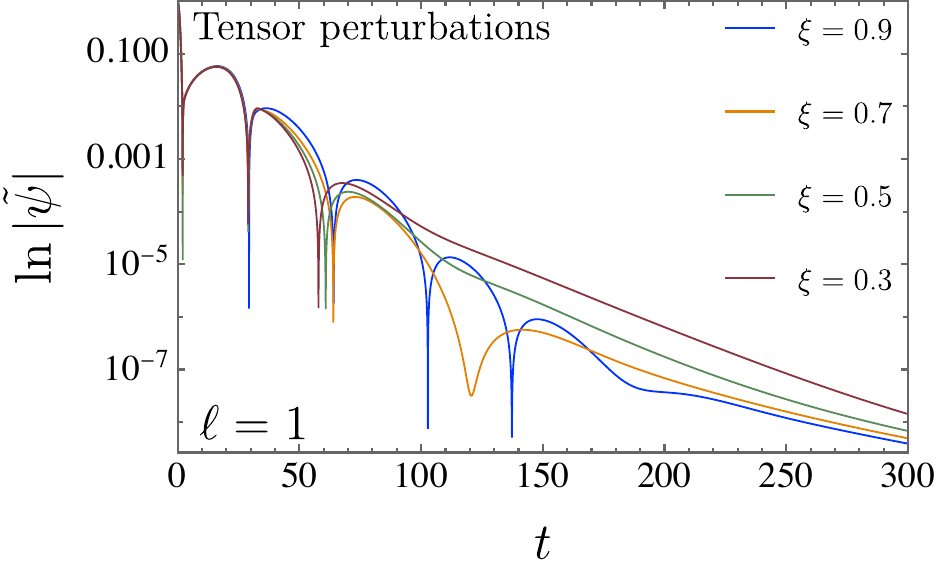}
    \includegraphics[scale=0.53]{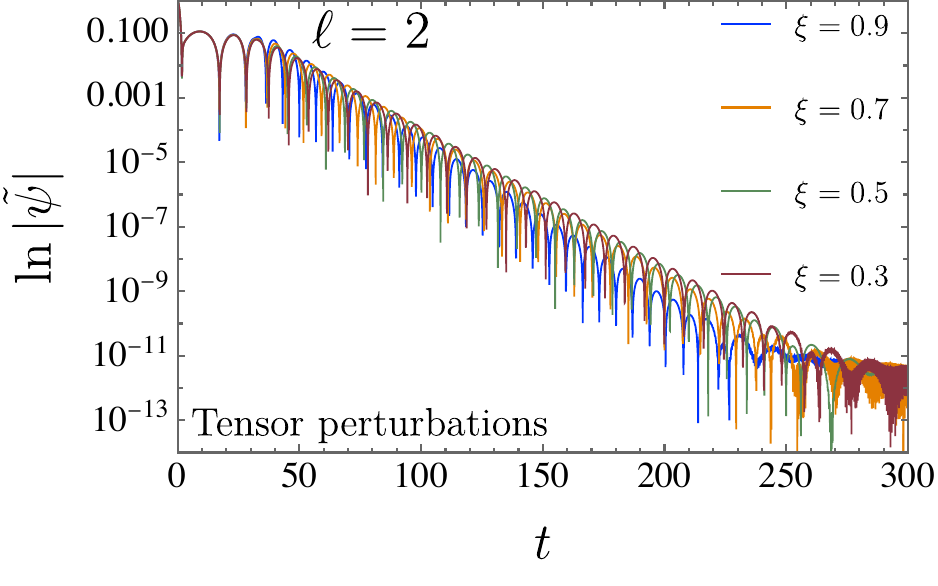}
    \includegraphics[scale=0.53]{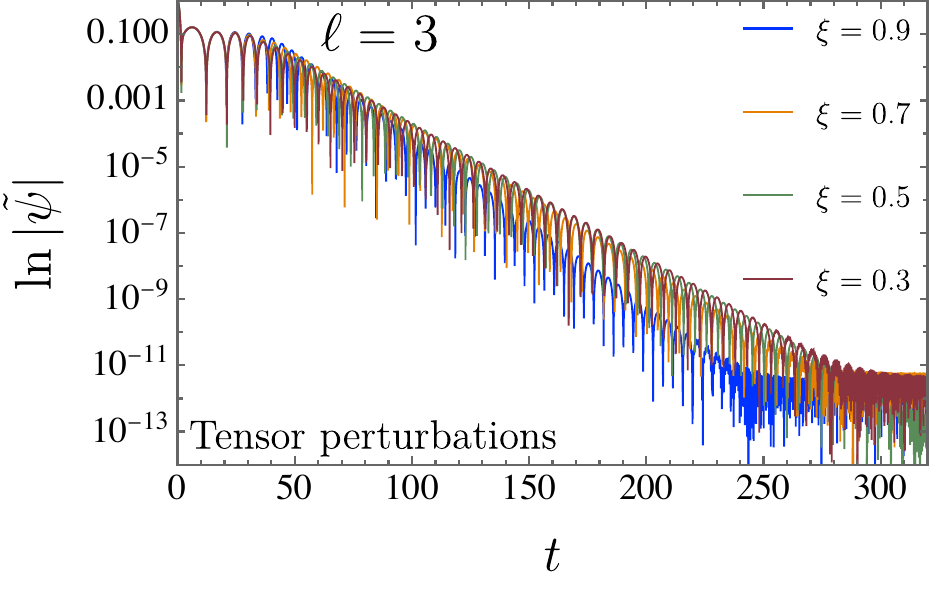}
    \caption{It is displayed the logarithmic time profile of the tensor perturbation, showing $\ln|\tilde{\psi}|$ as a function of $t$ for $M = 1$ and $\xi = 0.3, 0.5, 0.7,$ and $0.9$. The plots are displayed by multipole number: $\ell = 1$ (left), $\ell = 2$ (right), and $\ell = 3$ (bottom).}
    \label{lntpsi}
\end{figure}

\begin{figure}
    \centering
    \includegraphics[scale=0.53]{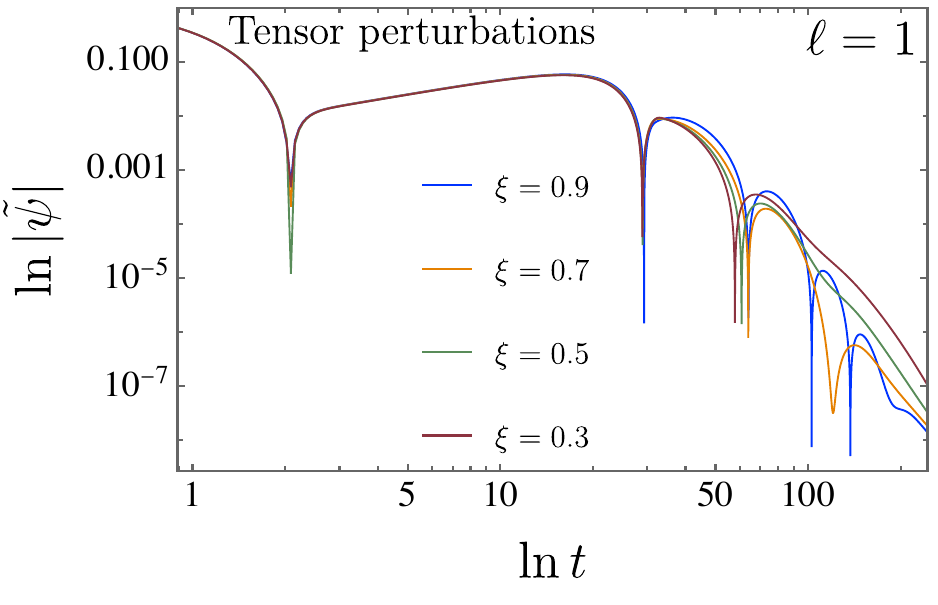}
    \includegraphics[scale=0.53]{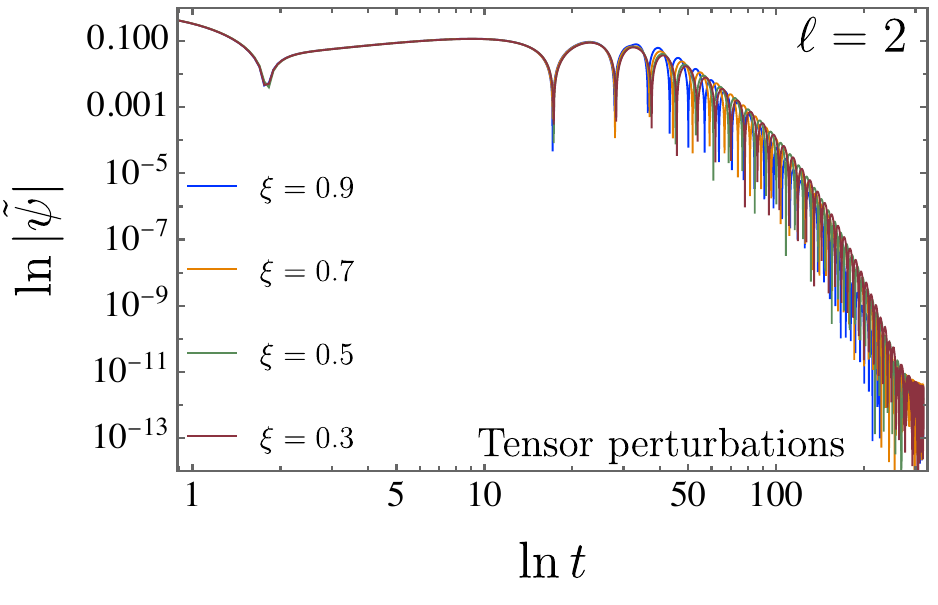}
    \includegraphics[scale=0.53]{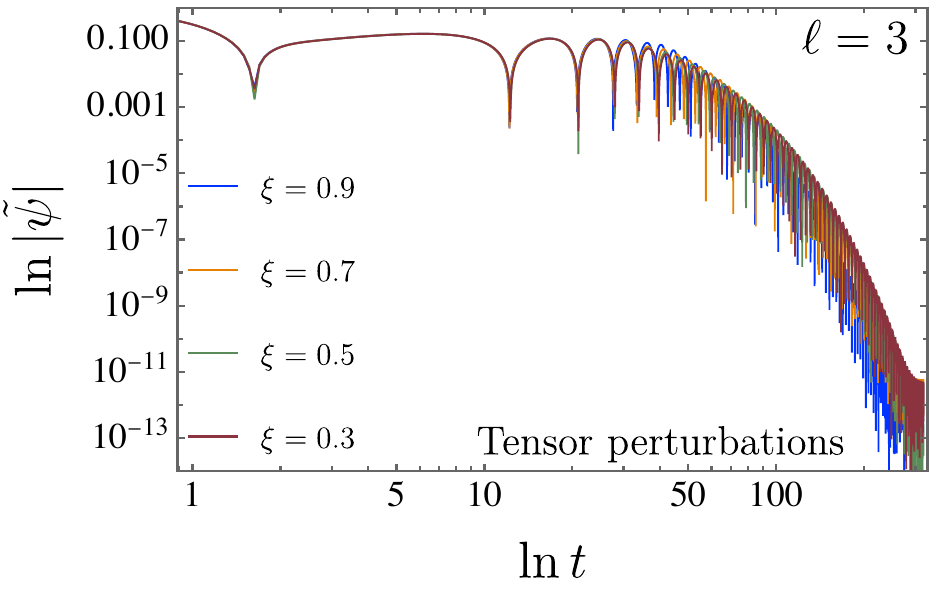}
    \caption{The asymptotic (late time) regime of tensor perturbations using a double--logarithmic representation, where $\ln|\tilde{\psi}|$ is plotted as a function of $\ln t$ for a black hole with $M = 1$. The analysis considers $\xi = 0.3, 0.5, 0.7,$ and $0.9$, with results separated by multipole number: $\ell = 1$ in the left panel, $\ell = 2$ in the right, and $\ell = 3$ in the bottom panel}
    \label{lnlntpsi}
\end{figure}

This subsection examines the time evolution of tensor fluctuations for our black hole. Fig.~\ref{tpsi} presents the waveform $\tilde{\psi}$ as a function of time for a fixed mass $M = 1$ and charge parameters $\xi = 0.3, 0.5, 0.7,$ and $0.9$. The results are displayed by angular mode, with $\ell = 1$ shown in the left panel, $\ell = 2$ in the right, and $\ell = 3$ in the bottom panel. To better visualize the damping behavior, Fig.~\ref{lntpsi} plots $\ln|\tilde{\psi}|$ as a function of $t$, clearly revealing the exponential decay followed by a slower attenuation phase. Furthermore, Fig.\ref{lnlntpsi} uses a double--logarithmic scale, $\ln|\tilde{\psi}|$ versus $\ln t$, to highlight the late--time regime and confirm therefore the characteristic power--law tail behavior for each choice of $\xi$ and $\ell$.

\subsection{Spinor field }

\begin{figure}
    \centering
    \includegraphics[scale=0.53]{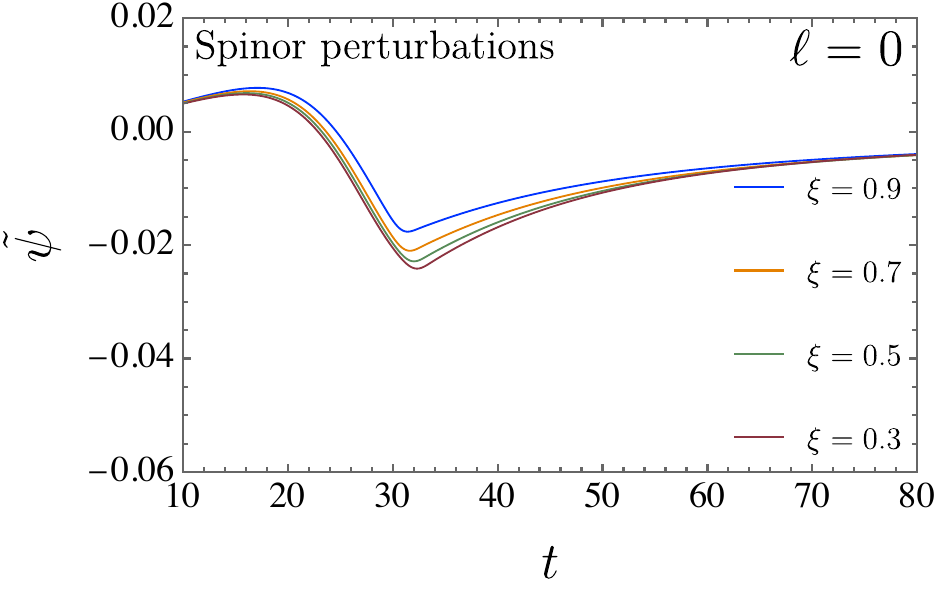}
    \includegraphics[scale=0.53]{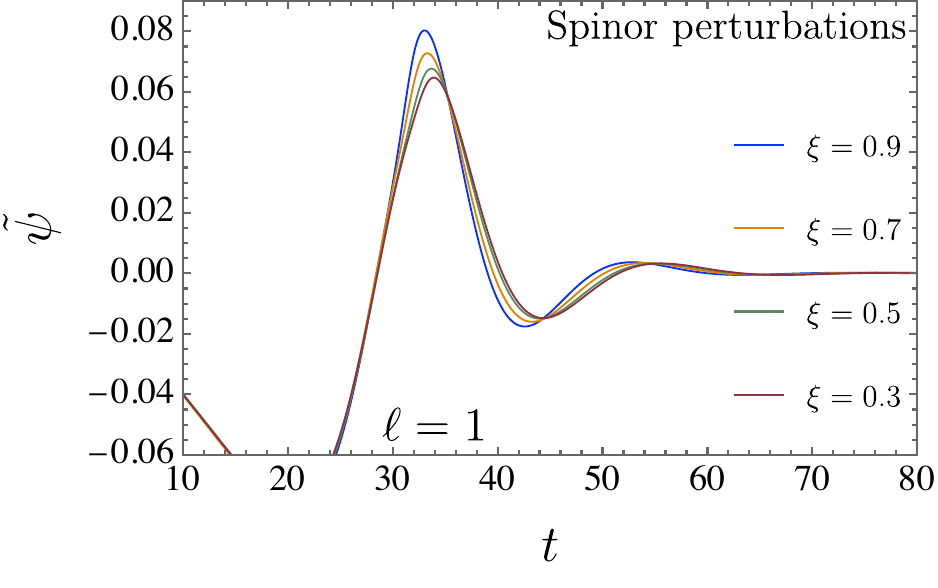}
    \includegraphics[scale=0.53]{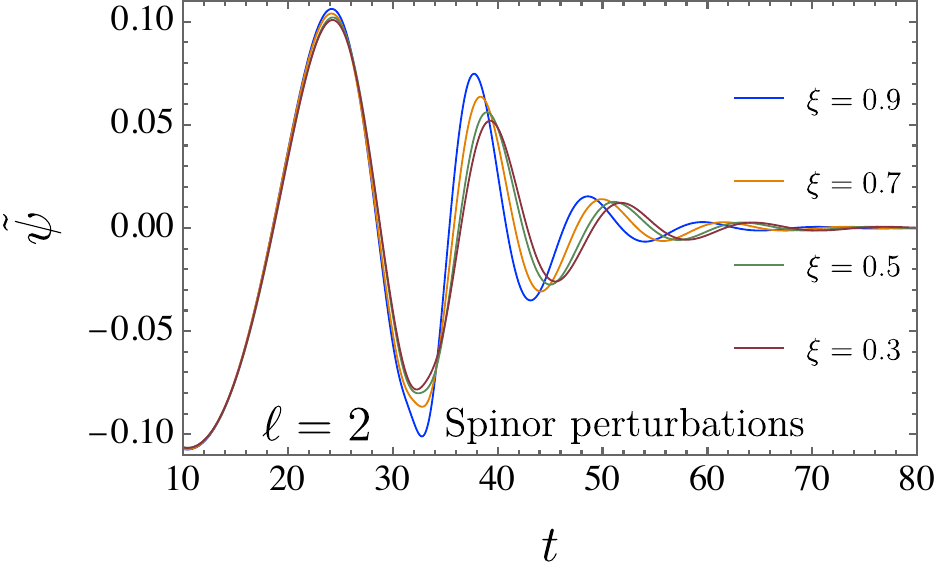}
    \caption{Temporal evolution of the tensor perturbation $\tilde{\psi}$ concerning the time $t$, for $M = 1$ and $\xi = 0.3, 0.5, 0.7,$ and $0.9$. The panels show the results for $\ell = 1$ (left), $\ell = 2$ (right), and $\ell = 3$ (bottom), showing the distinct waveform profiles for each multipole configuration.}
    \label{spinpsi}
\end{figure}

To complete the analysis, the final class of perturbations—those associated with spinor fields—is investigated in this subsection. Fig.~\ref{spinpsi} displays the time evolution of the spinor waveform $\tilde{\psi}$ for $M = 1$ and charge parameters $\xi = 0.3, 0.5, 0.7,$ and $0.9$. The panels correspond to the angular modes $\ell = 1$ (left), $\ell = 2$ (right), and $\ell = 3$ (bottom), showing how the signal evolves in each case.

\begin{figure}
    \centering
    \includegraphics[scale=0.53]{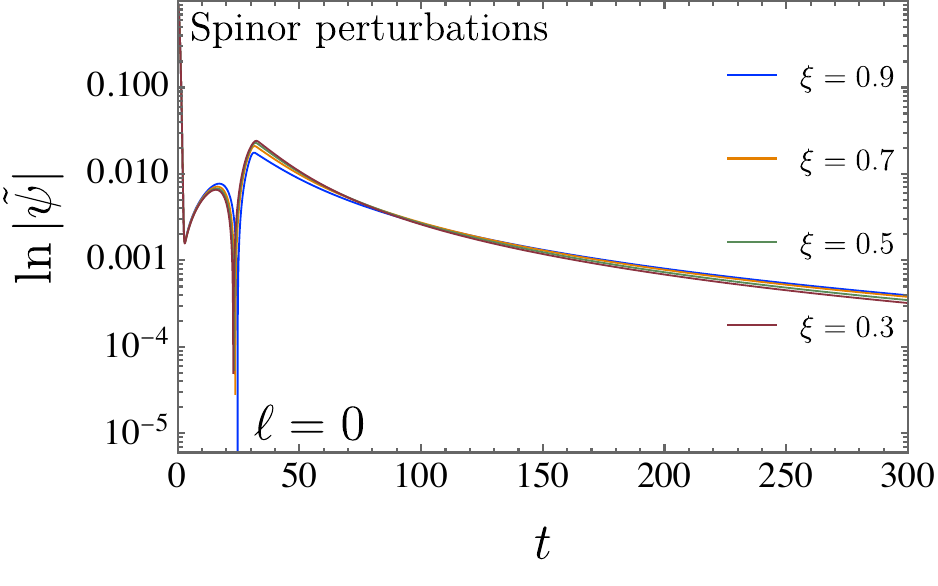}
    \includegraphics[scale=0.53]{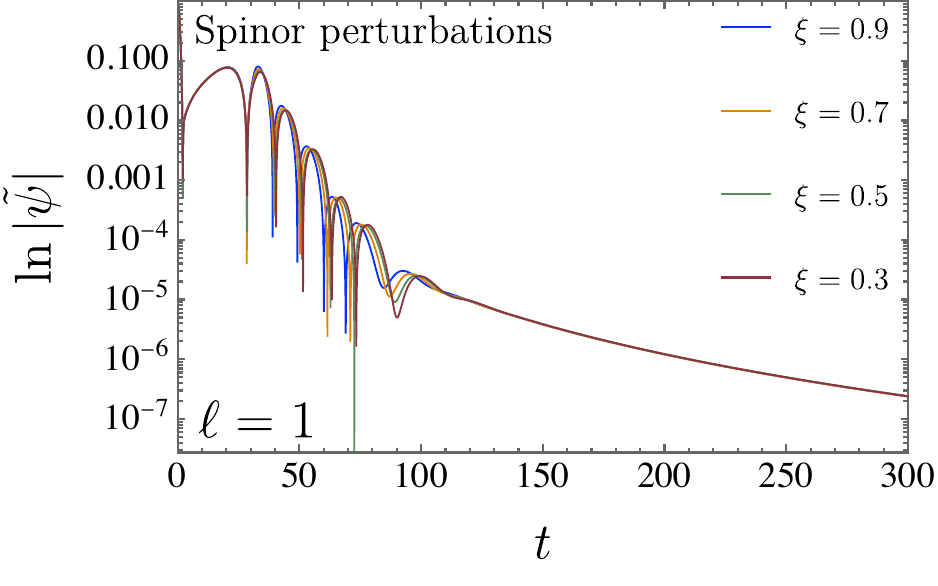}
    \includegraphics[scale=0.53]{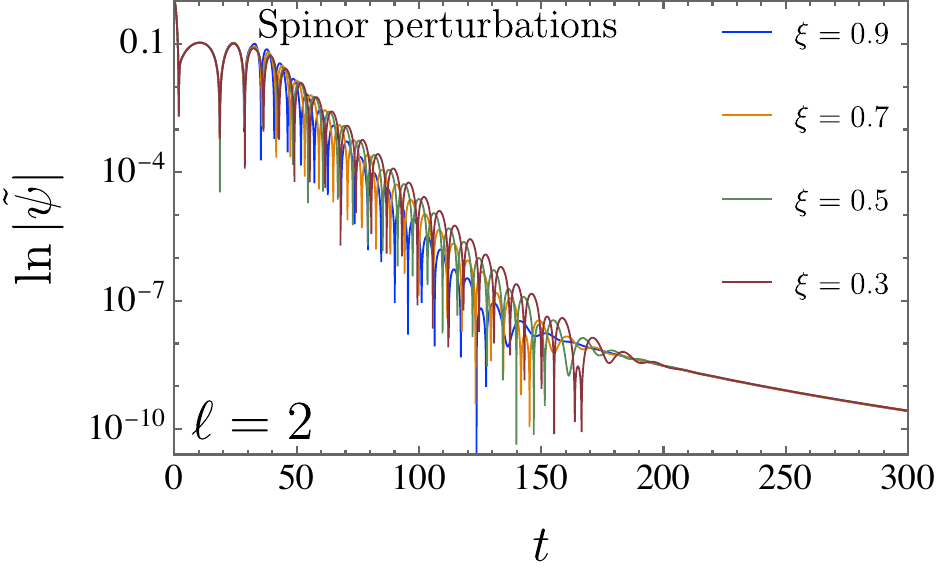}
    \caption{Logarithmic representation of the tensor perturbation, with $\ln|\tilde{\psi}|$ shown as a function of $t$ for $M = 1$ and $\xi = 0.3, 0.5, 0.7,$ and $0.9$. Results are arranged by multipole number: $\ell = 1$ in the left panel, $\ell = 2$ in the right, and $\ell = 3$ in the bottom panel.}
    \label{lnspinpsi}
\end{figure}

\begin{figure}
    \centering
    \includegraphics[scale=0.53]{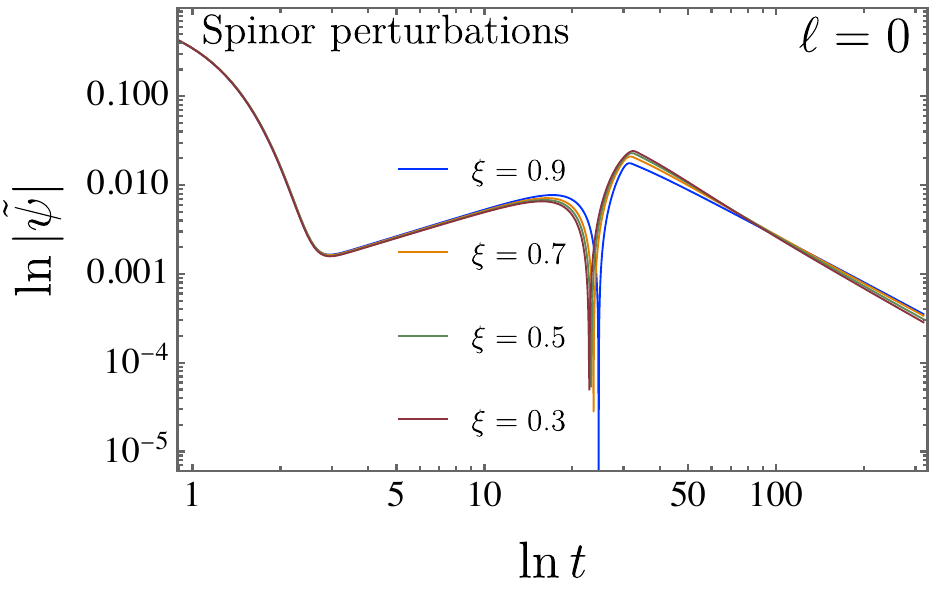}
    \includegraphics[scale=0.53]{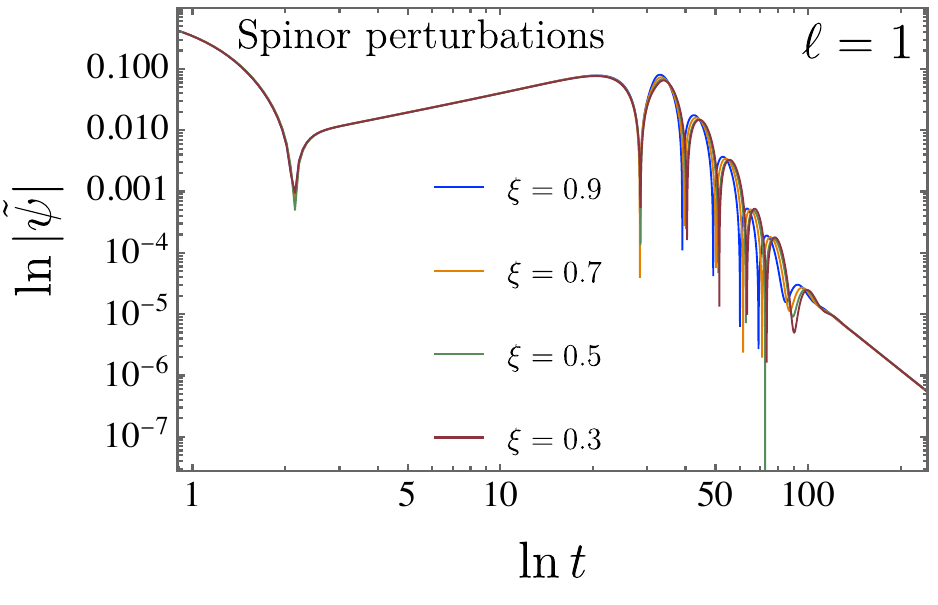}
    \includegraphics[scale=0.53]{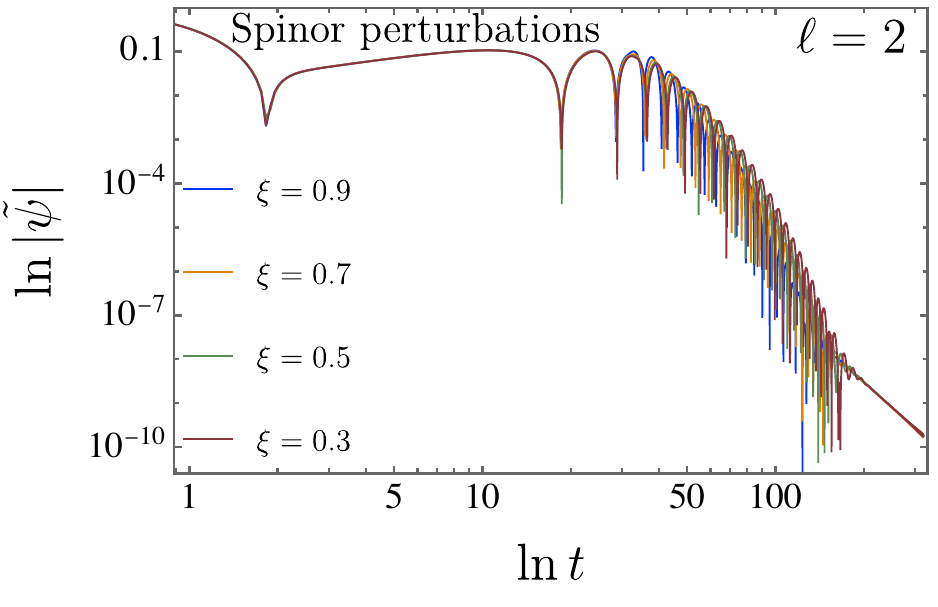}
    \caption{Late--time behavior of tensor perturbations shown in a double–logarithmic plot, with $\ln|\tilde{\psi}|$ versus $\ln t$ for $M = 1$ and $\xi = 0.3, 0.5, 0.7,$ and $0.9$. The panels correspond to: $\ell = 1$ (left), $\ell = 2$ (right), and $\ell = 3$ (bottom).}
    \label{lnlnspinpsi}
\end{figure}

For a clearer view of the damping behavior, as we have done in the previous subsections, Fig.\ref{lnspinpsi} presents $\ln|\tilde{\psi}|$ as a function of time, making the quasinormal ringing and subsequent decay phase evident. Also, Fig.\ref{lnlnspinpsi} adopts a double--logarithmic representation, plotting $\ln|\tilde{\psi}|$ against $\ln t$, which highlights therefore the asymptotic regime and demonstrates the emergence of the expected power--law tails for all values of $\xi$ and $\ell$.

Here, a remark is in order in light of the cases discussed above: at late times, the decay of all perturbations is characterized by power--law tails, which arise from the backscattering of the fields off the asymptotic spacetime curvature. This behavior is governed by the far--region structure of the corresponding effective potentials. In our present model, the effective potentials associated with scalar, vector, tensor, and spinor perturbations can be expressed in the generic form
\begin{equation}
\mathcal{V}(r,\ell,\xi)=A(r,\xi)\left[\frac{\nu}{r^{2}}+\Delta(r,\xi)\right],
\end{equation}
where $\nu=\ell(\ell+1)$ for the bosonic sectors and $\nu=(\ell+\tfrac{1}{2})^{2}$ for the spinor sector, while $\Delta(r,\xi)$ collects derivative contributions involving the metric functions. Using the corresponding metric components $A(r,\xi)$ and $B(r,\xi)$,
we find that, for all perturbation types, the effective potentials admit the asymptotic expansion
\begin{equation}
\mathcal{V}(r,\ell,\xi)=\frac{\nu}{r^{2}}+\mathcal{O}(r^{-3})+\mathcal{O}(\xi\,r^{-4})+\cdots,
\qquad r\to\infty.
\end{equation}
Therefore, the higher--order curvature parameter $\xi$ enters only through subleading terms in the far region, while the leading long--range behavior responsible for the late--time tails remains unchanged. As a consequence, the power--law decay exponents are expected to coincide with those of the corresponding general relativity case, with $\xi$ primarily affecting the amplitude of the tails and the transition time between the quasinormal ringing and the late--time regime. Moreover, notice that this interpretation is consistent with the numerical time--domain profiles obtained for all perturbation sectors.

\section{Geodesics}

In gravitational physics, geodesics play a essential role by connecting the geometry of spacetime to the motion of free particles. When considered in the setting of higher--order curvature--scalar gravity, analyzing geodesic motion becomes a powerful tool to explore how the presence of curvature--scalar couplings reshapes the underlying geometry and influences particle trajectories. In the case of null geodesics, this approach is particularly effective for examining how the HOCG parameter $\xi$ modifies light propagation and the causal structure of the spacetime.

Thereby, the geodesic equation takes the general form
\begin{equation}
\frac{\mathrm{d}^{2}x^{\mu}}{\mathrm{d}\mathrm{t}^{2}} + \Gamma^\mu_{\nu \lambda} \frac{\mathrm{d}x^{\nu}}{\mathrm{d}\mathrm{t}}\frac{\mathrm{d}x^{\lambda}}{\mathrm{d}\mathrm{t}} = 0.
\label{1g1e1o1d1e1sic1f1u1ll}
\end{equation}
Here, $\Gamma$ denotes the Christoffel symbols and $\mathrm{t}$ is an affine parameter along the geodesic. The main objective is to investigate how the parameter $\xi$ influences the motion of massless particles. This requires solving the coupled system of differential equations derived from Eq.\eqref{1g1e1o1d1e1sic1f1u1ll}. The equation produces four separate relations, one for each spacetime coordinate, which must be integrated simultaneously using the metric given in Eq.\eqref{mainmetric}.

\begin{align}
&\frac{\mathrm{d}t^{\prime}}{\mathrm{d}\mathrm{t}} =
-\frac{2 r' t' \left(M r-\xi ^2\right)}{r \left(-2 M r+\xi ^2+r^2\right)},\\
&\frac{\mathrm{d}r^{\prime}}{\mathrm{d}\mathrm{t}}  = \frac{\left(2 M \xi ^{3/2}-2 M r^3+r^4\right) \left(\left(t'\right)^2 \left(\xi ^2-M r\right)+r^4 \left(\left(\theta '\right)^2+\sin ^2(\theta ) \left(\varphi '\right)^2\right)\right)}{r^7}\\ \nonumber
&~~~~~+\frac{M \left(r^3-4 \xi ^{3/2}\right) \left(r'\right)^2}{-2 M r^4+2 M \xi ^{3/2} r+r^5},\\
&\frac{\mathrm{d}\theta^{\prime}}{\mathrm{d}\mathrm{t}} = \sin (\theta ) \cos (\theta ) \left(\varphi '\right)^2-\frac{2 \theta ' r'}{r},\\
&\frac{\mathrm{d}\varphi^{\prime}}{\mathrm{d}\mathrm{t}} = -\frac{2 \varphi ' \left(r'+r \theta ' \cot (\theta )\right)}{r},
\end{align}
with the prime ($'$) indicating derivatives taken with respect to the affine parameter.

 \begin{figure}
    \centering
    \includegraphics[scale=0.65]{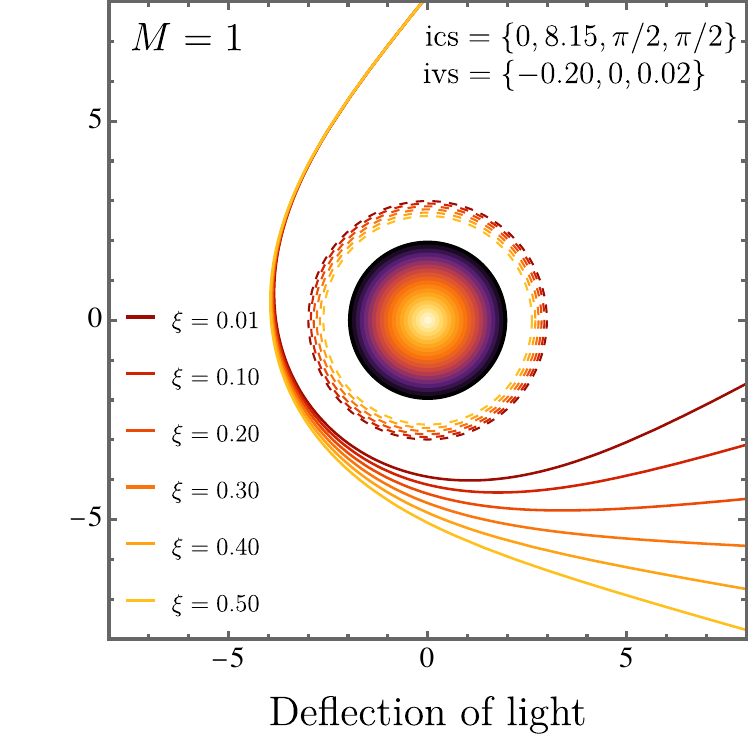}
    \includegraphics[scale=0.65]{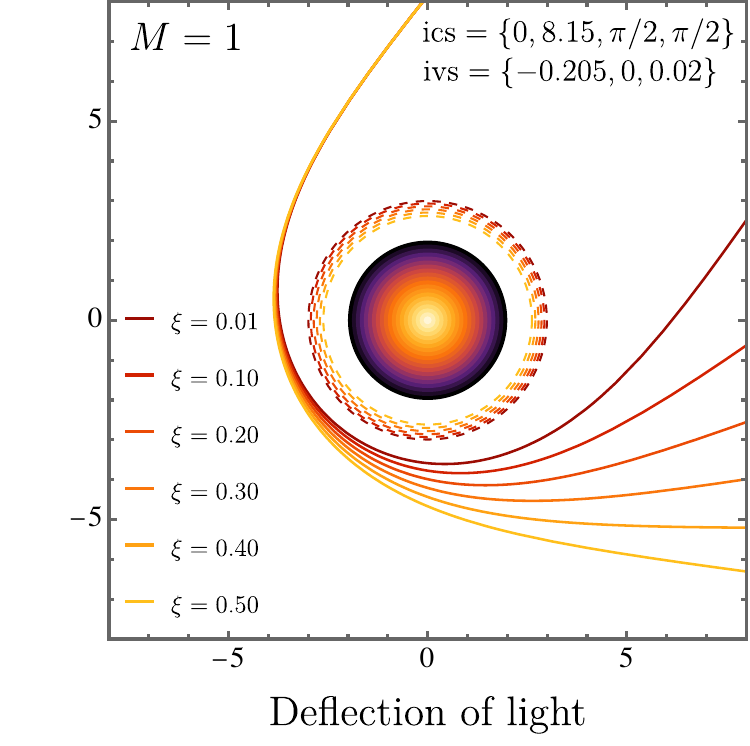}
    \includegraphics[scale=0.65]{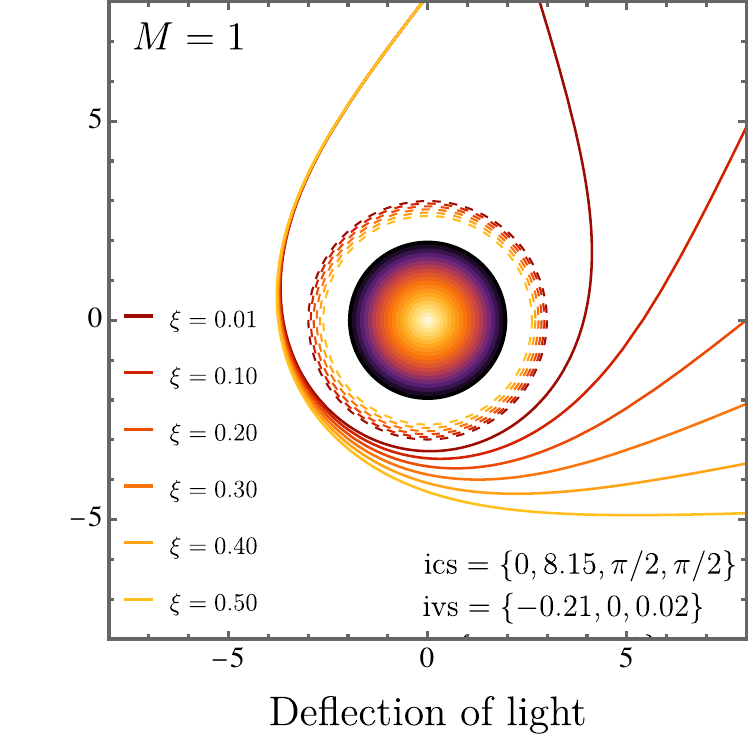}
    \includegraphics[scale=0.65]{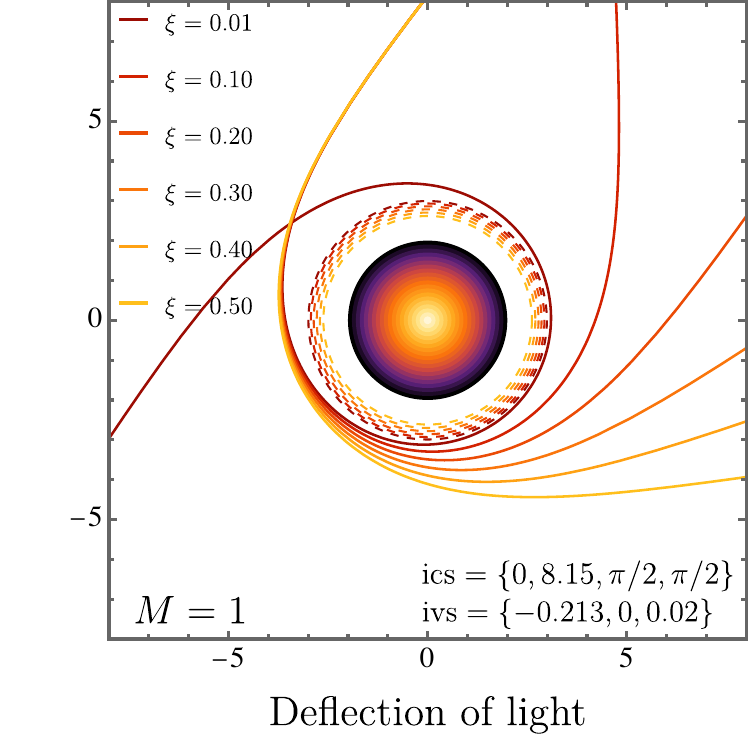}
    \caption{The geodesic trajectories are obtained through numerical integration using $M = 1$ and several choices of $\xi$. Dashed circles in the plots mark the locations of the corresponding photon spheres.}
    \label{geodespath}
\end{figure}


Fig.~\ref{geodespath} displays the numerically integrated null geodesics for $M = 1$ with $\xi$ varying from $0$ to $0.5$. The filled disk marks the event horizon, the dashed circles indicate the photon sphere, and the solid curves trace the light trajectories.

The results illustrate how photon paths respond to changes in the HOCG parameter $\xi$. As $\xi$ decreases, light rays experience stronger bending, whereas larger values of $\xi$ lead to trajectories that are less deflected, reflecting a weaker effective curvature near the black hole. This behavior signals that the parameter $\xi$ directly modulates the gravitational field felt by photons, subtly modifying their propagation and thus influencing observable lensing patterns. These implications for gravitational lensing will be analyzed in detail in the next section.

\section{Critical orbits and shadows }

The study of black hole shadows has become a major focus in modern gravitational research \cite{hamil2023thermodynamics,araujo2024charged,zeng2022shadows,araujo2025remarks}, receiving renewed attention after the groundbreaking Event Horizon Telescope (EHT) images of $Sgr A^{*}$ and $M87$ \cite{ball2019first, gralla2021can, akiyama2019first}. These observations have transformed the shadow into a powerful manner for testing strong--field gravity and probing the near--horizon structure of compact objects for instance.

To investigate the shadow in the present context, we take the background metric given in Eq.~(\ref{mainmetric}) as the starting point and analyze the propagation of photons through the Lagrangian formalism, expressed as
\ie
\mathcal{L} = \frac{1}{2}{g_{\mu \nu }}{{\dot x}^\mu }{{\dot x}^\nu }.
\fe
Equivalently, it can be written as
\ie
\label{lagrangian}
\mathcal{L} = \frac{1}{2}\Big[ - A(r,\xi){{\dot t}^2} + \frac{1}{B(r,\xi)}{{\dot r}^2} + C(r,\xi){{\dot \theta }^2} + D(r,\xi){{\mathop{\rm \sin}\nolimits} ^2}\, \theta {{\dot \varphi }^2}\Big].
\fe

Using the Euler--Lagrange equations and restricting motion to the equatorial plane ($\theta = \pi/2$), one obtains two constants of motion: the conserved energy $E$ and angular momentum $L$, as it is commonly reported in the literature. Their explicit forms are derived as
\ie
\label{eqdcons}
E = A(r,\xi)\dot t \quad\mathrm{and}\quad L = D(r,\xi)\dot \varphi,
\end{equation}
and taking into account solely the massless scenario, we obtain
\ie
\label{light-like}
- A(r,\xi){{\dot t}^2} +  \frac{1}{B(r,\xi)}{{\dot r}^2} + D(r,\xi){{\dot \varphi }^2} = 0.
\fe
Substituting the conserved quantities from Eq.(\ref{eqdcons}) into the null geodesic condition of Eq.(\ref{light-like}) and simplifying leads to the following expression:
\begin{equation}\label{rdot}
\frac{{{{\dot r}^2}}}{{{{\dot \varphi }^2}}} = {\left(\frac{{\mathrm{d}r}}{{\mathrm{d}\varphi }}\right)^2} = {D(r,\xi)}B(r,\xi) \left(\frac{{D(r,\xi)}}{{A(r,\xi)}}\frac{{{E^2}}}{{{L^2}}} - 1\right).
\end{equation}
In addition, we also have
\ie
\frac{\mathrm{d}r}{\mathrm{d}\lambda} = \frac{\mathrm{d}r}{\mathrm{d}\varphi} \frac{\mathrm{d}\varphi}{\mathrm{d}\lambda}  = \frac{\mathrm{d}r}{\mathrm{d}\varphi}\frac{L}{D(r,\xi)}, 
\fe
with
\ie
\Dot{r}^{2} = \left( \frac{\mathrm{d}r}{\mathrm{d}\lambda} \right)^{2} =\left( \frac{\mathrm{d}r}{\mathrm{d}\varphi} \right)^{2} \frac{L^{2}}{D(r,\xi)^{2}},
\fe
which allows the effective potential $\mathrm{V}(\Theta,\ell)$ to be expressed in the form given by:
\ie
\label{ghhhjjjj}
\mathrm{V}(r,\xi,\ell) = D(r,\xi) B(r,\xi)\left(\frac{{D(r,\xi)}}{{A(r,\xi)}}\frac{{{E^2}}}{{{L^2}}} - 1\right)\frac{L^{2}}{ D(r,\xi)^{2}}.
\fe

Having set up the necessary framework, the next step is to locate the photon spheres. This is accomplished by applying the constraint
\ie
\mathrm{V}(r,\xi,\ell)=0, \quad\quad \frac{\mathrm{d} \,{\mathrm{V}(r,\xi,\ell)}}{\mathrm{d}r} = 0 .
\fe

Defining the critical impact parameter as $b_c = L/E$, the above requirement reduces to
\ie
\label{imppffff}
b_c=\sqrt{\frac{D(r,\xi)}{A(r,\xi)}}\Big|_{r=r_{\text{photon}}}.
\fe
Proceeding by inserting the expression for the impact parameter from Eq.(\ref{imppffff}) into the effective potential of Eq.(\ref{ghhhjjjj}) and then taking the derivative with respect to $r$, one arrives at:
\ie
\frac{\mathrm{d} \,{\mathrm{V}(r,\xi,\ell)}}{\mathrm{d}r} = \frac{ B(r,\xi) \, L^2 \Big[ A(r,\xi) D'(r,\xi)-D(r,\xi) A'(r,\xi)\Big]}{A(r,\xi)  D(r,\xi)^2}.
\fe

The next step involves imposing the condition $\frac{\mathrm{d}V(r,\xi,\ell)}{\mathrm{d}r}=0$ to determine the allowed radii. Solving this equation yields six possible roots; however, only one is physically relevant, as it is both real, positive, and located outside the event horizon. This admissible solution is given by
\ie
\label{photonsssss}
r_{\text{photon}} = \frac{1}{2} \left(3 M +\sqrt{9 M^2-8 \xi }\right),
\fe
where a restriction naturally arises due to the presence of the square root
\ie
9 M^2-8 \xi >0,
\fe
to ensure that the solution remains real and positive. In the limit $\xi \to 0$, the photon sphere coincides with the Schwarzschild value, $r_{\text{ph-Sch}} = 3M$, as expected. Moreover, Eq.~\ref{photonsssss} bears a close resemblance to the expression for the photon sphere radius in the Reissner--Nordström spacetime, given by $r_{\text{ph-RN}} = \frac{1}{2}\!\left(3M + \sqrt{9M^{2} - 8Q^{2}}\right)$.

\begin{table}[!h]
\begin{center}
\begin{tabular}{c c c || c c c} 
 \hline\hline\hline
 $\xi$ & $M$ &  $r_{\text{photon}}$ & $\xi$ & $M$ &  $r_{\text{photon}}$ \\ [0.2ex] 
 \hline
 0.1  & 1.0 & 2.93178 & 0.1  & 1.1 & 3.23824  \\ 

 0.2  & 1.0 & 2.86015 & 0.1  & 1.2 & 3.54356  \\
 
 0.3  & 1.0 & 2.78452 & 0.1  & 1.3 & 3.84803  \\
 
 0.4  & 1.0 & 2.70416 & 0.1  & 1.4 & 4.15183  \\
 
 0.5  & 1.0 & 2.61803 & 0.1  & 1.5 & 4.45511  \\ 
 [0.2ex] 
 \hline \hline \hline
\end{tabular}
\caption{\label{tabcritical} The critical photon orbit radii $r_{\text{photon}}$ are listed for various choices of $M$ and $\xi$.}
\end{center}
\end{table}

We now turn to the analysis of the black hole shadow radius. For the spacetime under consideration, it can be expressed as 
\ie
\begin{split}
R = & \,\sqrt{\frac{D(r,\xi)}{A(r,\xi)}}\, \Bigg|_{r=r_{\text{photon}}} = r_{\text{photon}}^2 \sqrt{\frac{1}{-2 M r_{\text{photon}}+\xi +r_{\text{photon}}^2}} \\
& \approx \,  3 \sqrt{3} M -\frac{\sqrt{3} \xi }{2 M} . 
\end{split}
\fe
It is simple to notice that the leading term in the expression corresponds to the Schwarzschild black hole result, while the remaining terms represent corrections arising from the presence of the parameter $\xi$ in the metric of Eq.~(\ref{mainmetric}).

\begin{figure}
    \centering
    \includegraphics[scale=0.51]{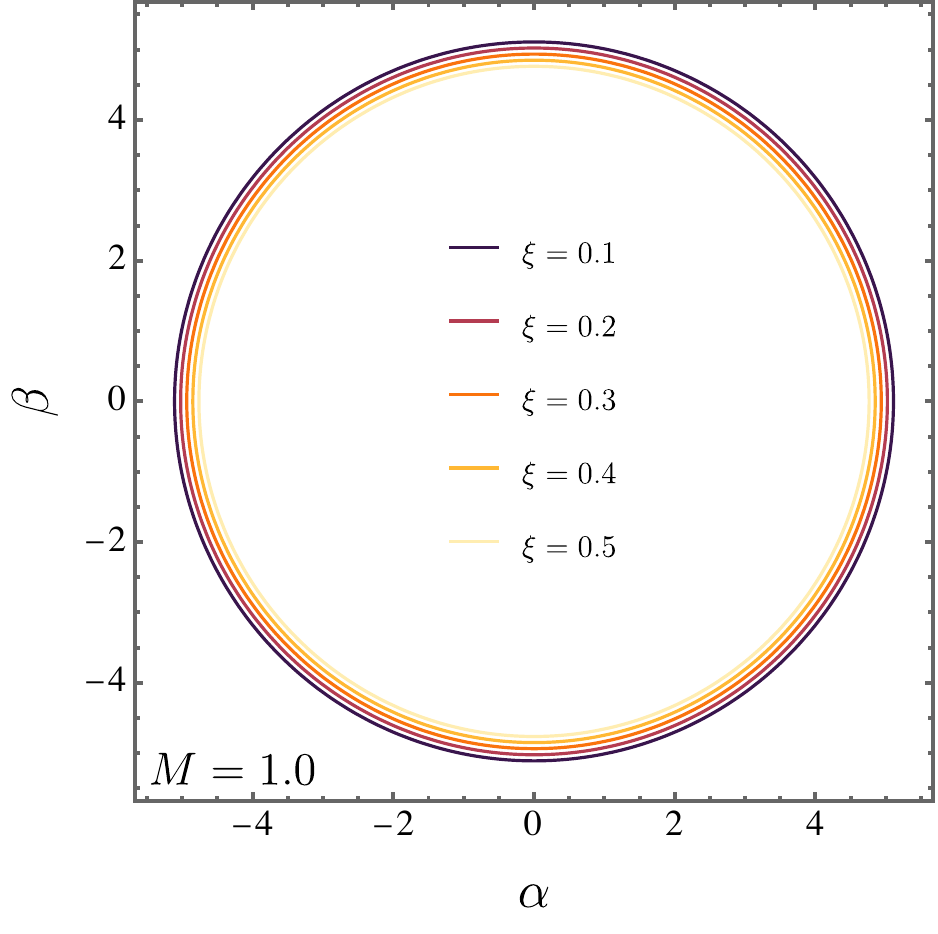}
    \includegraphics[scale=0.51]{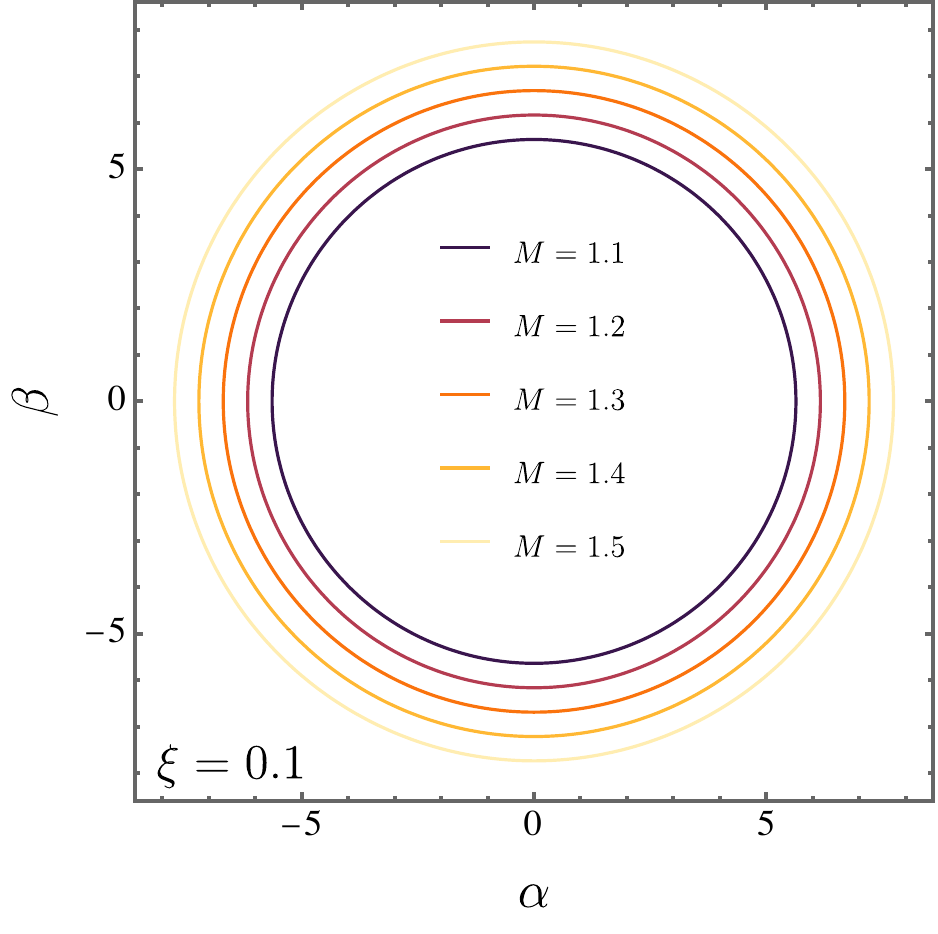}
    \caption{The shadow radius $R$ is plotted for several combinations of $M$ and $\xi$. In the left panel, $M$ is held fixed while $\xi$ ranges from $0.1$ to $0.5$, whereas in the right panel, $\xi$ is kept constant and $M$ is varied between $1.1$ and $1.5$.}
    \label{shaww}
\end{figure}

In Tab. \ref{tabshadd}, we present the values of the shadows radii by taking into account different values of $M$ and $\xi$. In a general panorama, we notice that when increasing $\xi$ (maintaining $M$ fixed), the values of $R$ decreases. On the other hand, when we keep $\xi$ fixed, and vary $M$, $R$ turns out to increase. In addition, to corroborate our interpretations, in Fig. \ref{shaww}, we display the polar projection for the shadow radii, which are shown for diverse values of $M$ and $\xi$. On the left hand, it is considered $M$ fixed and varies $\xi$ from $0.1$ to $0.5$. On the right side, $\xi$ is kept constant while $M$ varies from $1.1$ to $1.5$.

\begin{table}[!h]
\begin{center}
\begin{tabular}{c c c || c c c} 
 \hline\hline\hline
 $\xi$ & $M$ &  $R$ & $\xi$ & $M$ &  $R$ \\ [0.2ex] 
 \hline
 0.1  & 1.0 & 5.10955 & 0.1  & 1.1 & 5.63704  \\ 

 0.2  & 1.0 & 5.02295 & 0.1  & 1.2 & 6.16321  \\
 
 0.3  & 1.0 & 4.93634 & 0.1  & 1.3 & 6.68838  \\
 
 0.4  & 1.0 & 4.84974 & 0.1  & 1.4 & 7.21275 \\
 
 0.5  & 1.0 & 4.76314 & 0.1  & 1.5 & 7.73649  \\ 
 [0.2ex] 
 \hline \hline \hline
\end{tabular}
\caption{\label{tabshadd} The computed shadow radii $R$ are reported for different combinations of $M$ and $\xi$.}
\end{center}
\end{table}

\section{Lensing effects: weak field approximation }

This part of the work addresses gravitational lensing in the weak--deflection regime. The analysis is carried out using the Gauss--Bonnet method \cite{Gibbons:2008rj}, which provides the basis for computing the corresponding deflection angle.

We begin by examining the stability of the photon spheres described in Eq. (\ref{photonsssss}). To this end, we compute the Gaussian curvature, which is essential in assessing the nature of the critical orbits. As will be shown, the sign of the curvature determines the stability: positive curvature indicates stable orbits, while negative curvature corresponds to instability.

\subsection{Stability of the critical orbits }

The behavior of photon rings (or critical orbits) around black holes is governed by the geometry of the optical manifold, whose curvature dictates the stability of circular light paths. The existence of conjugate points plays a decisive role in identifying whether such orbits are stable or unstable. Small perturbations prevent photons from staying on perfectly circular paths: in unstable configurations, they either plunge into the black hole or escape to infinity, whereas in stable configurations, the photons remain confined near their initial orbit, repeatedly circling in a localized region \cite{qiao2022geometric,Heidari:2025iiv,qiao2022curvatures,araujo2025impact}.

The stability of photon trajectories can be reformulated within a geometric framework, where the intrinsic properties of the optical manifold play essential role. In particular, the Gaussian curvature $\mathcal{K}(r)$ dictates whether neighboring light rays converge or diverge along their paths. The Cartan--Hadamard theorem states that in regions with $\mathcal{K}(r) \leq 0$, conjugate points do not arise, implying that circular photon paths are inherently unstable. When $\mathcal{K}(r) > 0$, however, conjugate points may exist, opening the possibility of localized, stable photon orbits \cite{qiao2024existence}. Within this perspective, null geodesics satisfying $\mathrm{d}s^{2}=0$ can be reformulated and expressed as \cite{AraujoFilho:2024xhm}:
\ie
\mathrm{d}t^2 = \Tilde{\gamma}_{ij}\mathrm{d}x^i \mathrm{d}x^j = \frac{1}{A(r,\xi) \, B(r,\xi)}\mathrm{d}r^2  +\frac{\Bar{D}(r,\xi)}{A(r,\xi)}\mathrm{d}\varphi^2.
\fe

In this setup, the indices $i$ and $j$ run over the spatial coordinates $1$ to $3$, and $\tilde{\gamma}_{ij}$ denotes the metric components of the associated optical space. The quantity $\bar{D}(r,\xi)$ is introduced as the metric function restricted to the equatorial plane, namely $\bar{D}(r,\xi) = D(r,\xi,\theta = \pi/2)$. With these definitions, the intrinsic curvature of the optical manifold is fully described by the Gaussian curvature, which takes the form provided in \cite{qiao2024existence}:
\ie
\mathcal{K}(r,\xi) = \frac{R}{2} =  -\frac{ A(r,\xi) \sqrt{B(r,\xi)}}{\sqrt{ \,  \Bar{D}(r,\xi)}}  \frac{\partial}{\partial r} \left[  \frac{A(r,\xi) \sqrt{B(r,\xi)}}{2 \sqrt{  \Bar{D}(r,\xi) }}   \frac{\partial}{\partial r} \left(   \frac{\Bar{D}(r,\xi)}{A(r,\xi)}    \right)    \right].
\fe
Here, $R$ represents the Ricci scalar calculated for the two--dimensional optical geometry. Considering the regime in which $\xi$ is taken to be small, the curvature expression can be expanded accordingly, yielding
\ie
\begin{split}
\label{gbbhhh}
\mathcal{K}(r,\xi) =& \,\, \frac{3 M^2}{r^4}-\frac{2 M}{r^3} + \frac{2 \xi }{r^3 (r-2 M)}+  \frac{6 M^2 \xi }{r^5 (r-2 M)}-\frac{6 M \xi }{r^4 (r-2 M)}\\
& + \frac{24 M^3 \xi ^{3/2}}{r^7 (r-2 M)} -\frac{22 M^2 \xi ^{3/2}}{r^6 (r-2 M)}+\frac{4 M \xi ^{3/2}}{r^5 (r-2 M)},
\end{split}
\fe
where we have considered up to $\xi^{3/2}$. Two observations are in order here. First, the Gaussian curvature exhibits a divergence at $r = 2M$, which becomes evident when examining the denominator of several terms in Eq.~(\ref{gbbhhh}). Second, a closer inspection reveals that this singular behavior originates from the third contribution in $B(r,\xi)$, specifically the term $2M \xi^{3/2}/r^{4}$.

Previous works \cite{Heidari:2025iiv,araujo2025impact,qiao2024existence,qiao2022geometric,AraujoFilho:2025huk,qiao2022curvatures} emphasize that the Gaussian curvature $\mathcal{K}(r,\xi)$ is the quantity that dictates whether circular photon orbits persist or disperse under small perturbations. In other words, as briefly commented previously, positive curvature indicates that nearby geodesics converge, favoring stable closed photon loops, while negative curvature causes geodesic deviation to grow, destabilizing the orbit and leading to capture or escape.

\begin{figure}
    \centering
    \includegraphics[scale=0.61]{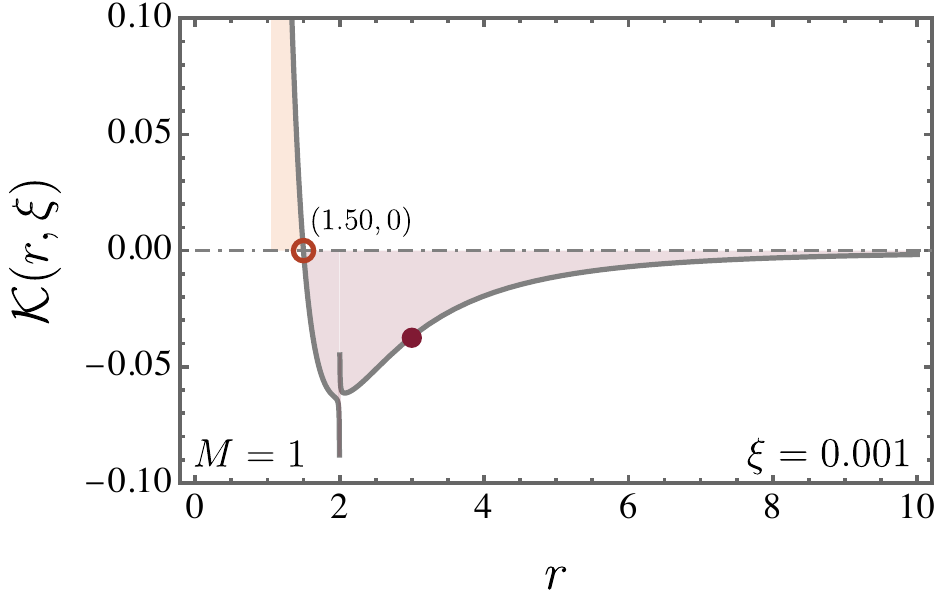}
    \caption{Gaussian curvature $\mathcal{K}(r,\xi)$ {as a function of radial coordinate $r$,} for $M = 1$ and $\xi = 0.001$. The wine circle marks the radius where $\mathcal{K}=0$, separating stable and unstable photon regions, while the wine dotted point indicates the photon sphere position $r_{\text{photon}}$, located in the unstable regime.}
    \label{cuvgahjj}
\end{figure}

This feature is visualized in Fig.~\ref{cuvgahjj}, where $\mathcal{K}(r,\xi)$ is plotted against $r$ for the representative case $M = 1$ and $\xi = 0.001$. The plot shows two clearly separated domains: a region of confinement (shaded light pink) where stability is possible and another region (shaded light orange) where instability dominates. The zero of $\mathcal{K}$ occurs at approximately $r \approx 1.50$, acting as the dividing line between these two regimes. Because the photon sphere radius lies outside this critical point, all circular photon paths in this background are unstable.

\subsection{Weak deflection angle }

The weak--field deflection angle is obtained by employing the Gauss--Bonnet theorem \cite{Gibbons:2008rj}, starting from the curvature expression provided in Eq.~(\ref{gbbhhh}). To implement this approach, the analysis is restricted to the equatorial plane, i.e., $\theta = \pi/2$, reducing the optical geometry to a two--dimensional surface. In this reduced setup, the corresponding area element can be written as:
\ie
\mathrm{d}S = \sqrt{\Tilde{\gamma}} \, \mathrm{d} r \mathrm{d}\varphi = \sqrt{\frac{1}{A(r,\xi)} \frac{1}{B(r,\xi)} \frac{D(r,\xi)}{A(r,\xi)} } \, \mathrm{d} r \mathrm{d}\varphi.
\fe

To perform the integration and compute the deflection angle, we adopt the same treatment for $\xi$ as outlined in this work and consider the impact parameter in the regime $b \gg 2M$, consistent with the original formulation in Ref.~\cite{Gibbons:2008rj}, where the Gauss--Bonnet theorem was first applied to gravitational lensing. In this context, the mass parameter $M$ is expanded up to second order, following the procedure commonly employed in related studies \cite{araujo2024effects,AraujoFilho:2025huk,araujo2025gravitadddtional,araujo2025geodesics}.

Using the previous relation as a starting point, the light--deflection angle is given by:
\ie
\begin{split}
\label{vfffdddd}
& \Tilde{\alpha} (b,\xi) =  - \int \int_{D} \mathcal{K} \mathrm{d}S = - \int^{\pi}_{0} \int^{\infty}_{\big(\frac{\sin (\varphi )}{\beta}+\frac{\Tilde{M} (1-\cos (\varphi ))^2}{ \beta^2}\big)^{-1}} \mathcal{K} \mathrm{d}S \\
\simeq & \,  \frac{4 M}{b}+\frac{15 \pi  M^2}{4 b^2} -\frac{\pi  \xi }{2 b^2}  -\frac{8 M \xi }{b^3}  -\frac{157 \pi  M^2 \xi }{16 b^4} -\frac{3 \pi  M \xi ^{3/2}}{8 b^4},
\end{split}
\fe
where, again, we have considered up to $\xi^{3/2}$.

Equation (\ref{vfffdddd}) shows that the first two contributions on the second line reproduce the standard light--bending result for a Schwarzschild black hole, while the third, fourth, and fifth terms coincide, up to appropriate modifications of the constant factors, the prediction for a Reissner--Nordstr\"{o}m spacetime. The remaining terms encode the effects of the higher--order curvature--scalar modifications introduced in this work, incorporating various powers of the parameter $\xi$.

\begin{figure}
    \centering
    \includegraphics[scale=0.6]{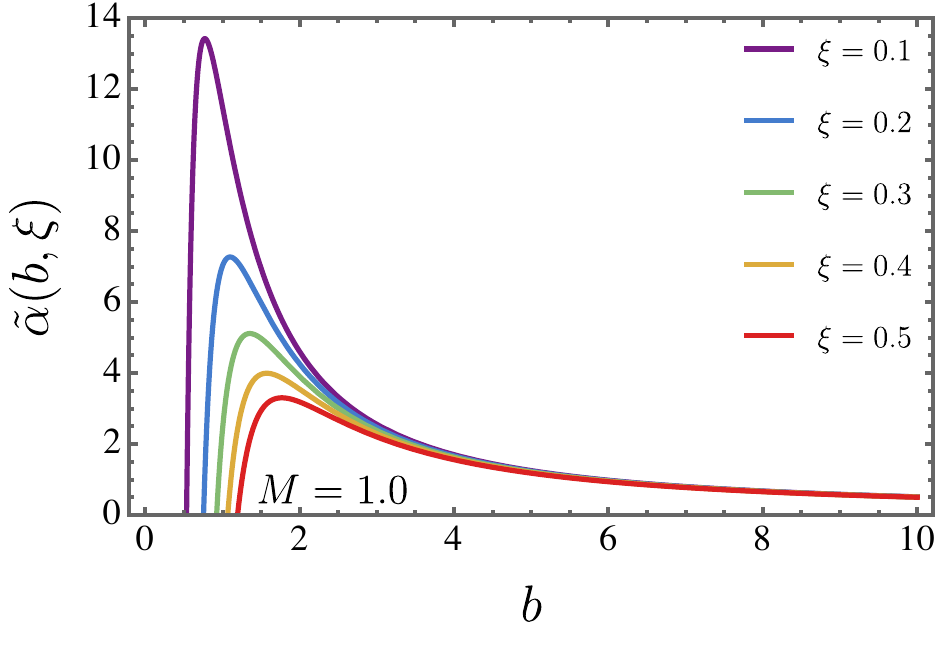}
    \caption{Deflection angle $\tilde{\alpha}(b,\xi)$ is plotted with respect to the impact parameter $b$ for several choices of $\xi$, with the black hole mass kept fixed at $M = 1$. }
    \label{tildealpha}
\end{figure}

Fig.~\ref{tildealpha} illustrates how the deflection angle $\tilde{\alpha}(b,\xi)$ varies with the black hole parameters. For a fixed impact parameter, e.g., $b = 2$, an increase in $\xi$ leads to a reduction in the deflection angle. This behavior is corroborated by the geodesic analysis shown in Fig.~\ref{geodespath}, where larger values of $\xi$ render the light trajectories progressively more “open.” Moreover, as shown in the next section, the same trend persists in the strong–deflection regime.

\section{Lensing effects: strong field approximation }

This section focuses on deriving the light--bending angle in the strong--field regime. Adopting the methodology used in several recent works \cite{nascimento2024gravitational,heidari2024absorption,araujo2024effects,araujo2025antisymmetric}, the computation was carried out assuming a static, spherically symmetric spacetime that becomes asymptotically flat at large $r$. The corresponding line element characterizing the background geometry takes the form \cite{tsukamoto2017deflection}:
\ie
\mathrm{d}s^{2} = - \Tilde{\mathrm{A}}(r) \mathrm{d}t^{2} + \Tilde{\mathrm{B}}(r) \mathrm{d}r^{2} + \Tilde{\mathrm{C}}(r)(\mathrm{d}\theta^2 + \sin^{2}\theta\mathrm{d}\phi^2).
\fe

To apply the analytical method proposed by Tsukamoto \cite{tsukamoto2017deflection}, the spacetime under consideration was required to satisfy asymptotic flatness. This condition enforces that, in the limit $r \to \infty$, the metric functions approach well--defined forms: $\tilde{\mathrm{A}}(r)$ and $\tilde{\mathrm{B}}(r)$ tend toward unity, while $\tilde{\mathrm{C}}(r)$ grows as $r^{2}$, namely,
\ie
\lim_{r \to \infty} \Tilde{\mathrm{A}}(r) = 1, \quad \lim_{r \to \infty} \Tilde{\mathrm{B}}(r) = 1, \quad \lim_{r \to \infty} \Tilde{\mathrm{C}}(r) = r^2.
\nonumber
\fe

The calculation of the strong–field deflection angle begins by introducing an auxiliary function $\tilde{\bar{D}}(r)$ to reformulate the radial dependence. This redefinition regularizes the expression close to the photon sphere and simplifies the analysis, allowing the light trajectory to be treated analytically without divergences dominating the result
\ie
\Tilde{\Bar{D}}(r) \equiv \frac{\Tilde{\mathrm{C}}^{\prime}(r)}{\Tilde{\mathrm{C}}(r)} - \frac{\Tilde{\mathrm{A}}^{\prime}(r)}{\Tilde{\mathrm{A}}(r)},
\fe
where the primes indicate differentiation with respect to $r$. The auxiliary function $\tilde{\bar{D}}(r)$ is constructed so that it possesses at least one positive root, and the photon sphere is determined by selecting the largest of these roots, denoted $r_{\text{photon}}$. To ensure the validity of the formalism in this regime, the metric functions $\tilde{\mathrm{A}}(r)$, $\tilde{\mathrm{B}}(r)$, and $\tilde{\mathrm{C}}(r)$ must stay smooth and strictly positive for all $r \geq r_{\text{photon}}$.

Because the background is invariant under time translations and rotations around the symmetry axis, null geodesics admit two constants of motion. The first is the energy, $E = \tilde{\mathrm{A}}(r)\dot{t}$, and the second is the angular momentum, $L = \tilde{\mathrm{C}}(r)\dot{\phi}$. When both quantities are nonzero, their ratio defines the impact parameter
\ie
b \equiv \frac{L}{E} = \frac{\Tilde{\mathrm{C}}(r)\Dot{\phi}}{\Tilde{\mathrm{A}}(r)\Dot{t}}.
\fe

Exploiting the axial symmetry of the metric allows the motion to be restricted to the equatorial plane by fixing $\theta = \pi/2$, which does not affect the generality of the solution. With this simplification, the equation describing the radial evolution of light--like geodesics takes the form:
\ie
\Dot{r}^{2} = V(r).
\fe
Now, let us define the effective potential for photon motion as
$$
V(r)=\frac{L^{2}\,\mathcal{R}(r)}{\tilde{\mathrm{B}}(r)\,\tilde{\mathrm{C}}(r)}, \qquad \text{and}
\qquad 
\mathcal{R}(r)=\frac{\tilde{\mathrm{C}}(r)}{\tilde{\mathrm{A}}(r)b^{2}}-1.
$$
This expression plays the role of a radial potential for massless particles. The physically accessible domain is determined by the requirement $V(r)\geq 0$. Because the geometry becomes flat at large $r$, the potential approaches $E^{2}$ as $r\to\infty$, meaning photons can escape to infinity. Additionally, $\mathcal{R}(r)=0$ is assumed to admit at least one positive real root, which sets the turning point of the trajectory.

In the context of gravitational lensing, the null geodesic of interest originates from spatial infinity, approaches the compact object, reaches a minimum radial distance $r_{\text{o}}$, and then travels back out to infinity. This turning point $r_{\text{o}}$ must lie outside the photon sphere radius $r_{\text{photon}}$, ensuring that the trajectory does not correspond to a circular orbit. Mathematically, $r_{\text{o}}$ is identified as the largest real root of $\mathcal{R}(r)=0$, provided that $\tilde{\mathrm{B}}(r)$ and $\tilde{\mathrm{C}}(r)$ are finite and positive there. At this radius, the effective potential satisfies $V(r_{\text{o}})=0$, making $\mathcal{R}(r_{\text{o}})=0$ the fundamental condition defining the closest approach
\ie
\Tilde{\mathrm{A}}_{\text{o}}\Dot{t}^{2}_{\text{o}} = \Tilde{\mathrm{C}}_{\text{o}}\Dot{\phi}^{2}_{\text{o}}.
\fe

Hereafter, the following quantities carrying the subscript “$\text{o}$” is to be considered at the turning point $r = r_{\text{o}}$. For the analysis of an individual photon trajectory, it is sufficient to consider a positive impact parameter $b$, as negative values merely reverse the direction of motion. Because $b$ remains constant along the entire geodesic, it can be expressed as
\ie
\label{impcss}
b(r_{\text{o}}) = \frac{L}{E} = \frac{\Tilde{\mathrm{C}}_{\text{o}}\Dot{\phi}_{\text{o}}}{\Tilde{\mathrm{A}}_{\text{o}}\Dot{t}_{\text{o}}} = \sqrt{\frac{\Tilde{\mathrm{C}}_{\text{o}}}{\Tilde{\mathrm{A}}_{\text{o}}}}.
\fe
It is important to note that $\mathcal{R}(r)$ can equivalently be expressed in the following form:
\ie
\mathrm{R}(r)= \frac{\Tilde{\mathrm{A}}_{\text{o}}\Tilde{\mathrm{C}}(r)}{\Tilde{\mathrm{A}}(r)\Tilde{\mathrm{C}}_{\text{o}}} - 1.
\fe

The condition that guarantees the presence of a circular null geodesic can be formulated using the approach described in Ref.~\cite{hasse2002gravitational}. Within this formalism, the photon trajectory is governed by the relation
\ie
\frac{\Tilde{\mathrm{B}}(r)\,\Tilde{\mathrm{C}}(r)\, \Dot{r}^{2}}{E^{2}} + b^{2} = \frac{\Tilde{\mathrm{C}}(r)}{\Tilde{\mathrm{A}}(r)},
\fe
in a such way that we can write
\ie
\ddot{r} + \frac{1}{2}\left( \frac{\Tilde{\mathrm{B}}(r)^{\prime}}{\Tilde{\mathrm{B}}(r)} + \frac{\Tilde{\mathrm{C}}(r)^{\prime}}{\Tilde{\mathrm{C}}(r)} \Dot{r}^{2} \right) = \frac{E^{2}\Tilde{\Bar{D}}(r)}{\Tilde{\mathrm{A}}(r)\Tilde{\mathrm{B}}(r)}. 
\fe

For radii satisfying $r \geq r_{\text{photon}}$, the metric components $\tilde{\mathrm{A}}(r)$, $\tilde{\mathrm{B}}(r)$, and $\tilde{\mathrm{C}}(r)$ are required to remain smooth and strictly positive. With $E>0$, the condition $\tilde{\bar{D}}(r)=0$ serves as the criterion for the presence of a circular photon orbit. Additionally, evaluating the derivative of $\mathcal{R}(r)$ at the photon sphere yields
$$
\mathcal{R}'_{\text{photon}}=\frac{\tilde{\bar{D}}_{\text{photon}}\tilde{\mathrm{C}}_{\text{photon}}\tilde{\mathrm{A}}_{\text{photon}}}{b^{2}}=0,
$$
where the subscript “photon” indicates that all quantities are taken at $r=r_{\text{photon}}$.

The next step is addressing the threshold value of the impact parameter, denoted $b_c$, which separates photons that are scattered from those that spiral toward the photon sphere:
\ie
b_{c}(r_{\text{photon}}) \equiv \lim_{r_{\text{o}} \to r_{\text{photon}}} \sqrt{\frac{\Tilde{\mathrm{C}}_{\text{o}}}{\Tilde{\mathrm{A}}_{\text{o}}}}.
\fe
Notice that this domain is identified as the strong--deflection regime. Taking the derivative of the effective potential $V(r)$ with respect to $r$ leads to the condition
\ie
V^{\prime}(r) = \frac{L^{2}}{\Tilde{\mathrm{B}}(r)\Tilde{\mathrm{C}}(r)} \left[ \mathrm{R}(r)^{\prime} + \left( \frac{\Tilde{\mathrm{C}}^{\prime}(r)}{\Tilde{\mathrm{C}}(r)} - \frac{\Tilde{\mathrm{B}}^{\prime}(r)}{\Tilde{\mathrm{B}}(r)}   \right)   \mathrm{R}(r)  \right].
\fe
Within such a regime, as the closest approach radius $r_{\text{o}}$ approaches the critical orbit $r_{\text{photon}}$, the $V(r_{\text{o}})$ and its derivative $V'(r_{\text{o}})$ vanish all at once. Under these circumstances, the null geodesic equation simplifies to
\ie
\left(  \frac{\mathrm{d}r}{\mathrm{d}\phi}     \right)^{2} = \frac{\mathrm{R}(r)\Tilde{\mathrm{C}}(r)}{\Tilde{\mathrm{B}}(r)}.
\fe

Consequently, the light--bending angle associated with a trajectory reaching the closest approach $r_{\text{o}}$ can be written in the form
\ie
\alpha(r_{\text{o}}) = I(r_{\text{o}}) - \pi,
\fe
in which $I(r_{\text{o}})$ reads
\ie
I(r_{\text{o}}) \equiv 2 \int^{\infty}_{r_{\text{o}}} \frac{\mathrm{d}r}{\sqrt{\frac{\mathrm{R}(r)\Tilde{\mathrm{C}}(r)}{\Tilde{\mathrm{B}}(r)}}}.
\fe

The evaluation begins by tackling the integral that defines the deflection angle—an operation known for its analytical complexity, as emphasized by Tsukamoto \cite{tsukamoto2017deflection}. To facilitate this calculation, we introduce the auxiliary quantity defined in \cite{tsukamoto2017deflection}:
\ie
z \equiv 1 - \frac{r_{\text{o}}}{r}.
\fe

This redefinition makes it possible to rewrite the integral in the following form:
\ie
I(r_{\text{o}}) = \int^{1}_{0} f(z,r_{\text{o}}) \mathrm{d}z,
\fe
with, in other words, 
\ie
f(z,z_{0}) \equiv \frac{2r_{\text{o}}}{\sqrt{G(z,r_{\text{o}})}}, \,\,\,\,\,\,\,\, \text{and} \,\,\,\,\,\,\,\,  G(z,r_{\text{o}}) \equiv \mathrm{R}(r) \frac{\Tilde{\mathrm{C}}(r)}{\Tilde{\mathrm{B}}(r)}(1-z)^{4}.
\fe

Alternatively, when rewritten using the variable $z$, the function $\mathcal{R}(r)$ becomes therefore
\ie
\mathrm{R}(r) = \Tilde{{\Bar{D}}}_{\text{o}} \, r_{\text{o}} z + \left[ \frac{r_{\text{o}}}{2}\left( \frac{\Tilde{\mathrm{C}}^{\prime\prime}_{\text{o}}}{\Tilde{\mathrm{C}}_{\text{o}}} - \frac{\Tilde{\mathrm{A}}_{\text{o}}^{\prime\prime}}{\Tilde{\mathrm{A}}_{\text{o}}}  \right) + \left( 1 - \frac{\Tilde{\mathrm{A}}_{\text{o}}^{\prime}r_{\text{o}}}{\Tilde{\mathrm{A}}_{\text{o}}}  \right) \Tilde{{\Bar{D}}}_{\text{o}}  \right] r_{\text{o}} z^{2} + \mathcal{O}(z^{3})+ ...    \,\,\,\,.
\fe

Expanding $G(z,r_{\text{o}})$ in a Taylor series about $z = 0$ yields the following expression:
\ie
G(z,r_{\text{o}}) = \sum^{\infty}_{n=1} c_{n}(r_{\text{o}})z^{n},
\fe
where we can identify $c_{1}(r)$ and $c_{2}(r)$ as
\ie
c_{1}(r_{\text{o}}) = \frac{\Tilde{\mathrm{C}}_{\text{o}}\Tilde{\Bar{D}}_{\text{o}}r_{\text{o}}}{\Tilde{\mathrm{B}}_{\text{o}}},
\fe
and
\ie
c_{2}(r_{\text{o}}) = \frac{\Tilde{\mathrm{C}}_{\text{o}}r_{\text{o}}}{\Tilde{\mathrm{B}}_{\text{o}}} \left\{ \Tilde{\Bar{D}}_{\text{o}} \left[ \left( \Tilde{\Bar{D}}_{\text{o}} - \frac{\Tilde{\mathrm{B}}^{\prime}_{\text{o}}}{\Tilde{\mathrm{B}}_{\text{o}}}  \right)r_{\text{o}} -3       \right] + \frac{r_{\text{o}}}{2} \left(  \frac{\Tilde{\mathrm{C}}^{\prime\prime}_{\text{o}}}{\Tilde{\mathrm{C}}_{\text{o}}} - \frac{\Tilde{\mathrm{A}}^{\prime\prime}_{\text{o}}}{\Tilde{\mathrm{A}}_{\text{o}}}  \right)                 \right\}.
\fe

Moreover, applying the strong--deflection approximation leads to the result
\ie
c_{1}(r_{\text{photon}}) = 0, \,\,\,\,\,\, \text{and} \,\,\,\,\,\, c_{2}(r_{\text{photon}}) =  \frac{\Tilde{\mathrm{C}}_{\text{photon}}r^{2}_{\text{photon}}}{2 \Tilde{\mathrm{B}}_{\text{photon}}}\Tilde{\Bar{D}}^{\prime}_{\text{photon}}, \,\,\,\,\,\,\, \text{with} \,\,\,\,\, \Tilde{\Bar{D}}^{\prime}_{\text{photon}} = \frac{\Tilde{\mathrm{C}}^{\prime\prime}}{\Tilde{\mathrm{C}}_{\text{photon}}} - \frac{\Tilde{\mathrm{A}}^{\prime\prime}}{\Tilde{\mathrm{A}}_{\text{photon}}},
\fe
with $G(z,r_{\text{o}})$ can be rewritten in a simplified form as
\ie
G_{\text{photon}}(z) = c_{2}(r_{\text{photon}})z^{2} + \mathcal{O}(z^{3}).
\fe

When the closest approach radius $r_{\text{o}}$ tends to the photon sphere $r_{\text{photon}}$, the function $f(z,r_{\text{o}})$ develops a singularity whose leading term behaves as $1/z$. This singular nature causes the integral $I(r_{\text{o}})$ to diverge logarithmically. To handle this, the integral is decomposed into two parts: a divergent contribution, $I_{\text{Div}}(r_{\text{o}})$, which isolates the $1/z$ behavior, and a regular part, $I_{\text{Reg}}(r_{\text{o}})$, which stays finite. The divergent term can then be written as
\ie
I_{_{\text{Div}}}(r_{\text{o}}) \equiv \int^{1}_{0} f_{_{\text{Div}}}(z,r_{\text{o}}) \mathrm{d}z, \,\,\,\,\,\,\, \text{with} \,\,\,\,\,\,f_{_{\text{Div}}}(z,r_{\text{o}}) \equiv \frac{2 r_{\text{o}}}{\sqrt{c_{1}(r_{\text{o}})z + c_{2}(r_{\text{o}})z^{2}}}.
\fe
Carrying out the integration yields the following expression:
\ie
I_{_{\text{Div}}} (r_{\text{o}}) = \frac{4 r_{\text{o}}}{\sqrt{c_{2}(r_{\text{o}})}} \ln \left[  \frac{\sqrt{c_{2}(r_{\text{o}})} + \sqrt{c_{1}(r_{\text{o}}) + c_{2}(r_{\text{o}})     }  }{\sqrt{c_{1}(r_{\text{o}})}}  \right].
\fe

Furthermore, a Taylor expansion of $c_{1}(r_{\text{o}})$ and $b(r_{\text{o}})$ about $r_{\text{o}} = r_{\text{photon}}$ gives
\ie
c_{1}(r_{\text{o}}) = \frac{\Tilde{\mathrm{C}}_{\text{photon}}r_{\text{photon}}\Tilde{\Bar{D}}^{\prime}_{\text{photon}}}{\Tilde{\mathrm{B}}_{\text{photon}}} (r_{\text{o}}-r_{\text{photon}}) + \mathcal{O}((r_{\text{o}}-r_{\text{photon}})^{2}),
\fe
and
\ie
b(r_{\text{o}}) = b_{c}(r_{\text{photon}}) + \frac{1}{4} \sqrt{\frac{\Tilde{\mathrm{C}}_{\text{photon}}}{\Tilde{\mathrm{A}}_{\text{photon}}}}  \Tilde{\Bar{D}}^{\prime}_{\text{photon}}(r_{\text{o}}-r_{\text{photon}})^{2} + \mathcal{O}((r_{\text{o}}-r_{\text{photon}})^{3}),
\fe
which results in:
\ie
\lim_{r_{\text{o}} \to r_{\text{photon}}} c_{1}(r_{\text{o}})  =  \lim_{b \to b_{c}} \frac{2 \Tilde{\mathrm{C}}_{\text{photon}} r_{\text{photon}} \sqrt{\Tilde{{\Bar{D}}}^{\prime}}}{\Tilde{\mathrm{B}}_{\text{photon}}} \left(  \frac{b}{b_{c}} -1  \right)^{1/2}.
\fe

With these expansions, the divergent part of the integral $I_{\text{Div}}(b)$ can be expressed as
\ie
I_{_{\text{Div}}}(b) = - \frac{r_{\text{photon}}}{\sqrt{c_{2}(r_{\text{photon}})}} \ln\left[ \frac{b}{b_{c}} - 1 \right] + \frac{r_{\text{photon}}}{\sqrt{c_{2}(r_{\text{photon}})}}\ln \left[ r^{2}\Tilde{{\Bar{D}}}^{\prime}_{\text{photon}}\right] + \mathcal{O}[(b-b_{c})\ln(b-b_{c})].
\fe

Moreover, the finite (regular) contribution is defined by
\ie
I_{_{\text{Reg}}}(b) = \int^{0}_{1} f_{\text{Reg}}(z,b_{c})\mathrm{d}z + \mathcal{O}[(b-b_{c})\ln(b-b_{c})].
\fe
Introduce the function $f_{\text{Reg}}$ by subtracting the divergent part from the full expression,
$$
f_{\text{Reg}} = f(z,r_{\text{o}}) - f_{\text{Div}}(z,r_{\text{o}}).
$$
Using this regularized quantity and working within the strong--deflection approximation, the resulting expression for the deflection angle becomes
\ie
\label{strgffddd}
a(b) = - \Tilde{a} \ln \left[ \frac{b}{b_{c}}-1    \right] + \Tilde{b} + \mathcal{O}[(b-b_{c})\ln(b-b_{c})],
\fe
with we have particularly considered 
\ie
\Tilde{a} = \sqrt{\frac{2 \,\Tilde{\mathrm{B}}_{\text{photon}}\Tilde{\mathrm{A}}_{\text{photon}}}{\Tilde{\mathrm{C}}^{\prime\prime}_{\text{photon}}\Tilde{\mathrm{A}}_{\text{photon}} - \Tilde{\mathrm{C}}_{\text{photon}}\Tilde{\mathrm{A}}^{\prime\prime}_{\text{photon}}}}, 
\fe
and
\ie
\Tilde{b} = \Tilde{a} \ln\left[ r^{2}_{\text{photon}}\left( \frac{\Tilde{\mathrm{C}}_{\text{photon}}^{\prime\prime}}{\Tilde{\mathrm{C}}_{\text{photon}}}  -  \frac{\Tilde{\mathrm{A}}^{\prime\prime}_{\text{photon}}}{\Tilde{\mathrm{C}}_{\text{photon}}} \right)   \right] + I_{_{\text{Reg}}}(r_{\text{photon}}) - \pi.
\fe

In the following subsection, the formalism developed above is applied to the specific black hole geometry given in Eq.~(\ref{mainmetric}).


\subsection{Bending angle of a black hole in higher-order curvature scalar gravity }

Here, it is worth emphasizing that, as employed in the previous sections, all calculations are carried out up to order $\xi^{3/2}$. With the general formalism in place, the next step is to specialize the analysis to the spacetime defined by Eq.(\ref{mainmetric}). Substituting this metric into Eq.(\ref{impcss}) yields the explicit expression for the impact parameter:
\ie
b_{c} \approx \,\, 3 \sqrt{3} M -\frac{\sqrt{3} \xi }{2 M} ,
\fe
In this derivation, the result was expanded perturbatively, keeping terms up to $\xi^{2}$.
Additionally, the parameters $\tilde{a}$ and $\tilde{b}$ are expressed explicitly as
\ie
\Tilde{a} = \, 1 + \frac{5 \xi }{18 M^2} -\frac{\xi ^{3/2}}{27 M^3} .
\fe
Consequently, the relation can be rewritten in the following form:
\ie
\begin{split}
\Tilde{b} = &  \left(1 +\frac{5 \xi }{18 M^2} -\frac{\xi ^{3/2}}{27 M^3} \right) \left( \ln [6] -\frac{\xi }{9 M^2} \right) \\
& + I_{_{\text{Reg}}}(r_{\text{photon}}) - \pi.
\end{split}
\fe

Unlike the Schwarzschild scenario, where $\tilde{a}$ has a simpler dependence, here its value is largely dictated by the effects of ascribed to $\xi$. Moreover, the expression for the regular part of the integral, evaluated at $r = r_{\text{photon}}$, can be written as
\ie
\begin{split}
 & I_{_{\text{Reg}}}(r_{\text{photon}}) =   \\
 & \int_{0}^{1} \mathrm{d}z \left\{   \frac{2}{\sqrt{1-\frac{2 z}{3}} z} +\frac{\xi  (4 z (2 z-5)+15)}{3 \sqrt{3} M^2 (3-2 z)^{3/2} z (2 z+1)}-\frac{5 \xi }{9 M^2 z} -\frac{2 \xi ^{3/2} (z-1)^4}{9 M^3 \sqrt{9-6 z} z (2 z+1)}  -\frac{2}{z} \frac{2 \xi ^{3/2}}{27 M^3 z}    \right\} \\
 &  =  \ln[36]+\ln \left[7-4 \sqrt{3}\right] -\frac{3 \sqrt{3} \xi ^{3/2} \coth ^{-1}[2]}{8 M^3}-\frac{\xi }{18 M^2}+\frac{\xi }{6 \sqrt{3} M^2}+\frac{5 \xi  \ln [3]}{9 M^2}\\
 &  +\frac{10 \xi  \ln \left[\sqrt{3}-1\right]}{9 M^2}-\frac{3 \sqrt{3} \xi  \ln \left[\sqrt{3}+2\right]}{4 M^2}+\frac{3 \sqrt{3} \xi  \coth ^{-1}[2]}{4 M^2} \\
 & + \frac{3 \sqrt{3} \xi ^{3/2}}{20 M^3}-\frac{11 \xi ^{3/2}}{20 M^3}+\frac{3 \sqrt{3} \xi ^{3/2} \ln \left[\sqrt{3}+2\right]}{8 M^3}-\frac{2 \xi ^{3/2} \ln [3]}{27 M^3}-\frac{4 \xi ^{3/2} \ln \left[\sqrt{3}-1\right]}{27 M^3} .
\end{split}
\fe

This procedure yields a closed--form analytic result. An additional observation is that the regular contribution $I_{\text{Reg}}(r_{\text{photon}})$ in this spacetime coincides with its Schwarzschild counterpart plus extra terms proportional to $\xi$, as anticipated. Using this expression in Eq.~(\ref{strgffddd}), the deflection angle in the strong--field regime is finally obtained as
\ie
\begin{split}
& a (b,\xi)  =  -\left( 1 + \frac{5 \xi }{18 M^2} -\frac{\xi ^{3/2}}{27 M^3}  \right)   \times \, \ln \left[ \frac{b}{3 \sqrt{3} M -\frac{\sqrt{3} \xi }{2 M} } - 1  \right] \\
& + \left(1 +\frac{5 \xi }{18 M^2} -\frac{\xi ^{3/2}}{27 M^3} \right) \left( \ln [6] -\frac{\xi }{9 M^2}   \right)  - \pi\\
&  +  \ln[36]+\ln \left[7-4 \sqrt{3}\right] -\frac{3 \sqrt{3} \xi ^{3/2} \coth ^{-1}[2]}{8 M^3}-\frac{\xi }{18 M^2}+\frac{\xi }{6 \sqrt{3} M^2}+\frac{5 \xi  \ln [3]}{9 M^2}\\
 &  +\frac{10 \xi  \ln \left[\sqrt{3}-1\right]}{9 M^2}-\frac{3 \sqrt{3} \xi  \ln \left[\sqrt{3}+2\right]}{4 M^2}+\frac{3 \sqrt{3} \xi  \coth ^{-1}[2]}{4 M^2} \\
 & + \frac{3 \sqrt{3} \xi ^{3/2}}{20 M^3}-\frac{11 \xi ^{3/2}}{20 M^3}+\frac{3 \sqrt{3} \xi ^{3/2} \ln \left[\sqrt{3}+2\right]}{8 M^3}-\frac{2 \xi ^{3/2} \ln [3]}{27 M^3}-\frac{4 \xi ^{3/2} \ln \left[\sqrt{3}-1\right]}{27 M^3}  \\
 & + \mathcal{O}\Bigg\{\Bigg[b- \left(3 \sqrt{3} M -\frac{\sqrt{3} \xi }{2 M} \right) \Bigg]  \times \, \ln\Bigg[b- \left(3 \sqrt{3} M -\frac{\sqrt{3} \xi }{2 M}  \right) \Bigg] \Bigg\}.
\end{split}
\fe

\begin{figure}
    \centering
    \includegraphics[scale=0.6]{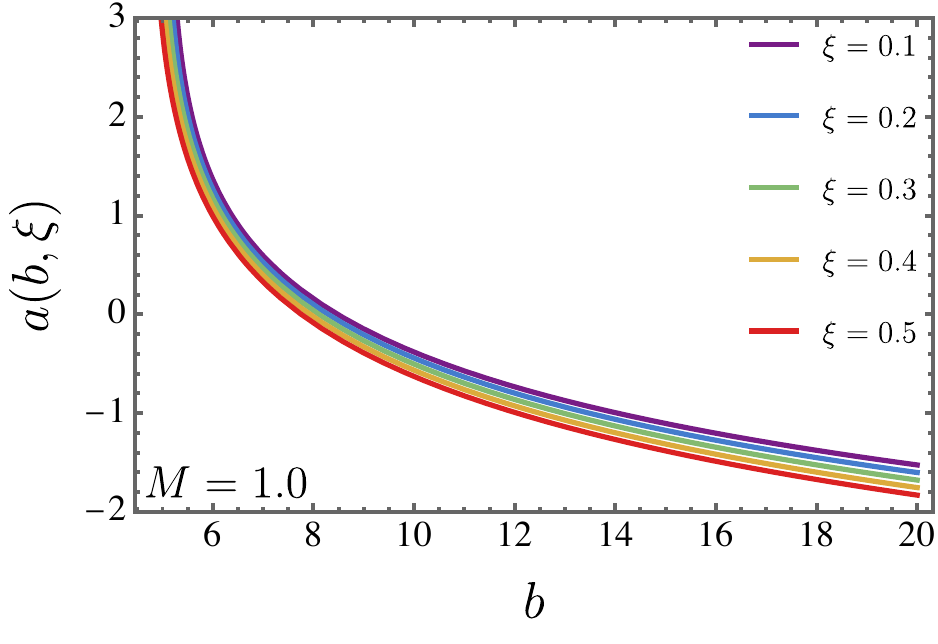}
    \caption{Deflection angle $a(b,\xi)$ is plotted with respect to the impact parameter $b$ for several choices of $\xi$, with the mass parameter fixed at $M = 1$.}
    \label{dflcstrr}
\end{figure}

For better visualization, Fig. \ref{dflcstrr} shows how the deflection angle varies with the impact parameter $b$ for several parameter choices. The plots reveal that larger values of $\xi$ fundamentally reduce $a(b,\xi)$. This behavior aligns with the geodesic results in Fig. \ref{geodespath}, where light rays move far away to the photon sphere as $\xi$ grows. Additionally, to reinforce the conclusions drawn in the strong--deflection regime, the next subsection turns to phenomenological aspects, emphasizing observable quantities that can be compared with Event Horizon Telescope (EHT) measurements.

\section{Lensing observables}

The shadow observations of $Sgr A^*$ and $M87^*$ by the Event Horizon Telescope (EHT) \cite{akiyama2022firstSgr,akiyama2022firstSgrA,Akiyama2019,akiyama2019first} provide a natural testing ground for higher--order curvature--scalar gravity. In this framework, the HOCG parameter $\xi$ modifies the photon sphere structure and therefore the shadow diameter. By confronting the theoretical predictions with the EHT data, one can delineate the observationally consistent range of $\xi$.

One of the relevant observables is the angular shadow diameter, $\Omega_{\text{sh}}$, given in terms of the critical impact parameter $b_c$ and the observer’s distance $\mathcal{D}$ as \cite{perlick2022calculating,afrin2023tests,kumar2020rotating}

\begin{equation}
\Omega_{\text{sh}} = \frac{2b_c}{\mathcal{D}}.  
\end{equation}  
In observational units, it can be expressed as 
\cite{xu2025optical,heydari2024effect}  
\begin{equation}\label{omega}
\Omega_{\text{sh}} = \frac{6.191165 \times 10^{-8}\gamma}{\pi{\mathcal{D}/{\text{Mpc}}}} \frac{b_c}{M} \, (\mu\text{as}),  
\end{equation}  
where $\gamma$ is the mass ratio of the black hole to the Sun. Now we check the impact of the HOCG parameter on the angular shadow diameter with both $Sgr A^*$ and $M87^*$ data.


\subsection{Constraints with image of $M87^*$}

The EHT collaboration reports
the mass and distance of $M87^*$ to be $M \simeq 6.5 \times 10^9 M_\odot$ and $\mathcal{D} = 16.8$ Mpc, respectively \cite{blakeslee2009acs,gebhardt2011black,bird2010inner}. 
Moreover, the angular diameter of the supermassive black hole $M87^*$ has been measured by EHT as $42 \pm 3~\mu\text{as}$ \cite{akiyama2019firstL6,akiyama2019first}. Utilizing $M87^*$ parameters in Eq. \eqref{omega} and expanding up to order $\xi^{3/2}$, yields to the follwing equation
\ie
\Omega^{\scriptscriptstyle{M87^*}}_{\text{sh}}=39.61199-6.60199\left(\frac{ \xi }{M^2}\right).
\fe

\begin{figure}[ht!]
	\centering
	\includegraphics[height=85mm]{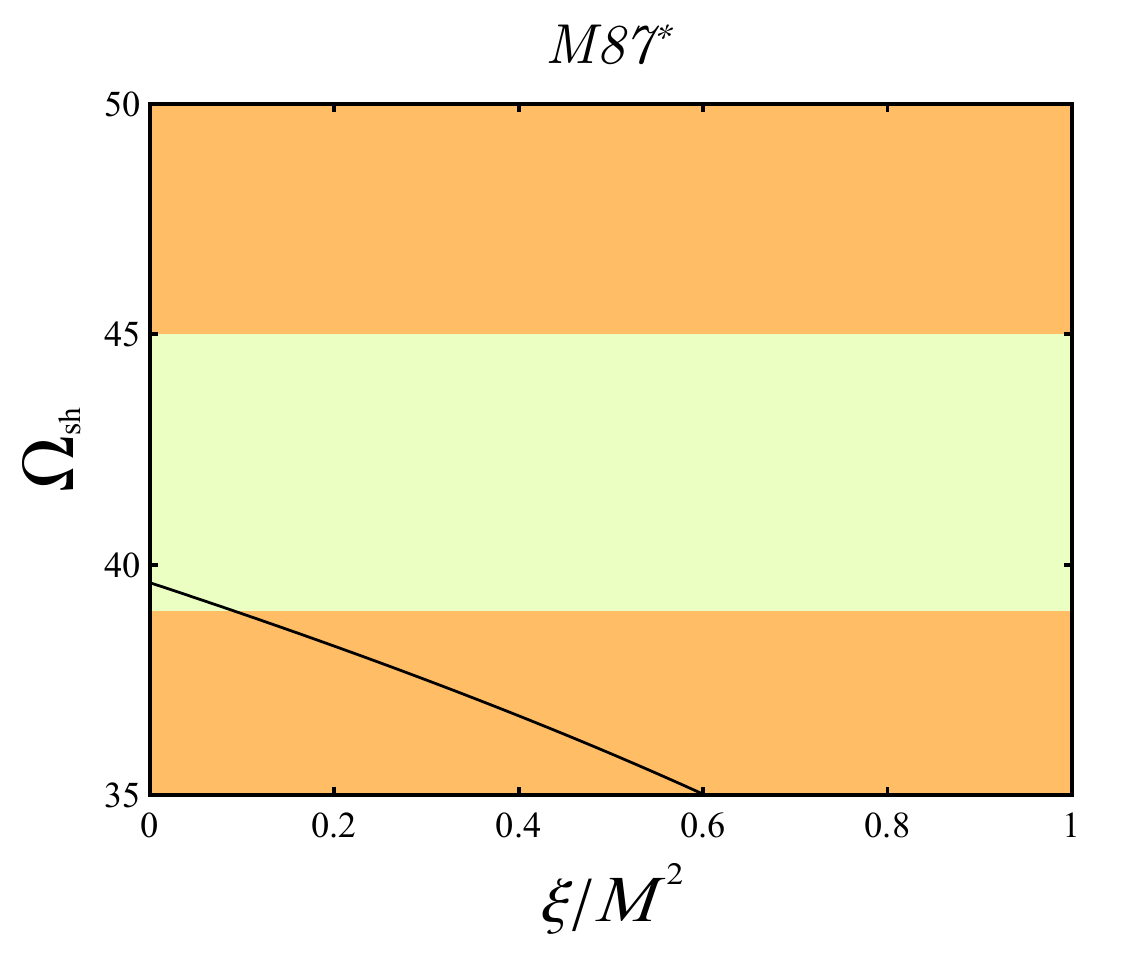}
	\caption{Angular shadow diameter $\Omega_{\text{sh}}$ of $M87^*$ as a function of the scaled HOCG parameter $\xi/M^2$. The shaded band indicates the EHT observational bounds.}
	\label{fig:M87}
\end{figure}  

The relationship between $\Omega_{\text{sh}}$ and $\xi/M^2$, shown in Fig.~\ref{fig:M87}, reveals a decreasing angular shadow size for increasing values of the HOCG parameter. The theoretical prediction falls below the EHT observational limit of $39.00~\mu\text{as}$ for $\xi/M^2 > 0.091$. Consequently, to remain consistent with the EHT results, the parameter $\xi/M^2$ is constrained to the range $0 \leq \xi/M^2 \lesssim 0.091$ for M87$^*$.


\subsection{Constraints with image of $Sgr A^*$}

Based on the most recent measurements, the Gravity collaboration reports the mass of $Sgr A^*$ to be $M = 4 \times 10^6 M_\odot$ and its distance to be $\mathcal{D} = 8.15$ kpc \cite{abuter2022mass,abuter2020detection,banerjee2022shadows}. On the other hand, the EHT collaboration reports $\Omega_{\text{sh}} = 48.7 \pm 7~\mu\text{as}$ \cite{akiyama2022firstSgr,akiyama2022firstSgrA}, corresponding to an observational window of $41.7 \leq \Omega_{\text{sh}} \leq 55.7~\mu\text{as}$. 
Substituting the $Sgr A^*$ properties, in Eq. \eqref{omega} and expanding it up to order $\xi^{3/2}$, results in the following expression
\ie
\Omega^{\scriptscriptstyle{Sgr A^*}}_{\text{sh}}=53.23368-8.87226\left(\frac{\xi }{M^2}\right)
.
\fe
 
\begin{figure}[ht!]
	\centering
	\includegraphics[height=85mm]{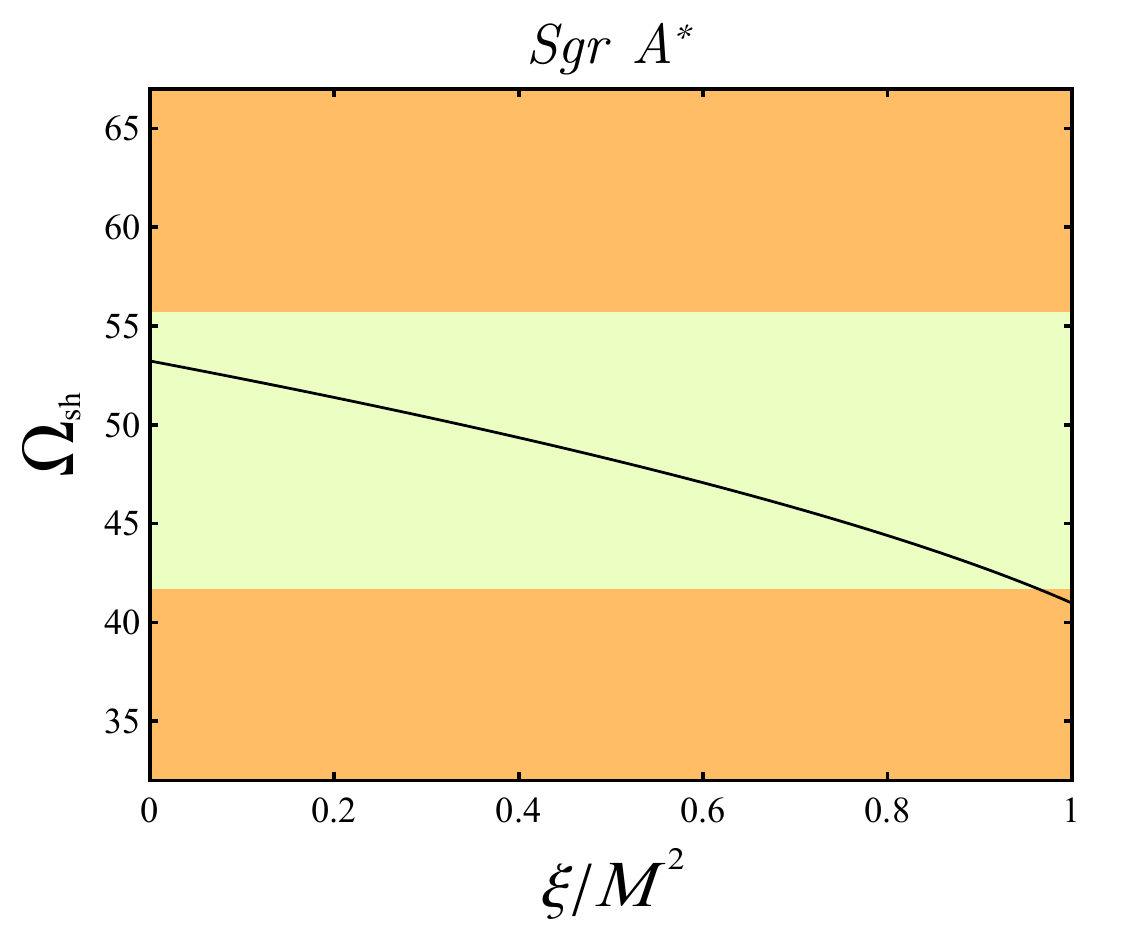}
	\caption{Angular shadow diameter $\Omega_{\text{sh}}$ of $Sgr A^*$ as a function of the scaled HOCG parameter $\xi/M^2$. The green band corresponds to the EHT measurement range.}
	\label{fig:SgrA}
\end{figure}  

The theoretical prediction of $\Omega^{\scriptscriptstyle{Sgr A^*}}_{\text{sh}}$ with respect to HOCG parameter in the mass unit, is displayed in Fig.~\ref{fig:SgrA}. The allowed range of angular diameter based on $Sgr A^*$ is represented by the green region.
We find that $\Omega_{\text{sh}}$ decreases monotonically with increasing $\xi/M^2$. A critical restriction point occurs at  $\xi/M^2 \simeq 0.963$. For $\xi > 0.963 M^2$, the predicted shadow size drops below the EHT lower bound of $41.7~\mu\text{as}$ and is therefore observationally excluded. This yields the constraint $0 \leq \xi/M^2 \lesssim 0.963$ for $Sgr A^*$.


\section{\label{Sec12}Bounds Inferred from Solar System Observations}

Einstein’s theory achieved its first confirmation through Solar System observations, where the Sun was modeled as a static, spherically symmetric source described by the Schwarzschild metric. These classical tests inevitably carry experimental uncertainties, which leave room for small deviations that could, in principle, be attributed to corrections to General Relativity. Taking advantage of this observational tolerance, we now turn to Eq. (\ref{mainmetric}) and explore how the parameter $\xi$ modifies the geometry. The objective is to determine the size of this effect and ensure that such modifications remain compatible with present--day measurements.

For particle trajectories restricted to the equatorial plane ($\theta = \pi/2$), the starting point is the Lagrangian formalism. This framework encodes the dynamics of the system and provides the equations of motion needed to study geodesics in the spacetime under consideration
\ie
\label{eq:lagr1}
A(r,\xi){{\dot t}^2} - B(r,\xi)^{-1} \, {{\dot r}^2} - r^2{{\dot \varphi }^2} = \eta\, .
\fe

To constrain the motion, the four--velocity must satisfy a normalization rule, which is implemented by fixing the Lagrangian to
$L(x,\dot{x}) = -\eta/2$.
The symbol $\eta$ encodes the nature of the trajectory: setting $\eta = 0$ selects null geodesics, relevant for lightlike motion, whereas $\eta = 1$ singles out timelike paths traced by massive particles, whose evolution is described as a function of their proper time $\lambda$.

With this normalization in place, the spacetime symmetries lead directly to conserved quantities. Exploiting the time--translation and rotational invariance of the metric tensor, one defines two constants of motion: the energy $E$ and the angular momentum $L$ (as we introduced in the previous sections of the paper), both of which follow from the conjugate momenta associated with the coordinates $t$ and $\phi$, namely,
\ie
\label{constant}
E = A(r,\xi)\dot t \quad\mathrm{and}\quad L = D(r,\xi)\dot \varphi.
\fe

Starting from the Lagrangian in Eq.~\eqref{eq:lagr1} and using the relations that define the conserved energy and angular momentum given in Eq.~\eqref{constant}, the resulting combination leads to the following expression:
\ie
\label{massive}
    \left[\frac{\mathrm{d}}{\mathrm{d}\varphi}\left(\frac{1}{r}\right)\right]^2=r^{-4}D^2(r,\xi)\left[\frac{E^2}{A(r,\xi)B(r,\xi)^{-1} \, L^2}-\frac{1}{B(r,\xi)^{-1} \, L^2}\left(\eta+\frac{L^2}{r^2\sin^2(\theta)}\right)\right]\, .
\fe

By defining the auxiliary quantity $u = \dfrac{L^{2}}{M r}$ and taking the derivative of Eq.~\eqref{massive} with respect to the azimuthal coordinate $\varphi$, one isolates the leading contributions that depend on the parameter $\xi$, which can then be expressed as follows:
\begin{align}\label{eq:u-massive}
&\frac{\mathrm{d}^2 u}{\mathrm{d}\varphi^2} = \eta - u + \frac{3 M^2 u^2}{L^2}\nonumber\\
&+\frac{E^2 \xi  \left(L^2 u-M^2 u^2\right)}{\left(A^2-2 M^2 u\right)^2}
+\frac{2 M^3 \xi ^{3/2}}{L^8 \left(L^2-2 M^2 u\right)^2}\left(2 L^6 E^2 u^3-2 L^6 \eta  u^3-3 L^4 E^2 M^2 u^4+8 L^4 \eta  M^2 u^4\right.\nonumber\\
&\left.-3 L^4 M^2 u^5-8 L^2 \eta  M^4 u^5+12 L^2 M^4 u^6-12 M^6 u^7\right).
\end{align}

When both $\xi$ and $M$ are treated as perturbatively small, Eq.~\eqref{eq:u-massive} can be expanded and reorganized so that its leading contribution takes the simplified form:
\begin{align}\label{eq:u}
     u''(\varphi)= \, \eta-\left(1-\frac{E^2 \xi }{L^2}\right)u+\frac{3 M^2  \left(L^4+L^2 E^2 \xi \right)}{L^6}u^2\, .
\end{align}

From the reorganized form of the equation, a few distinctive aspects can be highlighted. First, the parameter $\xi$ couples explicitly with the test particle’s energy, a feature often predicted in approaches inspired by quantum gravity \cite{Addazi:2021xuf,AlvesBatista:2023wqm}. Second, there is the fact that some of the $\xi$--dependent contributions remain completely independent of the central mass $M$, leading to direct modifications of the Newtonian regime. Such mass--independent corrections are consistent with the predictions of several quantum gravity frameworks, in which Planck--scale effects are expected to deform the classical Galilean and Minkowski limits \cite{Amelino-Camelia:2008aez}.


\subsection{The precession of Mercury’s orbit}

A classic test of any alternative to General Relativity is provided by Mercury’s perihelion precession, observed as an angular shift accumulated over a hundred years. In this treatment, it is important to mention that Mercury is set as a massive particle ($\eta = 1$) traveling through the Sun’s gravitational field, approximated by a static, spherically symmetric geometry. Setting $\xi = 0$ retrieves the usual general relativistic contribution, where the main post-Newtonian effect originates from the $u$--linear term in Eq.~\eqref{eq:u}.

To streamline the calculation, the mass and deformation parameter are rescaled by defining $m = M/L$ and $\epsilon = \xi E^{2}/L^{2}$. Expressed with these reparametrized quantities, the principal deviation from the Newtonian prediction becomes:
\ie
\label{eq:merc}
     u''(\varphi)= \, 1 -u+\epsilon u+3 m^2 u^2\, .
\fe

Advancing the calculation requires rewriting the solution as a perturbative series in the small quantities $M/L$ and $\xi$. For this purpose, we expand $u$ in the form
$u = u_{0} + m^{2}u_{m} + \epsilon u_{\epsilon}$,
where each term captures a distinct order of correction. The piece $u_{0}$ corresponds to the Newtonian result, including its leading adjustment, and can be explicitly expressed as:
\ie
    u_0=1+e \cos(\varphi)\, .
\fe
Substituting the perturbative decomposition of $u$ back into Eq.~\eqref{eq:merc} and rearranging the resulting terms yields the differential equation in the form:
\ie
\label{eq:pert_hay_merc}
    m^2 \left(3 (e \cos (\varphi )+1)^2-u_m''-u_m\right)+\epsilon \left(1+e \cos (\varphi )-u_{\epsilon}''-u_{\epsilon}\right)=0\, .
\fe
After neglecting the mixed terms proportional to $m^{2}\epsilon$ and $m^{2}\lambda$, the surviving contribution proportional to $m^{2}$ corresponds to the standard post--Newtonian correction from General Relativity. This term can be written in the form:
\ie
    u_m=3m^2\left[\left(1+\frac{e^2}{2}\right)-\frac{e^2}{6}\cos\left(2\varphi\right)+e\varphi\sin\left(\varphi\right)\right]\, .
\fe

The terms that are either constant or purely periodic inside the brackets do not contribute to any long--term displacement of the perihelion. The constant piece simply shifts the reference level without evolving, while the oscillatory component cancels out over a complete revolution because of its symmetric nature. These contributions are therefore discarded. What remains significant is the part that grows linearly with $\varphi$, as it accumulates over successive orbits and leads to a measurable precession.

For the perturbation proportional to $\epsilon$ in Eq.~\eqref{eq:pert_hay_merc}, consistency of the expansion demands that it introduce only higher--order corrections. The trigonometric terms generated in this step mostly average to zero when integrated over $0 \leq \varphi \leq 2\pi$, except for the secular piece $-\frac{1}{2}e\varphi\sin\varphi$, which steadily shifts the orbital configuration. After combining this effect with the Newtonian baseline and the post--Newtonian contribution from General Relativity, and rewriting everything in terms of the original parameters, the final expression for $u(\varphi)$ is obtained as:
\ie
    u(\varphi)=1 +e \cos (\varphi )+\frac{3M^2}{L^2}\left(1+\epsilon\frac{L^2}{6M^2}\right)e\varphi\sin(\varphi)\, .
\fe

Because the contribution proportional to $\varphi\sin\varphi$ is extremely small, it is omitted from the approximation. After dropping this term, the two surviving pieces can be combined using standard trigonometric relations, which leads to the compact expression:
\ie
u(\varphi)\approx 1+e\cos\left[\left(1-\frac{3M^2}{L^2}\left(1+\epsilon\frac{L^2}{6M^2}\right)\right)\varphi\right]\doteq 1+e\cos\left[\left(1-\frac{3\widetilde{M}^2}{L^2}\right)\varphi\right]\, .
\fe

This result can be interpreted as if the central mass were replaced by an effective quantity
$\widetilde{M}^{2} = M^{2}\!\left(1 + \frac{\epsilon L^{2}}{6M^{2}}\right)$,
which modifies the standard general relativistic prediction. Using this redefined mass parameter, the extra term responsible for the perihelion shift is obtained as:
\ie
\Delta\Phi=6\pi\frac{\widetilde{M}^2}{L^2}=6\pi\frac{M^2}{L^2}\left(1+\epsilon\frac{L^2}{6M^2}\right)\, .
\fe

From this formulation, a dimensionless measure of the departure from the standard general relativistic result can be written as
$\delta_{\text{Perih}} = \dfrac{\epsilon L^{2}}{6 M^{2}}$.
Since Mercury completes roughly one revolution every 88 days, it performs approximately
$415 \approx 100 \times 365.25/88$
orbits in a century. Multiplying this number by the per-orbit angular shift gives the net precession accumulated over 100 years. In the framework of General Relativity, the predicted value is
$\Delta\Phi_{\text{GR}} = 42.9814''$ per century,
which is in excellent agreement with the observed result
$\Delta\Phi_{\text{Exp}} = (42.9794 \pm 0.0030)''$/century \cite{Casana:2017jkc,Yang:2023wtu}. This near-perfect match not only confirms Einstein’s theory but also allows one to place stringent limits on the parameter $\epsilon$, and consequently on the corrections induced by $\xi$.

To express the orbital quantities explicitly, the angular momentum is written as
$L^{2} = M a (1 - e^{2})$,
where $a$ and $e$ denote the semi-major axis and eccentricity of the orbit, respectively. The specific orbital energy is given by
$E = - M/(2a)$ \cite{Goldstein:2002}.
Adopting natural units and Mercury’s orbital parameters, one sets
$M = M_{\odot} = 9.138 \times 10^{37}$,
$a = 3.583 \times 10^{45}$,
$e = 0.2056$,
which yields
$L = 5.600 \times 10^{41}$,
confirming that $M^{2}/L^{2}$ is indeed very small, validating a perturbative expansion. The corresponding energy squared is
$E^{2} = 1.627 \times 10^{-16}$,
implying that any energy-dependent correction is negligible.

Finally, inserting these values into
$\epsilon = \xi \dfrac{E^{2}}{L^{2}}$
leads to the allowed range for the deformation parameter:
$-9.15 \times 10^{18}\,\mathrm{m}^{2} \leq \xi \leq 1.83 \times 10^{18}\,\mathrm{m}^{2}$.


\subsection{Deflection of light }

When a photon travels near a gravitating object, its path no longer follows a straight line, producing an apparent shift in the position of the distant source as seen by the observer. This effect, usually referred to as gravitational light deflection, is described mathematically by the null geodesics of the spacetime. In this case, one enforces $\eta = 0$ in Eq.~\eqref{eq:u} to represent massless particles. For convenience, the radial coordinate is inverted through the change of variable $u = 1/r$, which transforms the equation into the form:
\ie
\label{eq:light}
    u''(\varphi)=\frac{-L^2+E^2\xi}{L^2}u+\frac{3 M \left(L^2+\xi E^2 \right)}{L^2}u^2 \, .
\fe

Within this framework, the quantity $b = L/E$ is identified as the impact parameter, characterizing how closely the photon approaches the central object, as we have introduced in the previous sections. To examine exclusively the influence of the deformation parameter $\xi$, one separates its contributions from those depending on $M$. Interestingly, the $\xi$--dependent terms survive even in the limit where the central mass vanishes, indicating that these corrections deform the light path independently of the gravitational potential and therefore alter the Newtonian prediction at the very first order. After redefining the mass contribution for convenience, the dominant terms take the form:
\ie
\label{eq:redef_light_theta}
     u''(\varphi)+\left(1-\frac{\xi}{b^2}\right)u=3\widetilde{M}u^2\, .
\fe

In this formulation, the parameter of mass may be rewritten in an effective form, $\widetilde{M}=M\!\left(1+\frac{\xi}{b^{2}}\right),$
which encodes the deformation effects. By imposing that the left-hand side of the equation vanishes, one recovers the standard Newtonian expression, and the deviation from the classical prediction appears as an additional correction term \cite{Yang:2023wtu}
\ie
    u_0=b^{-1}\sin\left(\left(1-\frac{\xi}{2b^2}\right)\varphi \right)\, .
\fe
Choosing the initial condition $\varphi_{0}=0$ corresponds to a straight--line propagation with no bending. Inserting this zeroth--order path into Eq.~\eqref{eq:redef_light_theta} and expanding in the limit of very small angular deflections ($\varphi \ll 1$) leads to the first--order perturbed expression:
\ie
    u(\varphi)=\frac{1}{b}\sin\left(\left(1-\frac{\xi}{2b^2}\right)\varphi\right)+\frac{\widetilde{M}}{b^2(1-\xi/b^2)}\left[1+\cos^2\left(\left(1-\frac{\xi}{2b^2}\right)\varphi\right)\right]\, .
\fe

When the photon escapes to large distances, its trajectory asymptotically satisfies $u \to 0$ (equivalently $r \to \infty$). The entry and exit angles of the trajectory can be obtained by imposing this asymptotic condition and solving for $\varphi$. Allowing small fluctuations in both the angular coordinate and the deformation parameters gives the approximate results
$\varphi_{\text{in}} = -\frac{2\bar{M}}{b}, \qquad  
\varphi_{\text{ex}} = \pi + \frac{2\bar{M}}{b},$
where the effective mass is given by  $\bar{M}=M\!\left(1+\frac{5\xi}{2b^{2}}\right)$.
The total change in direction of the light ray is obtained from the difference between the incoming and outgoing angles, giving the deflection angle $\delta = -2\varphi_{u \to 0}$
\ie
    \delta = \frac{4\bar{M}}{b}=4\frac{M}{b}\left(1 + \frac{5\xi}{2b^2}\right)\, .
\fe

For a photon skimming the solar limb, the impact parameter is effectively the solar radius,
$b \simeq R_{\odot} = 4.305 \times 10^{43},$
with the Sun’s mass being regarded as
$M = M_{\odot} = 9.138 \times 10^{37}.$
The effect of the deformed parameter $\xi$ appears through the multiplicative factor
$1 + \frac{5\xi}{2b^{2}}$,
which is remarkable in that it does not depend explicitly on $M$ but rather on the impact parameter, and hence is tied to the solar radius.

In standard General Relativity, the expected deflection angle is
$\delta_{\text{GR}} = \frac{4M}{b} = 1.7516687''.$

Experimentally, the observed value is reported as
$\delta_{\text{Exp}} = \frac{1}{2}(1+\gamma)\times 1.7516687'',
\qquad 
\gamma = 0.99992 \pm 0.00012$ \cite{dsasdas}. 
To quantify the contribution from $\xi$, the above multiplicative factor is matched against the empirical ratio $(1+\gamma)/2$, resulting in the constraint $-1.94 \times 10^{13}\,\mathrm{m}^{2} \leq \xi \leq 3.87 \times 10^{12}\,\mathrm{m}^{2}.$


\subsection{Time delay of light }

The time delay of light (Shapiro effect) \cite{Shapiro:1964uw} describes the extra propagation time undergone by electromagnetic signals as they pass near a massive object before reaching their destination. In the case of planetary radar ranging, this means that signals released from Earth to an inner planet and reflected back take slightly longer to complete the round trip because spacetime is curved by the Sun’s gravitational field.

To quantify this delay, one studies null geodesics derived from Eq.~\eqref{massive}. Imposing the condition $\eta = 0$ for massless particles and using the conserved energy and angular momentum from Eq.~\eqref{constant}, the trajectory can be reformulated as shown below
\ie
  \left(  \frac{\mathrm{d}r}{\mathrm{d}t}\right)^2=\frac{A(r,\xi)r^2-\frac{L^2}{E^2}A(r,\xi)}{B(r,\xi)^{-1} \, r^2}\, .
\fe

Following the approach of Ref. \cite{Wang:2024fiz}, the constants of motion can be rewritten using the closest approach of the photon to the Sun, identified with the impact parameter $b$. This turning point is obtained by enforcing $\dot{r}=0$, which yields the relation
$\frac{L^{2}}{E^{2}} = \frac{D(r_{\text{min}},\xi)}{A(r_{\text{min}},\xi)}.$ With this identification, the travel time of the signal can properly be derived as a function of radial coordinate terms, resulting in the following integral representation:
\ie
\label{eq:shapiro_main}
    \mathrm{d} t=\pm \frac{1}{A(r,\xi)}\frac{1}{\sqrt{\frac{1}{A(r,\xi)B(r,\xi)^{-1}}-\frac{r_{\text{min}}^2/A(r_{\text{min},\xi})}{B(r,\xi)^{-1}r^2}}}\, .
\fe

To highlight the departure from Minkowski spacetime, subleading pieces are discarded and only the leading contributions proportional to $M$ and $\xi$ are retained. Under this perturbative treatment, integrating Eq.~\eqref{eq:shapiro_main} yields the compact result:
\begin{align}
    t&=\sqrt{r^2-r_{\text{min}}^2}+M\left(\sqrt{\frac{r-r_{\text{min}}}{r+r_{\text{min}}}}+2\ln\left(\frac{r+\sqrt{r^2-r_{\text{min}}^2}}{r_{\text{min}}}\right)\right)\\
    &+\xi\left[\frac{M}{r_{\text{min}}} \sqrt{\frac{r-r_{\text{min}}}{r+r_{\text{min}}}}\left(\frac{8}{r_{\text{min}}}+\frac{5}{r}\right)+\arctan\left(\frac{\sqrt{r^2-r_{\text{min}}^2}-r}{r_{\text{min}}}\right)\left(\frac{2}{r_{\text{min}}}-\frac{M}{r_{\text{min}}}\right)\right]\nonumber\\
    &-\xi^{3/2}\left[\frac{M}{2r_{\text{min}}r} \sqrt{\frac{r-r_{\text{min}}}{r+r_{\text{min}}}}\left(\frac{1}{r_{\text{min}}}+\frac{1}{r}\right)-\frac{M}{r_{\text{min}}^3}\arctan\left(\frac{\sqrt{r^2-r_{\text{min}}^2}-r}{r_{\text{min}}}\right)\right]\nonumber
\end{align}

For distances much larger than the point of closest approach ($r \gg r_{\text{min}}$), the expression simplifies significantly. In this asymptotic regime, the main contribution comes from the standard general relativistic piece together with the first-order correction induced by $\xi$, which together take the form:
\ie
\label{eq:sh_t_nc}
    t(r)=r+M+2M\ln\left(\frac{2r}{r_{\text{min}}}\right)+\xi\frac{8M}{r_{\text{min}}^2}+\xi^{3/2}\frac{M}{3r^3}\, .
\fe

Define $t(r_E)$ as the one--way propagation time from the emitter’s position until reaching the minimum solar distance, and $t(r_R)$ as the corresponding travel time from that point to the receiver. Using Eq.~\eqref{eq:sh_t_nc}, with $r_E$ and $r_R$ specifying the respective radial coordinates, these quantities can be evaluated explicitly. The total round–trip duration is then obtained by doubling both segments, giving
$T = 2\,t(r_E) + 2\,t(r_R). $
With this construction, the overall signal travel time can be compactly written as:
\ie
T= 2(r_E+r_R)+4M\left[1+\ln\left(\frac{4r_Rr_E}{r_{\text{min}}^2}\right)+\xi\frac{8}{r_{\text{min}}^2}+\xi^{3/2}\frac{1}{6}\left(\frac{1}{r_E^3}+\frac{1}{r_R^3}\right)\right]=T_{\text{flat}}+\delta T\, .
\fe

Thereby, the Shapiro delay represents the excess time taken by the signal compared to what would be measured if spacetime were flat. Without gravity, the round–trip duration would simply be
$T_{\text{flat}} = 2(r_E + r_R).$ In the language of the parametrized post--Newtonian framework, the correction produced by the curvature of spacetime is written as:
\ie
\delta T = 4M\left(1+\frac{1+\gamma}{2}\ln \left(\frac{4r_Rr_E}{r_{\text{min}}^2}\right)\right)\, .
\fe

Observations from the Cassini mission \cite{Bertotti:2003rm,Will:2014kxa} provided one of the tightest experimental tests of relativistic gravity, constraining the PPN parameter to
$|\gamma - 1| < 2.3 \times 10^{-5}.$ Working in natural units, the mean Earth--Sun distance is taken as one astronomical unit,
$r_E = 1\,\text{AU} = 2.457 \times 10^{45}.$

Throughout the observation period, the spacecraft was located at $r_R = 8.46\,\text{AU}, $
while the signal’s closest approach to the Sun was $r_{\min} = 1.6\,R_{\odot}, \, R_{\odot} = 4.305 \times 10^{43}.
$ Substituting these values into the expression for the Shapiro delay and keeping only the leading term that depends on $\xi$, one arrives at the constraint $ |\xi| \leq 2.04 \times 10^{14}\,\mathrm{m}^{2}. $

\begin{table}[h!]
\centering
\caption{Constraints on $\xi$ from Solar System Observables}
\label{tab:constr}
\begin{tabular}{lc}
\hline\hline
\textbf{Solar System Test} & Constraints ($\text{m}^2$)  \\
\hline
{\bf Mercury precession}   & \makecell{$-9.15 \times 10^{18} \, \text{m}^2 \leq \xi \leq 1.83 \times 10^{18} \, \text{m}^2$} \\
{\bf Light deflection}     & \makecell{$-1.94\times 10^{13}\, \text{m}^2 \leq\xi\leq 3.87\times 10^{12}\, \text{m}^2$}   \\
{\bf Shapiro time delay}   & \makecell{$-2.04 \times 10^{14} \, \text{m}^2 \leq \xi \leq 2.04 \times 10^{14} \, \text{m}^2$} \\
\hline\hline
\end{tabular}
\end{table}

For better visualization, we represent all constraints obtained based on Solar System and observational data in Table. \ref{tab:allbounds}. We have converted all constraints to both dimensional (SI) and dimensionless 
forms for direct comparison. For Solar System,  $\xi/M^2$ is computed using $M = GM_\odot/c^2 = 1.477$ km, while for black holes, $\xi/M^2$ is calculated directly in geometric units. Since the masses involved in the constraints calculations, differ by 9--10 orders of magnitude ($M_{\text{Solar}}= M_\odot$ vs. $M_{\text{BH}} \sim 10^6$--$10^9 M_\odot$), the most meaningful comparison is in terms of the dimensional parameter $\xi$ (m$^2$) (assuming that $\xi$ is constant across scales). The table shows that the Solar System exploration provide tighter 
bounds on the absolute value of $\xi$ (in m$^2$) by light deflection method. Our black hole constraints, when expressed in SI units as $\xi \leq 3.36 \times 10^{19}$ m$^2$ (\textit{Sgr A}$^*$) and $\xi \leq  8.38 \times 10^{24}$ m$^2$ (\textit{M87}$^*$), are indeed much weaker than all the Solar System bounds. {It is worth noting that the comparison between the Solar System and black hole constraints depends on whether $\xi$ is assumed to have the same value at all scales. Under this assumption, the Solar System tests provide the tighter bounds. The black-hole results should therefore be regarded as complementary probes of the strong-field regime. A possible variation of $\xi$ between Solar System and black hole scales is not considered here and would require a separate theoretical justification.}
\begin{table}[h!]
	\centering
	\caption{Comprehensive constraints on the deformation parameter $\xi$ from various gravitational aspects.  For Solar System tests, $\xi_{\text{SI}}$ is in SI units; $\xi/M^2$ is computed using 
		$M = GM_\odot/c^2 = 1.477$ km. For black holes, $\xi/M^2$ is calculated directly in geometric units.}
	\label{tab:allbounds}
	\begin{tabular}{|l|c|c|c|}
		\hline
		\textbf{System/Test} & \textbf{Mass ($M_\odot$)} & \textbf{$\xi_{\text{SI}}$ (m$^2$)} & \textbf{$\xi/M^2$ (dimensionless)}\\
		\hline
		
		\multicolumn{4}{|l|}{\textbf{Solar System Tests}} \\
		\hline
		Light deflection & 1 & $-1.94 \times 10^{13} \leq \xi \leq 3.87 \times 10^{12}$ & $-8.90 \times 10^{6} \leq \xi/M^2 \leq 1.77 \times 10^{6}$ \\
		\hline
		Mercury precession & 1 & $-9.15 \times 10^{18} \leq \xi \leq 1.83 \times 10^{18}$ & $-4.20 \times 10^{12} \leq \xi/M^2 \leq 8.39 \times 10^{11}$ \\
		\hline
		Shapiro time delay & 1 & $-2.04 \times 10^{14} \leq \xi \leq 2.04 \times 10^{14}$ & $-9.35 \times 10^{7} \leq \xi/M^2 \leq 9.35 \times 10^{7}$ \\
		\hline
		
		\multicolumn{4}{|l|}{\textbf{Black Hole Lensing}} \\
		\hline
		\textit{Sgr A}$^*$ & $4.0 \times 10^{6}$ & $0 \leq \xi \leq 3.36 \times 10^{19}$ & $0 \leq \xi/M^2 \lesssim 0.963$ \\
		\hline
		\textit{M87}$^*$ & $6.5 \times 10^{9}$ & $0 \leq \xi \leq 8.38 \times 10^{24}$ & $0 \leq \xi/M^2 \lesssim 0.091$ \\
		\hline
	\end{tabular}
\end{table}


\section{Conclusion }

In this work, we explored the gravitational features of a recently proposed black hole solution within the framework of higher--order curvature--scalar gravity \cite{Nashed:2025ebr}. The resulting spacetime could be interpreted as a ``mix geometry": its $g_{tt}$ component resembled the Reissner--Nordström solution, while its $g_{rr}$ component was similar to that of a Loop Quantum Gravity–inspired black hole.

As a first step, we determined the event horizon, which was approximately $r_{h} \approx 2M - \frac{\xi^{3/2}}{4 M^{2}}$. Solving $1/g_{rr}=0$ also revealed an additional real and positive root, corresponding to the Cauchy horizon, given by $r_{\text{cau}} \approx \sqrt{\xi} + \frac{\xi}{6M} + \frac{\xi^{3/2}}{12M^{2}} + \frac{35 \xi^{2}}{648 M^{3}}$. We observed opposite behaviors: increasing $\xi$ reduced $r_{h}$ while enlarged $r_{\text{cau}}$.

The analysis then turned to quasinormal modes. We considered perturbations of all spins—scalar, vector, tensor, and spinorial—by solving the corresponding massless field equations (Klein–Gordon, Proca--like, Dirac, etc.). After separating variables, the effective potentials were extracted and the WKB method was employed to compute the complex frequencies. For all perturbations, larger values of $\xi$ led to less damped oscillations. These results were confirmed by numerical simulations in the time domain.

Next, we studied null geodesics, photon spheres, and black hole shadows. Light--like geodesics were solved numerically, showing that larger $\xi$ values produced weaker light deflection. The photon sphere radius was obtained analytically as $r_{\text{photon}} = \frac{1}{2} \bigl( 3M + \sqrt{9M^{2} - 8\xi} \bigr)$, in close analogy with the Reissner–Nordström case. The shadow radius was found to be $R = 3\sqrt{3}M - \frac{\sqrt{3}\xi}{2M} - \frac{7\xi^{2}}{24 \sqrt{3} M^{3}}$, which decreased as $\xi$ grew.

The gravitational lensing analysis was carried out in both weak and strong deflection regimes. In the weak--field case, we applied the Gauss--Bonnet theorem to compute the deflection angle $\tilde{\alpha}(b,\xi)$, which increased with $\xi$. In the strong deflection limit, however, the deflection angle $a(b,\xi)$ decreased as $\xi$ grew. This behavior was confirmed by analyzing geodesics near the photon sphere. We also estimated the corresponding lensing observables using Event Horizon Telescope data for $Sgr A^{*}$ and $M87^{*}$.

Furthermore, we constrained the parameter $\xi$ using Solar System tests, obtaining
Mercury perihelion precession: $-9.15 \times 10^{18}\,\mathrm{m}^{2} \leq \xi \leq 1.83 \times 10^{18}\,\mathrm{m}^{2}$,
light deflection: $-1.94 \times 10^{13}\,\mathrm{m}^{2} \leq \xi \leq 3.87 \times 10^{12}\,\mathrm{m}^{2}$,
Shapiro time delay: $-2.04 \times 10^{14}\,\mathrm{m}^{2} \leq \xi \leq 2.04 \times 10^{14}\,\mathrm{m}^{2}$.
Finally, the constraints on the HOCG parameter, obtained by different gravitational methods and observational data, have been compared and discussed.

As a natural continuation of this study, it seems to be interesting to analyze the metric presented here within the framework of ensemble theory, in line with recent developments reported in the literature \cite{araujo2023thermodsddynamical,furtado2023thermal,araujo2022does,filho2025modified}. In parallel, a comprehensive investigation of greybody factors for scalar, vector, tensor, and spinor perturbations, together with their connection to quasinormal modes and particle creation processes for both bosonic and fermionic fields, is already underway. This latter work, developed following the methodology outlined in Refs.~\cite{araujo2025particle,araujo2025does,araujo2025particleasasa}, is in its final stage of revision and is expected to be published shortly on arXiv.

\section*{Acknowledgments}
\hspace{0.5cm}

A. A. Araújo Filho is supported by Conselho Nacional de Desenvolvimento Cient\'{\i}fico e Tecnol\'{o}gico (CNPq) and Fundação de Apoio à Pesquisa do Estado da Paraíba (FAPESQ), project numbers 150223/2025-0 and 1951/2025. I. P. L. acknowledges partial financial support from the National Council for Scientific and Technological Development (CNPq), under grant No. 312547/2023-4. In addition, I. P. L. would like to acknowledge the networking support of the COST Actions BridgeQG (CA23130), RQI (CA23115) and FuSe (CA24101), both funded by COST — the European Cooperation in Science and Technology. N. H. is supported by Conselho Nacional de Desenvolvimento Cient\'{\i}fico e Tecnol\'{o}gico — CNPq, project number 152891/2025-0. N. H. is grateful for the support provided by three COST Actions: CA21106 (COSMIC WISPers in the Dark Universe: Theory, Astrophysics and Experiments), CA21136 (Addressing Observational Tensions in Cosmology with Systematics and Fundamental Physics, also known as CosmoVerse), and CA23130 (Bridging High and Low Energies in Search of Quantum Gravity, or BridgeQG). In addition, the authors thank C.~F.~S.~Pereira and M.~V.~S.~Silva for the fruitful discussions on gravitational lensing incorporated into the revised version of the paper.


	\bibliography{main}

\begin{thebibliography}{100}

\bibitem{Penrose1965}
Roger Penrose.
\newblock Gravitational collapse and spacetime singularities.
\newblock {\em Physical Review Letters}, 14(3):57, 1965.

\bibitem{Hawking1973}
Stephen~W Hawking and George~FR Ellis.
\newblock {\em The large scale structure of space-time}.
\newblock Cambridge university press, 2023.

\bibitem{Joshi2012}
Pankaj~S Joshi.
\newblock {\em Gravitational collapse and spacetime singularities}.
\newblock Cambridge University Press, 2007.

\bibitem{Abbott2017}
Benjamin~P Abbott, Rich Abbott, Thomas~D Abbott, Fausto Acernese, Kendall
  Ackley, Carl Adams, Thomas Adams, Paolo Addesso, Rana~X Adhikari, Vaishali~B
  Adya, et~al.
\newblock Gw170817: observation of gravitational waves from a binary neutron
  star inspiral.
\newblock {\em Physical review letters}, 119(16):161101, 2017.

\bibitem{Akiyama2019}
The Event Horizon~Telescope Collaboration.
\newblock First {M87$^*$} event horizon telescope results. i. the shadow of the
  supermassive black hole.
\newblock {\em The Astrophysical Journal Letters}, 875(1):L1, 2019.

\bibitem{Akiyama2022}
Kazunori Akiyama, Antxon Alberdi, Walter Alef, Juan~Carlos Algaba, Richard
  Anantua, Keiichi Asada, Rebecca Azulay, Uwe Bach, Anne-Kathrin Baczko, David
  Ball, et~al.
\newblock First sagittarius a$^*$ event horizon telescope results. i. the
  shadow of the supermassive black hole in the center of the milky way.
\newblock {\em The Astrophysical Journal Letters}, 930(2):L12, 2022.

\bibitem{Addazi:2021xuf}
P~Jetzer, J~Alvarez-Muniz, R~Alves~Batista, G~Amelino-Camelia, V~Antonelli,
  M~Arzano, M~Asorey, JL~Atteia, S~Bahamonde, F~Bajardi, et~al.
\newblock Quantum gravity phenomenology at the dawn of the multi-messenger
  era—a review.
\newblock {\em Progress in Particle and Nuclear Physics}, 125:103948, 2022.

\bibitem{AlvesBatista:2023wqm}
R.~Alves~Batista et~al.
\newblock {White paper and roadmap for quantum gravity phenomenology in the
  multi-messenger era}.
\newblock {\em Classical and Quantum Gravity}, 42(3):032001, 2025.

\bibitem{DeFelice2010}
Antonio De~Felice and Shinji Tsujikawa.
\newblock {$f(R)$} theories.
\newblock {\em Living Reviews in Relativity}, 13(1):1--161, 2010.

\bibitem{Sotiriou2010}
Thomas~P Sotiriou and Valerio Faraoni.
\newblock {$f(R)$} theories of gravity.
\newblock {\em Reviews of Modern Physics}, 82(1):451--497, 2010.

\bibitem{Nojiri2006}
Shin'Ichi Nojiri and Sergei~D Odintsov.
\newblock Introduction to modified gravity and gravitational alternative for
  dark energy.
\newblock {\em International Journal of Geometric Methods in Modern Physics},
  4(01):115--145, 2007.

\bibitem{Cai2016}
Yi-Fu Cai, Salvatore Capozziello, Mariafelicia De~Laurentis, and Emmanuel~N
  Saridakis.
\newblock {f(T)} teleparallel gravity and cosmology.
\newblock {\em Reports on Progress in Physics}, 79(10):106901, 2016.

\bibitem{Nojiri2017}
Shin'ichi Nojiri, Sergei~D. Odintsov, and Vasilis~K. Oikonomou.
\newblock Modified gravity theories on a nutshell: Inflation, bounce and
  late-time evolution.
\newblock {\em Physics Reports}, 692:1--104, 2017.

\bibitem{Nojiri2019a}
Shin'ichi Nojiri, Sergei~D. Odintsov, and Vasilis~K. Oikonomou.
\newblock k-essence {$f(R)$} gravity inflation.
\newblock {\em Nuclear Physics B}, 941:11--27, 2019.

\bibitem{Nashed2014}
Gamal~GL Nashed.
\newblock Schwarzschild solution in extended teleparallel gravity.
\newblock {\em Europhysics Letters}, 105(1):10001, 2014.

\bibitem{Faraoni2011}
Valerio Faraoni and Salvatore Capozziello.
\newblock {\em Beyond Einstein gravity: a survey of gravitational theories for
  cosmology and astrophysics}.
\newblock Springer, 2011.

\bibitem{Capozziello2011}
Salvatore Capozziello and Mariafelicia De~Laurentis.
\newblock Extended theories of gravity.
\newblock {\em Physics Reports}, 509(4-5):167--321, 2011.

\bibitem{ElHanafy2016}
W~El~Hanafy and Gamal~GL Nashed.
\newblock Reconstruction of {f(T)}-gravity in the absence of matter.
\newblock {\em Astrophysics and Space Science}, 361(6):197, 2016.

\bibitem{Nashed2002}
Gamal~GL Nashed.
\newblock Vacuum non--singular black hole solutions in tetrad theory of
  gravitation.
\newblock {\em General Relativity and Gravitation}, 34(7):1047--1058, 2002.

\bibitem{Awad2018a}
Adel Awad, W~El~Hanafy, GGL Nashed, and Emmanuel~N Saridakis.
\newblock Phase portraits of general f (t) cosmology.
\newblock {\em Journal of Cosmology and Astroparticle Physics}, 2018(02):052,
  2018.

\bibitem{Awad2018b}
Adel Awad, W~El~Hanafy, GGL Nashed, SD~Odintsov, and VK~Oikonomou.
\newblock Constant-roll inflation in {f(T)} teleparallel gravity.
\newblock {\em Journal of Cosmology and Astroparticle Physics}, 2018(07):026,
  2018.

\bibitem{nashed2015kerr}
Gamal~GL Nashed.
\newblock Kerr-nut black hole thermodynamics in {f(T)} gravity theories.
\newblock {\em The European Physical Journal Plus}, 130(7):124, 2015.

\bibitem{Zubair2016}
M~Zubair and G~Abbas.
\newblock Analytic models of anisotropic strange stars in f (t) gravity with
  off-diagonal tetrad.
\newblock {\em Astrophysics and Space Science}, 361(1):27, 2016.

\bibitem{Zubair2016b}
M~Zubair, G~Abbas, and I~Noureen.
\newblock Possible formation of compact stars in {f(R,T)} gravity.
\newblock {\em Astrophysics and Space Science}, 361(1), 2015.

\bibitem{Harko2011}
Tiberiu Harko, Francisco~SN Lobo, Shin’ichi Nojiri, and Sergei~D Odintsov.
\newblock f (r, t) gravity.
\newblock {\em Physical Review D}, 84(2):024020, 2011.

\bibitem{Saleem2020}
Rabia Saleem, Faisal Kramat, and M~Zubair.
\newblock Interior solutions of compact stars in f (t, t) gravity under
  karmarkar condition.
\newblock {\em Physics of the Dark Universe}, 30:100592, 2020.

\bibitem{Cognola2006}
Guido Cognola, Emilio Elizalde, Shin’ichi Nojiri, Sergei~D Odintsov, and
  Sergio Zerbini.
\newblock Dark energy in modified gauss-bonnet gravity: Late-time acceleration
  and<? format?> the hierarchy problem.
\newblock {\em Physical Review D}, 73(8):084007, 2006.

\bibitem{rois2025charged}
Gabriel~I R{\'o}is, Jos{\'e} Tarciso~SS Junior, Francisco~SN Lobo, Manuel~E
  Rodrigues, and Tiberiu Harko.
\newblock Charged black hole solutions in {f(R,T)} gravity coupled to nonlinear
  electrodynamics.
\newblock {\em Physical Review D}, 111(12):124044, 2025.

\bibitem{araujo2025gasdsadravitational}
A.~A Ara{\'u}jo~Filho, N~Heidari, I.~P Lobo, and VB~Bezerra.
\newblock Gravitational signatures of a nonlinear electrodynamics in f (r, t)
  gravity.
\newblock {\em Journal of Cosmology and Astroparticle Physics}, 2025(09):015,
  2025.

\bibitem{de2015constraining}
Ivan De~Martino, Mariafelicia De~Laurentis, and Salvatore Capozziello.
\newblock Constraining {$f(R)$} gravity by the large-scale structure.
\newblock {\em Universe}, 1(2):123--157, 2015.

\bibitem{Starobinsky1979}
AA~Starobinskii.
\newblock Spectrum of relict gravitational radiation and the early state of the
  universe.
\newblock {\em JETP Letters}, 30(11):682--685, 1979.

\bibitem{astashenok2016neutron}
Artyom~V Astashenok.
\newblock Neutron and quark stars in {$f(R)$} gravity.
\newblock In {\em International Journal of Modern Physics: Conference Series},
  volume~41, page 1660130. World Scientific, 2016.

\bibitem{Astashenok2014}
Artyom~V. Astashenok, Salvatore Capozziello, and Sergei~D. Odintsov.
\newblock Maximal neutron star mass and the resolution of the hyperon puzzle in
  modified gravity.
\newblock {\em Physical Review D}, 89(10):103509, 2014.

\bibitem{Astashenok2013}
Artyom~V Astashenok, Salvatore Capozziello, and Sergei~D Odintsov.
\newblock Further stable neutron star models from {$f(R)$} gravity.
\newblock {\em Journal of Cosmology and Astroparticle Physics}, 2013(12):040,
  2013.

\bibitem{Astashenok2017}
Artyom~V Astashenok, Sergei~D Odintsov, and Alvaro De~la Cruz-Dombriz.
\newblock The realistic models of relativistic stars in {$f(R)= R+ \alpha R^2$}
  gravity.
\newblock {\em Classical and Quantum Gravity}, 34(20):205008, 2017.

\bibitem{Astashenok2015}
Artyom~V. Astashenok, Salvatore Capozziello, and Sergei~D. Odintsov.
\newblock Nonperturbative models of quark stars in {$f(R)$} gravity.
\newblock {\em Physics Letters B}, 742:160--166, 2015.

\bibitem{Nashed2019}
Gamal~GL Nashed, W~El~Hanafy, Sergei~D Odintsov, and Vasilis~K Oikonomou.
\newblock Thermodynamical correspondence of {$f(R)$} gravity in the jordan and
  einstein frames.
\newblock {\em International Journal of Modern Physics D}, 29(13):2050090,
  2020.

\bibitem{Nashed2021}
Gamal~GL Nashed and Salvatore Capozziello.
\newblock Anisotropic compact stars in {$f(R)$} gravity.
\newblock {\em The European Physical Journal C}, 81(5):481, 2021.

\bibitem{Stabile2013}
An~Stabile and S~Capozziello.
\newblock Conformal transformations and weak field limit of scalar-tensor
  gravity.
\newblock {\em Physical Review D}, 88(12):124011, 2013.

\bibitem{Capozziello2018}
Salvatore Capozziello, Carlo~Alberto Mantica, and Luca~Guido Molinari.
\newblock Cosmological perfect-fluids in {$f(R)$} gravity.
\newblock {\em International Journal of Geometric Methods in Modern Physics},
  16(01):1950008, 2019.

\bibitem{Odintsov2019a}
SD~Odintsov and VK~Oikonomou.
\newblock Effects of spatial curvature on the {$f(R)$} gravity phase space: no
  inflationary attractor?
\newblock {\em Classical and Quantum Gravity}, 36(6):065008, 2019.

\bibitem{Odintsov2019b}
SD~Odintsov and VK~Oikonomou.
\newblock {$f(R)$} gravity inflation with string-corrected axion dark matter.
\newblock {\em Physical Review D}, 99(6):064049, 2019.

\bibitem{Shah2019}
Parth Shah and Gauranga~C. Samanta.
\newblock {Stability analysis for cosmological models in {$f(R)$} gravity using
  dynamical system analysis}.
\newblock {\em The European Physical Journal C}, 79(5):414, 2019.

\bibitem{Miranda2019}
Tays Miranda, Celia Escamilla-Rivera, Oliver~F. Piattella, and J{\'u}lio~C.
  Fabris.
\newblock {Generic slow-roll and non-gaussianity parameters in {$f(R)$}
  theories}.
\newblock {\em Journal of Cosmology and Astroparticle Physics,}, 05:028, 2019.

\bibitem{Nascimento2019}
J.~R. Nascimento, Gonzalo~J. Olmo, P.~J. Porfirio, A.~Yu. Petrov, and A.~R.
  Soares.
\newblock {Global Monopole in Palatini {$f(R)$} gravity}.
\newblock {\em Physical Review D}, 99(6):064053, 2019.

\bibitem{Elizalde2019a}
Emilio Elizalde, Sergei~D. Odintsov, Tanmoy Paul, and Diego
  S{\'a}ez-Chill{\'o}n~G{\'o}mez.
\newblock {Inflationary universe in {$F(R)$} gravity with antisymmetric tensor
  fields and their suppression during its evolution}.
\newblock {\em Physical Review D}, 99(6):063506, 2019.

\bibitem{Elizalde2019b}
E.~Elizalde, S.~D. Odintsov, V.~K. Oikonomou, and Tanmoy Paul.
\newblock {Logarithmic-corrected $R^2$ Gravity Inflation in the Presence of
  Kalb-Ramond Fields}.
\newblock {\em Journal of Cosmology and Astroparticle Physics,}, 02:017, 2019.

\bibitem{Chen2019}
Long Chen.
\newblock {Dynamical analysis of loop quantum $R^2$ cosmology}.
\newblock {\em Physical Review D}, 99(6):064025, 2019.

\bibitem{Sbisa2019}
Fulvio Sbis{\`a}, Oliver~F. Piattella, and Sergio~E. Jor{\'a}s.
\newblock {Pressure effects in the weak-field limit of {$f(R)=R+\alpha R^2$}
  gravity}.
\newblock {\em Physical Review D}, 99(10):104046, 2019.

\bibitem{Samanta2019}
Gauranga~C. Samanta and Nisha Godani.
\newblock {Validation of energy conditions in wormhole geometry within viable
  {$f(R)$} gravity}.
\newblock {\em The European Physical Journal C}, 79(7):623, 2019.

\bibitem{Bombacigno2019}
Flavio Bombacigno and Giovanni Montani.
\newblock {Big bounce cosmology for Palatini $R^2$ gravity with a
  Nieh{\textendash}Yan term}.
\newblock {\em The European Physical Journal C}, 79(5):405, 2019.

\bibitem{Astashenok2019}
Artyom~V Astashenok, Karim Mosani, Sergey~D Odintsov, and Gauranga~C Samanta.
\newblock Gravitational collapse in general relativity and in r 2-gravity: A
  comparative study.
\newblock {\em International Journal of Geometric Methods in Modern Physics},
  16(03):1950035, 2019.

\bibitem{Nashed2023}
GGL Nashed.
\newblock Charged solution with equal metric ansatz in gauss--bonnet theory
  coupled to scalar field.
\newblock {\em Physics of the Dark Universe}, 41:101260, 2023.

\bibitem{Nojiri2021}
S.~Nojiri, S.~D. Odintsov, V.~K. Oikonomou, and Arkady~A. Popov.
\newblock {Ghost-free $F(R,\mathcal G)$ gravity}.
\newblock {\em Nuclear Physics B}, 973:115617, 2021.

\bibitem{deHaro2023}
Jaume de~Haro, Shin’ichi Nojiri, Sergei~D Odintsov, Vasilis~K Oikonomou, and
  Supriya Pan.
\newblock Finite-time cosmological singularities and the possible fate of the
  universe.
\newblock {\em Physics Reports}, 1034:1--114, 2023.

\bibitem{Capozziello2023}
Salvatore Capozziello and G.~G.~L. Nashed.
\newblock Charged spherically symmetric black holes in scalar-tensor
  gauss–bonnet gravity.
\newblock {\em Classical and Quantum Gravity}, 40(20):205023, 2023.

\bibitem{Millano2023}
Alfredo~D Millano, Genly Leon, and Andronikos Paliathanasis.
\newblock Global dynamics in einstein-gauss-bonnet scalar field cosmology with
  matter.
\newblock {\em Physical Review D}, 108(2):023519, 2023.

\bibitem{Nojiri2024a}
Shin'ichi Nojiri and Sergei~D Odintsov.
\newblock F (q) f(q) gravity with gauss--bonnet corrections: From early-time
  inflation to late-time acceleration.
\newblock {\em Fortschritte der Physik}, 72(9-10):2400113, 2024.

\bibitem{Ilyas2023}
M~Ilyas and Kazuharu Bamba.
\newblock Traversable wormholes with static spherical symmetry and their
  stability in higher-curvature gravity.
\newblock {\em Journal of Cosmology and Astroparticle Physics}, 2023(10):038,
  2023.

\bibitem{Paul:2024rto}
Souvik Paul, Sunandan Gangopadhyay, and Ashis Saha.
\newblock {Gauss{\textendash}Bonnet AdS planar and spherical black hole
  thermodynamics and holography}.
\newblock {\em Classical Quantum and Gravity}, 41(23):235010, 2024.

\bibitem{Nashed2018a}
G.~G.~L. Nashed.
\newblock Spherically symmetric charged black holes in {$f(R)$} gravitational
  theories.
\newblock {\em The European Physical Journal Plus}, 133(1):18, 2018.

\bibitem{Nashed2018c}
G.~G.~L. Nashed.
\newblock Higher dimensional charged black hole solutions in {$f(R)$}
  gravitational theories.
\newblock {\em Advances in High Energy Physics}, 2018:7323574, 2018.

\bibitem{Multamaki2006}
T~Multam{\"a}ki and Iiro Vilja.
\newblock Spherically symmetric solutions of modified field equations in
  {$f(R)$} theories of gravity.
\newblock {\em Physical Review D}, 74(6):064022, 2006.

\bibitem{Nashed2018b}
GGL Nashed.
\newblock Rotating charged black hole spacetimes in quadratic {$f(R)$}
  gravitational theories.
\newblock {\em International Journal of Modern Physics D}, 27(07):1850074,
  2018.

\bibitem{Capozziello2012}
S~Basilakos, Salvatore Capozziello, M~De~Laurentis, A~Paliathanasis, and
  M~Tsamparlis.
\newblock Noether symmetries and analytical solutions in {f(T)} cosmology: A
  complete study.
\newblock {\em Physical Review D}, 88(10):103526, 2013.

\bibitem{Capozziello2008}
Salvatore Capozziello and Mauro Francaviglia.
\newblock Extended theories of gravity and their cosmological and astrophysical
  applications.
\newblock {\em General Relativity and Gravitation}, 40(2):357--420, 2008.

\bibitem{Nashed2019b}
Gamal~GL Nashed and Salvatore Capozziello.
\newblock Charged spherically symmetric black holes in {$f(R)$} gravity and
  their stability analysis.
\newblock {\em Physical Review D}, 99(10):104018, 2019.

\bibitem{delaCruzDombriz2009}
A.~de~la Cruz-Dombriz, A.~Dobado, and A.~L. Maroto.
\newblock Black holes in {$f(R)$} theories.
\newblock {\em Physical Review D}, 80(12):124011, 2009.

\bibitem{Sultana2018}
Joseph Sultana and Demosthenes Kazanas.
\newblock A no-hair theorem for spherically symmetric black holes in {$R^2$}
  gravity.
\newblock {\em General Relativity and Gravitation}, 50(11):137, 2018.

\bibitem{Canate2018a}
Pedro Ca{\~n}ate.
\newblock A no-hair theorem for black holes in {$f(R)$} gravity.
\newblock {\em Classical and Quantum Gravity}, 35(2):025018, 2017.

\bibitem{Yu2018}
Shuang Yu, Chang-Jun Gao, and Ming-Jun Liu.
\newblock On static and spherically symmetric solutions of starobinsky model.
\newblock {\em Research in Astronomy and Astrophysics}, 18(12):157, 2018.

\bibitem{Kehagias2015}
Alex Kehagias, Costas Kounnas, Dieter L{\"u}st, and Antonio Riotto.
\newblock Black hole solutions in {$R^2$} gravity.
\newblock {\em Journal of High Energy Physics}, 2015(5):1--20, 2015.

\bibitem{Nelson2010}
William Nelson.
\newblock Static solutions for fourth order gravity.
\newblock {\em Physical Review D}, 82(10):104026, 2010.

\bibitem{Canate2016}
Pedro Ca{\~n}ate, Luisa~G Jaime, and Marcelo Salgado.
\newblock Spherically symmetric black holes in {$f(R)$} gravity: is geometric
  scalar hair supported?
\newblock {\em Classical and Quantum Gravity}, 33(15):155005, 2016.

\bibitem{AparicioResco2016}
Miguel~Aparicio Resco, {\'A}lvaro de~la Cruz-Dombriz, Felipe J~Llanes Estrada,
  and V{\'\i}ctor~Zapatero Castrillo.
\newblock On neutron stars in {$f(R)$} theories: Small radii, large masses and
  large energy emitted in a merger.
\newblock {\em Physics of the dark universe}, 13:147--161, 2016.

\bibitem{Staykov2018}
Kalin~V Staykov, Dimitar Popchev, Daniela~D Doneva, and Stoytcho~S Yazadjiev.
\newblock Static and slowly rotating neutron stars in scalar--tensor theory
  with self-interacting massive scalar field.
\newblock {\em The European Physical Journal C}, 78(7):586, 2018.

\bibitem{Doneva2016}
Daniela~D Doneva and Stoytcho~S Yazadjiev.
\newblock Rapidly rotating neutron stars with a massive scalar
  field—structure and universal relations.
\newblock {\em Journal of Cosmology and Astroparticle Physics}, 2016(11):019,
  2016.

\bibitem{Arapoglu2011}
Sava{\c{s}} Arapo{\u{g}}lu, Cemsinan Deliduman, and K~Yavuz Ek{\c{s}}i.
\newblock Constraints on perturbative {$f(R)$} gravity via neutron stars.
\newblock {\em Journal of Cosmology and Astroparticle Physics}, 2011(07):020,
  2011.

\bibitem{Feng2017}
Wei-Xiang Feng, Chao-Qiang Geng, Win-Fun Kao, and Ling-Wei Luo.
\newblock Equation-of-state of neutron stars with junction conditions in the
  starobinsky model.
\newblock {\em International Journal of Modern Physics D}, 27(01):1750186,
  2018.

\bibitem{Ganguly2014}
Apratim Ganguly, Radouane Gannouji, Rituparno Goswami, and Subharthi Ray.
\newblock Neutron stars in the starobinsky model.
\newblock {\em Physical Review D}, 89(6):064019, 2014.

\bibitem{Yazadjiev2016}
Stoytcho~S Yazadjiev, Daniela~D Doneva, and Dimitar Popchev.
\newblock Slowly rotating neutron stars in scalar-tensor theories with a
  massive scalar field.
\newblock {\em Physical Review D}, 93(8):084038, 2016.

\bibitem{Yazadjiev2015}
Stoytcho~S Yazadjiev, Daniela~D Doneva, and Kostas~D Kokkotas.
\newblock Rapidly rotating neutron stars in r-squared gravity.
\newblock {\em Physical Review D}, 91(8):084018, 2015.

\bibitem{Yazadjiev2014}
Stoytcho~S Yazadjiev, Daniela~D Doneva, Kostas~D Kokkotas, and Kalin~V Staykov.
\newblock Non-perturbative and self-consistent models of neutron stars in
  r-squared gravity.
\newblock {\em Journal of Cosmology and Astroparticle Physics}, 2014(06):003,
  2014.

\bibitem{astashenok2013further}
Artyom~V Astashenok, Salvatore Capozziello, and Sergei~D Odintsov.
\newblock Further stable neutron star models from $f({R})$ gravity.
\newblock {\em Journal of Cosmology and Astroparticle Physics}, 2013(12):040,
  2013.

\bibitem{Orellana2013}
Mariana Orellana, Federico Garc{\'\i}a, Florencia~A Teppa~Pannia, and Gustavo~E
  Romero.
\newblock Structure of neutron stars in-squared gravity.
\newblock {\em General Relativity and Gravitation}, 45(4):771--783, 2013.

\bibitem{Capozziello2016}
Salvatore Capozziello, Mariafelicia De~Laurentis, Ruben Farinelli, and Sergei~D
  Odintsov.
\newblock Mass-radius relation for neutron stars in {$f(R)$} gravity.
\newblock {\em Physical Review D}, 93(2):023501, 2016.

\bibitem{Cooney2010}
Alan Cooney, Simon DeDeo, and Dimitrios Psaltis.
\newblock Neutron stars in {$f(R)$} gravity with perturbative constraints.
\newblock {\em Physical Review D—Particles, Fields, Gravitation, and
  Cosmology}, 82(6):064033, 2010.

\bibitem{OHanlon1972}
John O'Hanlon.
\newblock Intermediate-range gravity: a generally covariant model.
\newblock {\em Physical Review Letters}, 29(2):137, 1972.

\bibitem{Chiba2003}
Takeshi Chiba.
\newblock $1/{R}$ gravity and scalar-tensor gravity.
\newblock {\em Physics Letters B}, 575(1):1--3, 2003.

\bibitem{Chakraborty2017a}
Sumanta Chakraborty and Soumitra SenGupta.
\newblock Gravity stabilizes itself.
\newblock {\em The European Physical Journal C}, 77(3):166, 2017.

\bibitem{Brans1961}
Carl Brans and Robert~H. Dicke.
\newblock Mach's principle and a relativistic theory of gravitation.
\newblock {\em Physical Review}, 124(3):925, 1961.

\bibitem{Chakraborty2016}
Sumanta Chakraborty and Soumitra SenGupta.
\newblock Solving higher curvature gravity theories.
\newblock {\em The European Physical Journal C}, 76(10):552, 2016.

\bibitem{Nashed:2025ebr}
G.~G.~L. Nashed, Usman Zafar, and Kazuharu Bamba.
\newblock {An innovative black hole solution and thermodynamic properties in
  higher-order curvature gravity with a scalar field}.
\newblock {\em Physics of the Dark Universe}, 50:102061, 2025.

\bibitem{Kanti1996}
Panagiota Kanti, Nick~E. Mavromatos, John Rizos, Kyriakos Tamvakis, and
  Elizabeth Winstanley.
\newblock Dilatonic black holes in higher curvature string gravity.
\newblock {\em Physical Review D}, 54(8):5049--5058, 1996.

\bibitem{Nojiri2005}
Shin'ichi Nojiri and Sergei~D. Odintsov.
\newblock Modified gauss-bonnet theory as gravitational alternative for dark
  energy.
\newblock {\em Physics Letters B}, 631(1):1--6, 2005.

\bibitem{LIGOScientific:2016aoc}
et~al Abbott, Benjamin~P.
\newblock Observation of gravitational waves from a binary black hole merger.
\newblock {\em Physical review letters}, 116(6):061102, 2016.

\bibitem{019}
O~Contigiani.
\newblock Lensing efficiency for gravitational wave mergers.
\newblock {\em Monthly Notices of the Royal Astronomical Society},
  492(3):3359--3363, 2020.

\bibitem{020}
Suvodip Mukherjee, Benjamin~D Wandelt, and Joseph Silk.
\newblock Probing the theory of gravity with gravitational lensing of
  gravitational waves and galaxy surveys.
\newblock {\em Monthly Notices of the Royal Astronomical Society},
  494(2):1956--1970, 2020.

\bibitem{021}
Charles~Galton Darwin.
\newblock The gravity field of a particle.
\newblock {\em Proceedings of the Royal Society of London. Series A.
  Mathematical and Physical Sciences}, 249(1257):180--194, 1959.

\bibitem{022}
Robert~d'Escourt Atkinson.
\newblock On light tracks near a very massive star.
\newblock {\em Astronomical Journal, Vol. 70, p. 517}, 70:517, 1965.

\bibitem{mohan2025strong}
Gayatri Mohan, Ronit Karmakar, Rupam~Jyoti Borah, and Umananda~Dev Goswami.
\newblock Strong lensing effect and quasinormal modes of oscillations of black
  holes in {$f(R,T)$} gravity theory.
\newblock {\em arXiv preprint arXiv:2503.08402}, 2025.

\bibitem{Cunningham}
Christopher~T. Cunningham and James~M. Bardeen.
\newblock {The Optical Appearance of a Star Orbiting an Extreme Kerr Black
  Hole}.
\newblock {\em The Astrophysical Journal}, 183:237--264, 1973.

\bibitem{Falcke:1999pj}
Heino Falcke, Fulvio Melia, and Eric Agol.
\newblock Viewing the shadow of the black hole at the galacticcenter.
\newblock {\em The Astrophysical Journal}, 528(1):L13, 1999.

\bibitem{Khodadi:2024ubi}
Mohsen Khodadi, Sunny Vagnozzi, and Javad~T. Firouzjaee.
\newblock {Event Horizon Telescope observations exclude compact objects in
  baseline mimetic gravity}.
\newblock {\em Scientific Reports}, 14(1):26932, 2024.

\bibitem{Allahyari:2019jqz}
Alireza Allahyari, Mohsen Khodadi, Sunny Vagnozzi, and David~F. Mota.
\newblock {Magnetically charged black holes from non-linear electrodynamics and
  the Event Horizon Telescope}.
\newblock {\em Journal of Cosmology and Astroparticle Physics,}, 02:003, 2020.

\bibitem{Nojiri:2024txy}
Shin'ichi Nojiri and Sergei~D. Odintsov.
\newblock {Improving mimetic gravity with non-trivial scalar potential:
  Cosmology, black holes, shadow and photon sphere}.
\newblock {\em Physics of the Dark Universe}, 46:101669, 2024.

\bibitem{Afrin:2021imp}
Misba Afrin, Rahul Kumar, and Sushant~G Ghosh.
\newblock Parameter estimation of hairy kerr black holes from its shadow and
  constraints from {M87$^*$}.
\newblock {\em Monthly Notices of the Royal Astronomical Society},
  504(4):5927--5940, 2021.

\bibitem{Khodadi:2021gbc}
Mohsen Khodadi, Gaetano Lambiase, and David~F. Mota.
\newblock {No-hair theorem in the wake of Event Horizon Telescope}.
\newblock {\em Journal of Cosmology and Astroparticle Physics,}, 09:028, 2021.

\bibitem{Afrin:2021wlj}
Misba Afrin and Sushant~G Ghosh.
\newblock Testing horndeski gravity from eht observational results for rotating
  black holes.
\newblock {\em The Astrophysical Journal}, 932(1):51, 2022.

\bibitem{Khodadi:2022pqh}
Mohsen Khodadi and Gaetano Lambiase.
\newblock {Probing Lorentz symmetry violation using the first image of
  Sagittarius A*: Constraints on standard-model extension coefficients}.
\newblock {\em Physical Review D}, 106(10):104050, 2022.

\bibitem{Kumar:2020hgm}
Rahul Kumar, Sushant~G. Ghosh, and Anzhong Wang.
\newblock {Gravitational deflection of light and shadow cast by rotating
  Kalb-Ramond black holes}.
\newblock {\em Physical Review D}, 101(10):104001, 2020.

\bibitem{Afrin:2022ztr}
Misba Afrin, Sunny Vagnozzi, and Sushant~G Ghosh.
\newblock Tests of loop quantum gravity from the event horizon telescope
  results of {Sgr A$^*$}.
\newblock {\em The Astrophysical Journal}, 944(2):149, 2023.

\bibitem{Afrin:2023uzo}
Misba Afrin and Sushant~G Ghosh.
\newblock Eht observables as a tool to estimate parameters of supermassive
  black holes.
\newblock {\em Monthly Notices of the Royal Astronomical Society},
  524(3):3683--3691, 2023.

\bibitem{Fu:2021fxn}
Qi-Ming Fu and Xin Zhang.
\newblock Gravitational lensing by a black hole in effective loop quantum
  gravity.
\newblock {\em Physical Review D}, 105(6):064020, 2022.

\bibitem{Ghosh:2022kit}
Sushant~G Ghosh and Misba Afrin.
\newblock An upper limit on the charge of the black hole {Sgr A$^*$} from eht
  observations.
\newblock {\em The Astrophysical Journal}, 944(2):174, 2023.

\bibitem{Vagnozzi:2019apd}
Sunny Vagnozzi and Luca Visinelli.
\newblock {Hunting for extra dimensions in the shadow of {M87$^*$}}.
\newblock {\em Physical Review D}, 100(2):024020, 2019.

\bibitem{Liu:2024lve}
Wentao Liu, Di~Wu, and Jieci Wang.
\newblock {Shadow of slowly rotating Kalb-Ramond black holes}.
\newblock {\em Journal of Cosmology and Astroparticle Physics,}, 05:017, 2025.

\bibitem{Nojiri:2024qgx}
Shin'ichi Nojiri and S.~D. Odintsov.
\newblock {Black holes and their shadows in {$F(R)$} gravity}.
\newblock {\em Physics of the Dark Universe}, 47:101785, 2025.

\bibitem{Nojiri:2024nlx}
Shin'ichi Nojiri and S.~D. Odintsov.
\newblock {Black holes, photon sphere, and cosmology in ghost-free {$f(G)$}
  gravity}.
\newblock {\em Physics of the Dark Universe}, 46:101702, 2024.

\bibitem{Bambi:2019tjh}
Cosimo Bambi, Katherine Freese, Sunny Vagnozzi, and Luca Visinelli.
\newblock {Testing the rotational nature of the supermassive object {M87$^*$}
  from the circularity and size of its first image}.
\newblock {\em Physical Review D}, 100(4):044057, 2019.

\bibitem{Afrin:2024khy}
Misba Afrin, Sushant~G. Ghosh, and Anzhong Wang.
\newblock {Testing EGB gravity coupled to bumblebee field and black hole
  parameter estimation with EHT observations}.
\newblock {\em Physics of the Dark Universe}, 46:101642, 2024.

\bibitem{virbhadra2000schwarzschild}
Kumar~Shwetketu Virbhadra and George~FR Ellis.
\newblock Schwarzschild black hole lensing.
\newblock {\em Physical Review D}, 62(8):084003, 2000.

\bibitem{031}
Volker Perlick.
\newblock Theoretical gravitational lensing--beyond the weak-field small-angle
  approximation.
\newblock In {\em The Eleventh Marcel Grossmann Meeting: On Recent Developments
  in Theoretical and Experimental General Relativity, Gravitation and
  Relativistic Field Theories (In 3 Volumes)}, pages 680--699. World
  Scientific, 2008.

\bibitem{033}
Valerio Bozza, Salvatore Capozziello, Gerardo Iovane, and Gaetano Scarpetta.
\newblock Strong field limit of black hole gravitational lensing.
\newblock {\em General Relativity and Gravitation}, 33:1535--1548, 2001.

\bibitem{034}
Valerio Bozza.
\newblock Gravitational lensing in the strong field limit.
\newblock {\em Physical Review D}, 66(10):103001, 2002.

\bibitem{032}
Simonetta Frittelli, Thomas~P Kling, and Ezra~T Newman.
\newblock Spacetime perspective of schwarzschild lensing.
\newblock {\em Physical Review D}, 61(6):064021, 2000.

\bibitem{metcalf2019strong}
R~Benton Metcalf, MASSIMO Meneghetti, Camille Avestruz, Fabio Bellagamba,
  Cl{\'e}cio~R Bom, Emmanuel Bertin, R{\'e}mi Cabanac, F~Courbin, Andrew
  Davies, Etienne Decenci{\`e}re, et~al.
\newblock The strong gravitational lens finding challenge.
\newblock {\em Astronomy \& Astrophysics}, 625:A119, 2019.

\bibitem{grespan2023strong}
Margherita Grespan and Marek Biesiada.
\newblock Strong gravitational lensing of gravitational waves: A review.
\newblock {\em Universe}, 9(5):200, 2023.

\bibitem{Kuang:2022xjp}
Xiao-Mei Kuang and Ali {\"O}vg{\"u}n.
\newblock Strong gravitational lensing and shadow constraint from {M87$^*$} of
  slowly rotating kerr-like black hole.
\newblock {\em Annals of Physics}, 447:169147, 2022.

\bibitem{Pantig:2022ely}
Reggie~C Pantig and Ali {\"O}vg{\"u}n.
\newblock Testing dynamical torsion effects on the charged black hole’s
  shadow, deflection angle and greybody with {M87$^*$} and {Sgr A$^*$} from
  eht.
\newblock {\em Annals of Physics}, 448:169197, 2023.

\bibitem{Ovgun:2018fnk}
Ali \"Ovg\"un.
\newblock {Light deflection by Damour-Solodukhin wormholes and Gauss-Bonnet
  theorem}.
\newblock {\em Physical Review D}, 98(4):044033, 2018.

\bibitem{chakraborty2017strong}
Sumanta Chakraborty and Soumitra SenGupta.
\newblock Strong gravitational lensing—a probe for extra dimensions and
  kalb-ramond field.
\newblock {\em Journal of Cosmology and Astroparticle Physics}, 2017(07):045,
  2017.

\bibitem{araujo2025antisymmetric}
A.~A Ara{\'u}jo~Filho.
\newblock Antisymmetric tensor influence on charged black hole lensing
  phenomena and time delay.
\newblock {\em Journal of High Energy Astrophysics}, page 100401, 2025.

\bibitem{40}
Rajibul Shaikh and Sayan Kar.
\newblock Gravitational lensing by scalar-tensor wormholes and the energy
  conditions.
\newblock {\em Physical Review D}, 96(4):044037, 2017.

\bibitem{nascimento2024gravitational}
A.~A Araújo~Filho, J.~R Nascimento, A.~Yu Petrov, and P.~J Porf{\'\i}rio.
\newblock Gravitational lensing by a lorentz-violating black hole.
\newblock {\em arXiv preprint arXiv:2404.04176}, 2024.

\bibitem{heidari2023gravitational}
N~Heidari, H~Hassanabadi, A.~A Ara{\'u}jo~Filho, J~Kriz, S~Zare, and
  PJ~Porf{\'\i}rio.
\newblock Gravitational signatures of a non--commutative stable black hole.
\newblock {\em Physics of the Dark Universe}, page 101382, 2023.

\bibitem{ovgun2019exact}
Ali {\"O}vg{\"u}n, Kimet Jusufi, and {\.I}zzet Sakall{\i}.
\newblock Exact traversable wormhole solution in bumblebee gravity.
\newblock {\em Physical Review D}, 99(2):024042, 2019.

\bibitem{38.1}
Naoki Tsukamoto, Tomohiro Harada, and Kohji Yajima.
\newblock Can we distinguish between black holes and wormholes by their
  einstein-ring systems?
\newblock {\em Physical Review D}, 86(10):104062, 2012.

\bibitem{38.5}
Rajibul Shaikh, Pritam Banerjee, Suvankar Paul, and Tapobrata Sarkar.
\newblock Strong gravitational lensing by wormholes.
\newblock {\em Journal of Cosmology and Astroparticle Physics,}, 2019(07):028,
  2019.

\bibitem{38.2}
Gary~W Gibbons and Martin Vyska.
\newblock The application of weierstrass elliptic functions to schwarzschild
  null geodesics.
\newblock {\em Classical and Quantum Gravity}, 29(6):065016, 2012.

\bibitem{38.4}
Naoki Tsukamoto.
\newblock Retrolensing by a wormhole at deflection angles $\pi$ and 3 $\pi$.
\newblock {\em Physical Review D}, 95(8):084021, 2017.

\bibitem{Lobo:2020jfl}
Iarley~P. Lobo, Martin~G. Richarte, J.~P. Morais~Gra\c{c}a, and H.~Moradpour.
\newblock {Thin-shell wormholes in Rastall gravity}.
\newblock {\em Eur. Phys. J. Plus}, 135(7):550, 2020.

\bibitem{38.3}
Naoki Tsukamoto.
\newblock Strong deflection limit analysis and gravitational lensing of an
  ellis wormhole.
\newblock {\em Physical Review D}, 94(12):124001, 2016.

\bibitem{virbhadra2024conservation}
KS~Virbhadra.
\newblock Conservation of distortion of gravitationally lensed images.
\newblock {\em Physical Review D}, 109(12):124004, 2024.

\bibitem{Konoplya:2007zx}
R.~A. Konoplya and A.~Zhidenko.
\newblock {Decay of a charged scalar and Dirac fields in the Kerr-Newman-de
  Sitter background}.
\newblock {\em Physical Review D}, 76(8):084018, 2007.
\newblock [Erratum: Phys.Rev.D 90, 029901 (2014)].

\bibitem{Konoplya:2013rxa}
R.~A. Konoplya and A.~Zhidenko.
\newblock {Massive charged scalar field in the Kerr-Newman background I:
  quasinormal modes, late-time tails and stability}.
\newblock {\em Physical Review D}, 88:024054, 2013.

\bibitem{karmakar2024quasinormal}
Ronit Karmakar and Umananda~Dev Goswami.
\newblock {Quasinormal modes, thermodynamics and shadow of black holes in
  Hu--Sawicki {$f(R)$} gravity theory}.
\newblock {\em The European Physical Journal C}, 84(9):969, 2024.

\bibitem{Konoplya:2019hlu}
RA~Konoplya, A~Zhidenko, and AF~Zinhailo.
\newblock Higher order wkb formula for quasinormal modes and grey-body factors:
  recipes for quick and accurate calculations.
\newblock {\em Classical and Quantum Gravity}, 36(15):155002, 2019.

\bibitem{karmakar2022quasinormal}
Ronit Karmakar, Dhruba~Jyoti Gogoi, and Umananda~Dev Goswami.
\newblock Quasinormal modes and thermodynamic properties of gup-corrected
  schwarzschild black hole surrounded by quintessence.
\newblock {\em International Journal of Modern Physics A}, 37(28n29):2250180,
  2022.

\bibitem{Konoplya:2011qq}
Roman~A Konoplya and Alexander Zhidenko.
\newblock Quasinormal modes of black holes: From astrophysics to string theory.
\newblock {\em Reviews of Modern Physics}, 83(3):793--836, 2011.

\bibitem{Kokkotas:2010zd}
K.~D. Kokkotas, R.~A. Konoplya, and A.~Zhidenko.
\newblock {Quasinormal modes, scattering and Hawking radiation of Kerr-Newman
  black holes in a magnetic field}.
\newblock {\em Physical Review D}, 83:024031, 2011.

\bibitem{Jusufi:2020dhz}
Kimet Jusufi.
\newblock {Connection Between the Shadow Radius and Quasinormal Modes in
  Rotating Spacetimes}.
\newblock {\em Physical Review D}, 101(12):124063, 2020.

\bibitem{Konoplya:2024vuj}
RA~Konoplya and A~Zhidenko.
\newblock Correspondence between grey-body factors and quasinormal frequencies
  for rotating black holes.
\newblock {\em Physics Letters B}, 861:139288, 2025.

\bibitem{Konoplya:2024lir}
R.~A. Konoplya and A.~Zhidenko.
\newblock {Correspondence between grey-body factors and quasinormal modes}.
\newblock {\em Journal of Cosmology and Astroparticle Physics,}, 09:068, 2024.

\bibitem{Franchini:2023eda}
Nicola Franchini and Sebastian~H V{\"o}lkel.
\newblock {\em Testing general relativity with black hole quasi-normal modes},
  pages 361--416.
\newblock Springer, 2024.

\bibitem{Nojiri:2024hau}
Shin'ichi Nojiri and Sergei~D. Odintsov.
\newblock {F(Q) Gravity with Gauss{\textendash}Bonnet Corrections: From
  Early-Time Inflation to Late-Time Acceleration}.
\newblock {\em Fortsch. Phys.}, 72(9-10):2400113, 2024.

\bibitem{Nojiri:2023qgd}
Shin'ichi Nojiri and G.~G.~L. Nashed.
\newblock {Hayward black hole in scalar-Einstein-Gauss-Bonnet gravity in four
  dimensions}.
\newblock {\em Phys. Rev. D}, 108(2):024014, 2023.

\bibitem{Nashed:2022mij}
G.~G.~L. Nashed and Shin'ichi Nojiri.
\newblock {Multihorizons black hole solutions, photon sphere, and perihelion
  shift in weak ghost-free Gauss-Bonnet theory of gravity}.
\newblock {\em Phys. Rev. D}, 106(4):044024, 2022.

\bibitem{Lewandowski:2022zce}
Jerzy Lewandowski, Yongge Ma, Jinsong Yang, and Cong Zhang.
\newblock Quantum oppenheimer-snyder and swiss cheese models.
\newblock {\em Physical Review Letters}, 130(10):101501, 2023.

\bibitem{Nozari:2025veb}
Kourosh Nozari, Sara Saghafi, Milad Hajebrahimi, and Kimet Jusufi.
\newblock {Circular orbits and accretion disk around a deformed-Schwarzschild
  black hole in loop quantum gravity}.
\newblock {\em Physics of the Dark Universe}, 49:102027, 2025.

\bibitem{iyer1987black}
Sai Iyer and Clifford~M Will.
\newblock Black-hole normal modes: A wkb approach. i. foundations and
  application of a higher-order wkb analysis of potential-barrier scattering.
\newblock {\em Physical Review D}, 35(12):3621, 1987.

\bibitem{iyer1987black1}
Sai Iyer.
\newblock Black-hole normal modes: A wkb approach. ii. schwarzschild black
  holes.
\newblock {\em Physical Review D}, 35(12):3632, 1987.

\bibitem{konoplya2003quasinormal}
RA~Konoplya.
\newblock Quasinormal behavior of the d-dimensional schwarzschild black hole
  and the higher order wkb approach.
\newblock {\em Physical Review D}, 68(2):024018, 2003.

\bibitem{chandrasekhar1998mathematical}
Subrahmanyan Chandrasekhar.
\newblock {\em The mathematical theory of black holes}, volume~69.
\newblock Oxford university press, 1998.

\bibitem{Bouhmadi-Lopez:2020oia}
Mariam Bouhmadi-L\'opez, Suddhasattwa Brahma, Che-Yu Chen, Pisin Chen, and
  Dong-han Yeom.
\newblock {A consistent model of non-singular Schwarzschild black hole in loop
  quantum gravity and its quasinormal modes}.
\newblock {\em Journal of Cosmology and Astroparticle Physics,}, 07:066, 2020.

\bibitem{Gogoi:2023kjt}
Dhruba~Jyoti Gogoi, Ali \"Ovg\"un, and M.~Koussour.
\newblock {Quasinormal modes of black holes in f(Q) gravity}.
\newblock {\em The European Physical Journal C}, 83(8):700, 2023.

\bibitem{baruah2025quasinormal}
Anshuman Baruah, Yassine Sekhmani, Sunil~Kumar Maurya, Atri Deshamukhya, and
  Mahmood~Khalid Jasim.
\newblock {Quasinormal modes, greybody factors, and Hawking radiation sparsity
  of black holes influenced by a global monopole charge in Kalb-Ramond
  gravity}.
\newblock {\em Journal of Cosmology and Astroparticle Physics,}, 08:023, 2025.

\bibitem{g1}
Loyal Durand, Paul~M Fishbane, and LM~Simmons~Jr.
\newblock Expansion formulas and addition theorems for gegenbauer functions.
\newblock {\em Journal of Mathematical Physics}, 17(11):1933--1948, 1976.

\bibitem{g2}
Ala Amourah, A~Alamoush, and Mohammad Al-Kaseasbeh.
\newblock Gegenbauer polynomials and bi-univalent functions.
\newblock {\em Palestine Journal of Mathematics}, 10(2):625--632, 2021.

\bibitem{g3}
Georg Loh{\"o}fer.
\newblock Inequalities for legendre functions and gegenbauer functions.
\newblock {\em Journal of approximation theory}, 64(2):226--234, 1991.

\bibitem{g5}
Howard~S Cohl.
\newblock On a generalization of the generating function for gegenbauer
  polynomials.
\newblock {\em Integral Transforms and special functions}, 24(10):807--816,
  2013.

\bibitem{g6}
Wenjie Liu and Li-Lian Wang.
\newblock Asymptotics of the generalized gegenbauer functions of fractional
  degree.
\newblock {\em Journal of Approximation Theory}, 253:105378, 2020.

\bibitem{ashtekar2018quantum2}
Abhay Ashtekar, Javier Olmedo, and Parampreet Singh.
\newblock Quantum extension of the kruskal spacetime.
\newblock {\em Physical Review D}, 98(12):126003, 2018.

\bibitem{ashtekar2018quantum}
Abhay Ashtekar, Javier Olmedo, and Parampreet Singh.
\newblock Quantum transfiguration of kruskal black holes.
\newblock {\em Physical review letters}, 121(24):241301, 2018.

\bibitem{asasas2}
A.~A Ara{\'u}jo~Filho.
\newblock Analysis of a nonlinear electromagnetic generalization of the
  reissner--nordstr{\"o}m black hole.
\newblock {\em The European Physical Journal C}, 85(4):454, 2025.

\bibitem{chen2019gravitational}
Che-Yu Chen and Pisin Chen.
\newblock Gravitational perturbations of nonsingular black holes in conformal
  gravity.
\newblock {\em Physical Review D}, 99(10):104003, 2019.

\bibitem{araujo2025does}
A.~A Ara{\'u}jo~Filho.
\newblock How does non-metricity affect particle creation and evaporation in
  bumblebee gravity?
\newblock {\em Journal of Cosmology and Astroparticle Physics}, 2025(06):026,
  2025.

\bibitem{newman1962approach}
Ezra Newman and Roger Penrose.
\newblock An approach to gravitational radiation by a method of spin
  coefficients.
\newblock {\em Journal of Mathematical Physics}, 3(3):566--578, 1962.

\bibitem{chandrasekhar1984mathematical}
Subrahmanijan Chandrasekhar.
\newblock The mathematical theory of black holes.
\newblock In {\em General Relativity and Gravitation: Invited Papers and
  Discussion Reports of the 10th International Conference on General Relativity
  and Gravitation, Padua, July 3--8, 1983}, pages 5--26. Springer, 1984.

\bibitem{albuquerque2023massless}
Saulo Albuquerque, Iarley~P Lobo, and Valdir~B Bezerra.
\newblock Massless dirac perturbations in a consistent model of loop quantum
  gravity black hole: quasinormal modes and particle emission rates.
\newblock {\em Classical and Quantum Gravity}, 40(17):174001, 2023.

\bibitem{al2024massless}
Ahmad Al-Badawi and Sohan~Kumar Jha.
\newblock Massless dirac perturbations of black holes in f (q) gravity:
  quasinormal modes and a weak deflection angle.
\newblock {\em Communications in Theoretical Physics}, 76(9):095403, 2024.

\bibitem{arbey2021hawking}
Alexandre Arbey, J{\'e}r{\'e}my Auffinger, Marc Geiller, Etera~R Livine, and
  Francesco Sartini.
\newblock Hawking radiation by spherically-symmetric static black holes for all
  spins: Teukolsky equations and potentials.
\newblock {\em Physical Review D}, 103(10):104010, 2021.

\bibitem{devi2020quasinormal}
Saraswati Devi, Rittick Roy, and Sayan Chakrabarti.
\newblock Quasinormal modes and greybody factors of the novel four dimensional
  gauss--bonnet black holes in asymptotically de sitter space time: scalar,
  electromagnetic and dirac perturbations.
\newblock {\em The European Physical Journal C}, 80(8):760, 2020.

\bibitem{Gundlach:1993tp}
Carsten Gundlach, Richard~H. Price, and Jorge Pullin.
\newblock {Late time behavior of stellar collapse and explosions: 1. Linearized
  perturbations}.
\newblock {\em Physical Review D}, 49:883--889, 1994.

\bibitem{Skvortsova:2024wly}
Milena Skvortsova.
\newblock Ringing of extreme regular black holes.
\newblock {\em Gravitation and Cosmology}, 30(3):279--288, 2024.

\bibitem{Bolokhov:2024ixe}
S.~V. Bolokhov.
\newblock {Late time decay of scalar and Dirac fields around an asymptotically
  de Sitter black hole in the Euler\textendash{}Heisenberg electrodynamics}.
\newblock {\em The European Physical Journal C}, 84(6):634, 2024.

\bibitem{Guo:2023nkd}
Wen-Di Guo, Qin Tan, and Yu-Xiao Liu.
\newblock {Quasinormal modes and greybody factor of a Lorentz-violating black
  hole}.
\newblock {\em Journal of Cosmology and Astroparticle Physics,}, 07:008, 2024.

\bibitem{Yang:2024rms}
Zhen-Hao Yang, Cheng Xu, Xiao-Mei Kuang, Bin Wang, and Rui-Hong Yue.
\newblock Echoes of massless scalar field induced from hairy schwarzschild
  black hole.
\newblock {\em Physics Letters B}, 853:138688, 2024.

\bibitem{Baruah:2023rhd}
Anshuman Baruah, Ali \"Ovg\"un, and Atri Deshamukhya.
\newblock {Quasinormal modes and bounding greybody factors of GUP-corrected
  black holes in Kalb\textendash{}Ramond gravity}.
\newblock {\em Annals of Physics}, 455:169393, 2023.

\bibitem{Shao:2023qlt}
Cai-Ying Shao, Cong Zhang, Wei Zhang, and Cheng-Gang Shao.
\newblock {Scalar fields around a loop quantum gravity black hole in de Sitter
  spacetime: Quasinormal modes, late-time tails and strong cosmic censorship}.
\newblock {\em Physical Review D}, 109(6):064012, 2024.

\bibitem{hamil2023thermodynamics}
Bilel Hamil and BC~L{\"u}tf{\"u}o{\u{g}}lu.
\newblock Thermodynamics and shadows of quantum-corrected
  reissner--nordstr{\"o}m black hole surrounded by quintessence.
\newblock {\em Physics of the Dark Universe}, 42:101293, 2023.

\bibitem{araujo2024charged}
Adailton~Azevedo Ara{\'u}jo~Filho, Kimet Jusufi, Bertha Cuadros-Melgar, Genly
  Leon, Abdul Jawad, and CE~Pellicer.
\newblock Charged black holes with yukawa potential.
\newblock {\em Physics of the Dark Universe}, 46:101711, 2024.

\bibitem{zeng2022shadows}
Xiao-Xiong Zeng, Guo-Ping Li, and Ke-Jian He.
\newblock The shadows and observational appearance of a noncommutative black
  hole surrounded by various profiles of accretions.
\newblock {\em Nuclear Physics B}, 974:115639, 2022.

\bibitem{araujo2025remarks}
A.~A Ara{\'u}jo~Filho.
\newblock Remarks on a nonlinear electromagnetic extension in ads
  reissner-nordstr{\"o}m spacetime.
\newblock {\em Journal of Cosmology and Astroparticle Physics}, 2025(01):072,
  2025.

\bibitem{ball2019first}
David Ball, Chi-kwan Chan, Pierre Christian, Buell~T Jannuzi, Junhan Kim,
  Daniel~P Marrone, Lia Medeiros, Feryal Ozel, Dimitrios Psaltis, Mel Rose,
  et~al.
\newblock First {M87$^*$} event horizon telescope results. i. the shadow of the
  supermassive black hole.
\newblock {\em The Astrophysical Journal Letters}, 2019.

\bibitem{gralla2021can}
Samuel~E Gralla.
\newblock Can the eht {M87$^*$} results be used to test general relativity.
\newblock {\em Physical Review D}, 103(2):024023, 2021.

\bibitem{akiyama2019first}
Kazunori Akiyama, Antxon Alberdi, Walter Alef, Keiichi Asada, Rebecca Azulay,
  Anne-Kathrin Baczko, David Ball, Mislav Balokovi{\'c}, John Barrett, Dan
  Bintley, et~al.
\newblock First {M87$^*$} event horizon telescope results. v. physical origin
  of the asymmetric ring.
\newblock {\em The Astrophysical Journal Letters}, 875(1):L5, 2019.

\bibitem{Gibbons:2008rj}
G.~W. Gibbons and M.~C. Werner.
\newblock {Applications of the Gauss-Bonnet theorem to gravitational lensing}.
\newblock {\em Classical and Quantum Gravity}, 25:235009, 2008.

\bibitem{qiao2022geometric}
Chen-Kai Qiao and Ming Li.
\newblock Geometric approach to circular photon orbits and black hole shadows.
\newblock {\em Physical Review D}, 106(2):L021501, 2022.

\bibitem{Heidari:2025iiv}
N.~Heidari, A.~A. Ara{\'u}jo~Filho, and Iarley~P. Lobo.
\newblock {Non-commutativity in Hayward spacetime}.
\newblock 3 2025.

\bibitem{qiao2022curvatures}
Chen-Kai Qiao.
\newblock Curvatures, photon spheres, and black hole shadows.
\newblock {\em Physical Review D}, 106(8):084060, 2022.

\bibitem{araujo2025impact}
A.~A Ara{\'u}jo~Filho, N~Heidari, J.~A. A.~S Reis, and H~Hassanabadi.
\newblock The impact of an antisymmetric tensor on charged black holes:
  evaporation process, geodesics, deflection angle, scattering effects and
  quasinormal modes.
\newblock {\em Classical and Quantum Gravity}, 42(6):065026, 2025.

\bibitem{qiao2024existence}
Chen-Kai Qiao.
\newblock The existence and distribution of photon spheres near spherically
  symmetric black holes--a geometric analysis.
\newblock {\em arXiv preprint arXiv:2407.14035}, 2024.

\bibitem{AraujoFilho:2024xhm}
A.~A Ara{\'u}jo~Filho.
\newblock Analysis of a nonlinear electromagnetic generalization of the
  reissner--nordstr{\"o}m black hole.
\newblock {\em The European Physical Journal C}, 85(4):454, 2025.

\bibitem{AraujoFilho:2025huk}
A.~A. Ara{\'u}jo~Filho, N.~Heidari, Iarley~P. Lobo, and Yuxuan Shi.
\newblock {Optical Phenomena in a Non-Commutative Kalb-Ramond Black Hole
  Spacetime}.
\newblock 8 2025.

\bibitem{araujo2024effects}
A.~A Ara{\'u}jo~Filho, J.~R Nascimento, A~Yu Petrov, P.~J Porf{\'\i}rio, and
  Ali {\"O}vg{\"u}n.
\newblock Effects of non-commutative geometry on black hole properties.
\newblock {\em Physics of the Dark Universe}, 46:101630, 2024.

\bibitem{araujo2025gravitadddtional}
AA~Ara{\'u}jo~Filho, N~Heidari, IP~Lobo, and VB~Bezerra.
\newblock Gravitational signatures of a nonlinear electrodynamics in {f(R,T)}
  gravity.
\newblock {\em Journal of Cosmology and Astroparticle Physics}, 2025(09):015,
  2025.

\bibitem{araujo2025geodesics}
A.~A Ara{\'u}jo~Filho, N~Heidari, and Ali {\"O}vg{\"u}n.
\newblock Geodesics, accretion disk, gravitational lensing, time delay, and
  effects on neutrinos induced by a non-commutative black hole.
\newblock {\em Journal of Cosmology and Astroparticle Physics}, 2025(06):062,
  2025.

\bibitem{heidari2024absorption}
N.~Heidari, A.~A. Ara\'ujo~Filho, R.~C. Pantig, and A.~\"Ovg\"un.
\newblock {Absorption, scattering, geodesics, shadows and lensing phenomena of
  black holes in effective quantum gravity}.
\newblock {\em Physics of the Dark Universe}, 47:101815, 2025.

\bibitem{tsukamoto2017deflection}
Naoki Tsukamoto.
\newblock Deflection angle in the strong deflection limit in a general
  asymptotically flat, static, spherically symmetric spacetime.
\newblock {\em Physical Review D}, 95(6):064035, 2017.

\bibitem{hasse2002gravitational}
Wolfgang Hasse and Volker Perlick.
\newblock Gravitational lensing in spherically symmetric static spacetimes with
  centrifugal force reversal.
\newblock {\em Gen. Rel. Grav.}, 34:415--433, 2002.

\bibitem{akiyama2022firstSgr}
Kazunori Akiyama, Antxon Alberdi, Walter Alef, Juan~Carlos Algaba, Richard
  Anantua, Keiichi Asada, Rebecca Azulay, Uwe Bach, Anne-Kathrin Baczko, David
  Ball, et~al.
\newblock First sagittarius a* event horizon telescope results. iv.
  variability, morphology, and black hole mass.
\newblock {\em The Astrophysical Journal Letters}, 930(2):L15, 2022.

\bibitem{akiyama2022firstSgrA}
Kazunori Akiyama, Antxon Alberdi, Walter Alef, Juan~Carlos Algaba, Richard
  Anantua, Keiichi Asada, Rebecca Azulay, Uwe Bach, Anne-Kathrin Baczko, David
  Ball, et~al.
\newblock First sagittarius a* event horizon telescope results. vi. testing the
  black hole metric.
\newblock {\em The Astrophysical Journal Letters}, 930(2):L17, 2022.

\bibitem{perlick2022calculating}
Volker Perlick and Oleg~Yu Tsupko.
\newblock Calculating black hole shadows: Review of analytical studies.
\newblock {\em Physics Reports}, 947:1--39, 2022.

\bibitem{afrin2023tests}
Misba Afrin, Sunny Vagnozzi, and Sushant~G Ghosh.
\newblock Tests of loop quantum gravity from the event horizon telescope
  results of sgr a.
\newblock {\em The Astrophysical Journal}, 944(2):149, 2023.

\bibitem{kumar2020rotating}
Rahul Kumar and Sushant~G Ghosh.
\newblock Rotating black holes in 4d einstein-gauss-bonnet gravity and its
  shadow.
\newblock {\em Journal of Cosmology and Astroparticle Physics}, 2020(07):053,
  2020.

\bibitem{xu2025optical}
Mou Xu, Ruonan Li, Jianbo Lu, Shining Yang, and Shu-Min Wu.
\newblock Optical appearance and shadow of kalb--ramond black hole: effects of
  plasma and accretion models.
\newblock {\em The European Physical Journal C}, 85(6):676, 2025.

\bibitem{heydari2024effect}
Malihe Heydari-Fard.
\newblock Effect of quintessence dark energy on the shadow of hayward black
  holes with spherical accretion.
\newblock {\em Indian Journal of Physics}, 98(8):3019--3032, 2024.

\bibitem{blakeslee2009acs}
John~P Blakeslee, Andr{\'e}s Jord{\'a}n, Simona Mei, Patrick C{\^o}t{\'e},
  Laura Ferrarese, Leopoldo Infante, Eric~W Peng, John~L Tonry, and Michael~J
  West.
\newblock The acs fornax cluster survey. v. measurement and recalibration of
  surface brightness fluctuations and a precise value of the fornax--virgo
  relative distance.
\newblock {\em The Astrophysical Journal}, 694(1):556, 2009.

\bibitem{gebhardt2011black}
Karl Gebhardt, Joshua Adams, Douglas Richstone, Tod~R Lauer, SM~Faber, Kayhan
  G{\"u}ltekin, Jeremy Murphy, and Scott Tremaine.
\newblock The black hole mass in {M87$^*$} from gemini/nifs adaptive optics
  observations.
\newblock {\em The Astrophysical Journal}, 729(2):119, 2011.

\bibitem{bird2010inner}
Sarah Bird, William~E Harris, John~P Blakeslee, and Chris Flynn.
\newblock The inner halo of {M87$^*$}: a first direct view of the red-giant
  population.
\newblock {\em Astronomy \& Astrophysics}, 524:A71, 2010.

\bibitem{akiyama2019firstL6}
Kazunori Akiyama, Antxon Alberdi, Walter Alef, Keiichi Asada, Rebecca Azulay,
  Anne-Kathrin Baczko, David Ball, Mislav Balokovi{\'c}, John Barrett, Dan
  Bintley, et~al.
\newblock First {M87$^*$} event horizon telescope results. vi. the shadow and
  mass of the central black hole.
\newblock {\em The Astrophysical Journal Letters}, 875(1):L6, 2019.

\bibitem{abuter2022mass}
R~Abuter, N~Aimar, A~Amorim, J~Ball, M~Baub{\"o}ck, JP~Berger, H~Bonnet,
  G~Bourdarot, W~Brandner, V~Cardoso, et~al.
\newblock Mass distribution in the galactic center based on interferometric
  astrometry of multiple stellar orbits.
\newblock {\em Astronomy \& Astrophysics}, 657:L12, 2022.

\bibitem{abuter2020detection}
R~Abuter, A~Amorim, M~Baub{\"o}ck, JP~Berger, H~Bonnet, W~Brandner, V~Cardoso,
  Y~Cl{\'e}net, PT~De~Zeeuw, J~Dexter, et~al.
\newblock Detection of the schwarzschild precession in the orbit of the star s2
  near the galactic centre massive black hole.
\newblock {\em Astronomy \& Astrophysics}, 636:L5, 2020.

\bibitem{banerjee2022shadows}
Indrani Banerjee, Subhadip Sau, and Soumitra SenGupta.
\newblock Do shadows of {Sgr A$^*$} and {M87$^*$} indicate black holes with a
  magnetic monopole charge?
\newblock {\em arXiv preprint arXiv:2207.06034}, 2022.

\bibitem{Amelino-Camelia:2008aez}
Giovanni Amelino-Camelia.
\newblock {Quantum-Spacetime Phenomenology}.
\newblock {\em Living Rev. Rel.}, 16:5, 2013.

\bibitem{Casana:2017jkc}
R.~Casana, A.~Cavalcante, F.~P. Poulis, and E.~B. Santos.
\newblock {Exact Schwarzschild-like solution in a bumblebee gravity model}.
\newblock {\em Physical Review D}, 97(10):104001, 2018.

\bibitem{Yang:2023wtu}
Ke~Yang, Yue-Zhe Chen, Zheng-Qiao Duan, and Ju-Ying Zhao.
\newblock {Static and spherically symmetric black holes in gravity with a
  background Kalb-Ramond field}.
\newblock {\em Physical Review D}, 108(12):124004, 2023.

\bibitem{Goldstein:2002}
Herbert Goldstein, Charles Poole, and John Safko.
\newblock {\em {Classical Mechanics}}.
\newblock Addison-Wesley, San Francisco, USA, 3rd edition, 2002.

\bibitem{dsasdas}
SB~Lambert and Chr Le~Poncin-Lafitte.
\newblock Improved determination of $\gamma$ by vlbi.
\newblock {\em Astronomy \& Astrophysics}, 529:A70, 2011.

\bibitem{Shapiro:1964uw}
Irwin~I. Shapiro.
\newblock {Fourth Test of General Relativity}.
\newblock {\em Physical Review Letters}, 13:789--791, 1964.

\bibitem{Wang:2024fiz}
Rui-Bo Wang, Shi-Jie Ma, Jian-Bo Deng, and Xian-Ru Hu.
\newblock {Estimating the strength of Lorentzian distribution in
  non-commutative geometry by solar system tests}.
\newblock 11 2024.

\bibitem{Bertotti:2003rm}
B.~Bertotti, L.~Iess, and P.~Tortora.
\newblock {A test of general relativity using radio links with the Cassini
  spacecraft}.
\newblock {\em Nature}, 425:374--376, 2003.

\bibitem{Will:2014kxa}
Clifford~M Will.
\newblock The confrontation between general relativity and experiment.
\newblock {\em Living reviews in relativity}, 17(1):1--117, 2014.

\bibitem{araujo2023thermodsddynamical}
A.~A Ara{\'u}jo~Filho, J~Furtado, J.~A. A.~S Reis, and J.~E.~G Silva.
\newblock Thermodynamical properties of an ideal gas in a traversable wormhole.
\newblock {\em Classical and Quantum Gravity}, 40(24):245001, 2023.

\bibitem{furtado2023thermal}
J~Furtado, H~Hassanabadi, J.~A. A.~S Reis, et~al.
\newblock Thermal analysis of photon-like particles in rainbow gravity.
\newblock {\em arXiv preprint arXiv:2305.08587}, 2023.

\bibitem{araujo2022does}
A.~A Ara{\'u}jo~Filho and J.~A. A.~S Reis.
\newblock How does geometry affect quantum gases?
\newblock {\em International Journal of Modern Physics A}, 37(11n12):2250071,
  2022.

\bibitem{filho2025modified}
A.~A Ara{\'u}jo~Filho, J.~A. A.~S Reis, and Ali {\"O}vg{\"u}n.
\newblock Modified particle dynamics and thermodynamics in a traversable
  wormhole in bumblebee gravity.
\newblock {\em The European Physical Journal C}, 85(1):83, 2025.

\bibitem{araujo2025particle}
A.~A Ara{\'u}jo~Filho.
\newblock Particle creation and evaporation in kalb-ramond gravity.
\newblock {\em Journal of Cosmology and Astroparticle Physics}, 2025(04):076,
  2025.

\bibitem{araujo2025particleasasa}
A.~A Ara{\'u}jo~Filho.
\newblock Particle production induced by a lorentzian non--commutative
  spacetime.
\newblock {\em Annals of Physics}, page 170167, 2025.

\end{thebibliography}
	\bibliographystyle{unsrt}
	
\end{document}